\documentclass[11pt,a4paper]{article}
\usepackage{amssymb} % Mathematical symbols
\usepackage[a4paper, left=1in, right=1in, top=1in, bottom=1in, includehead, includefoot, foot = 0pt, head=0pt]{geometry} % Margins
\parskip=\medskipamount % Spacing between paragraphs
\usepackage{booktabs} % Tables
\usepackage{enumitem} % Enumerating lists
\usepackage[defaultcolor=black]{changes}

\usepackage[natbibapa]{apacite}
\defcitealias{camsdata}{INERIS et.~al, 2022}
\defcitealias{nasadataset}{CIESIN et.~al, 2018}
\defcitealias{meteoseasons}{NCEI, 2023}
\defcitealias{cds}{CDS, 2020}
\defcitealias{cds2}{CDS, 2021}
\usepackage{siunitx}
\usepackage[pdftex,colorlinks,citecolor=blue,urlcolor=blue,linkcolor=red]{hyperref} 
\PassOptionsToPackage{hyphens}{url}
\usepackage{xurl}
\usepackage{etoolbox}
\robustify\itshape
\robustify\bfseries

\usepackage{doi}

\usepackage{graphicx} % Figures
\usepackage[small,bf,hang]{caption}	% Caption labels
\usepackage{subcaption}
\usepackage{nicefrac}
\usepackage{tabularx}
\usepackage[export]{adjustbox}
\usepackage{afterpage}

\usepackage{amssymb} % Mathematical symbols
\usepackage{amsmath} % Math extras
\usepackage{accents}
\usepackage{bbm} % \mathbb for digits
\usepackage{amsbsy} % Bold symbols
\usepackage{array}
\usepackage{multirow} % Merge rows in tables
\usepackage[dvipsnames]{xcolor} % Extra colors
\usepackage{mathtools}
\usepackage{eurosym}

\usepackage{floatrow}
\newfloatcommand{capbtabbox}{table}[][\FBwidth]

\usepackage[ruled]{algorithm2e} % Algorithms
\SetAlCapFnt{\small} % Change the caption label font for the algorithms
\SetKw{KwTo}{in}
\definecolor{alizarin}{rgb}{0.82, 0.1, 0.26}

\usepackage{authblk} 
\usepackage{footnote}

\usepackage{fancyhdr}
\bibpunct[, ]{(}{)}{;}{a}{,}{,}

\usepackage{tikz,pgfplots}
\usetikzlibrary{fit, positioning, chains,shapes.arrows,fit}
\usetikzlibrary{shapes, arrows, arrows.meta}
\usetikzlibrary{decorations.pathreplacing}
\usetikzlibrary{calc} % for coordinate calculation
\usetikzlibrary{matrix} % Matrix library for the x-val scheme
\usepackage{pgf}
\usetikzlibrary{decorations.text}
\tikzset{cross/.style={cross out, draw=black, minimum size=2*(#1-\pgflinewidth), inner sep=0pt, outer sep=0pt},
cross/.default={1pt}}
\usetikzlibrary{shapes.misc}
\usepackage{environ}
\makeatletter
\newsavebox{\measure@tikzpicture}
\NewEnviron{scaletikzpicturetowidth}[1]{%
  \def\tikz@width{#1}%
  \begin{lrbox}{\measure@tikzpicture}%
  \BODY
  \end{lrbox}%
  \pgfmathparse{#1/\wd\measure@tikzpicture}%
  \BODY
}
\makeatother
\tikzdeclarecoordinatesystem{page}{
    \parsecomma#1\endparsecomma
    \pgfpointanchor{current page}{north east}
    \pgf@xc=\pgf@x%
    \pgf@yc=\pgf@y%
    \pgfpointanchor{current page}{south west}
    \pgf@xb=\pgf@x%
    \pgf@yb=\pgf@y%
    \pgfmathparse{(\pgf@xc-\pgf@xb)/2.*\page@x+(\pgf@xc+\pgf@xb)/2.}
    \expandafter\pgf@x\expandafter=\pgfmathresult pt
    \pgfmathparse{(\pgf@yc-\pgf@yb)/2.*\page@y+(\pgf@yc+\pgf@yb)/2.}
    \expandafter\pgf@y\expandafter=\pgfmathresult pt
}
\makeatother
\usetikzlibrary{arrows,automata}
\usepackage{ragged2e}

\tikzset{
    max width/.style args={#1}{
        execute at begin node={\begin{varwidth}{#1}},
        execute at end node={\end{varwidth}}
    }
}
\definecolor{newblue}{RGB}{86, 180, 233}
\definecolor{newred}{RGB}{248, 118, 109}
\definecolor{newgreen}{RGB}{163, 213, 150}

\colorlet{textcolor}{white}% color for the text inside the circles
\colorlet{bordercolor}{white}% color for the outer border of circles

\pgfdeclarelayer{background}
\pgfsetlayers{background,main}

\definecolor{airforceblue}{rgb}{0.36, 0.54, 0.66}
\definecolor{forestgreen}{rgb}{0.13, 0.55, 0.13}\definecolor{fulvous}{rgb}{0.86, 0.52, 0.0}
\definecolor{gray}{rgb}{0.5, 0.5, 0.5}
\definecolor{bistre}{rgb}{0.24, 0.17, 0.12}\definecolor{bostonuniversityred}{rgb}{0.8, 0.0, 0.0}
\definecolor{purpleheart}{rgb}{0.41, 0.21, 0.61}
\definecolor{lightsalmonpink}{rgb}{1.0, 0.6, 0.6}\definecolor{arrowcolor}{rgb}{0.92, 0.92, 0.92}
\tikzset{
inner/.style={
  on chain,
  circle,
  inner sep=4pt,
  fill=circlecolor,
  line width=1.5pt,
  draw=bordercolor,
  text width=1.2em,
  align=center,
  text height=1.25ex,
  text depth=0ex
},
on grid
}

\usepackage{regexpatch}
\allowdisplaybreaks

\usepackage{pbox}
\newcommand\drawarrow{% the arrow is placed in the background layer 

\node[on chain] (f) {};
\begin{pgfonlayer}{background}
\node[
  inner sep=10pt,
  single arrow,
  single arrow head extend=0.6cm,
  draw=none,
  fill=arrowcolor,
  fit= (c1) (f)
] (arrow) {};
\fill[white] % the decoration at the tail of the arrow
  (arrow.before tail) -- (c1|-arrow.west) -- (arrow.after tail) -- cycle;
\end{pgfonlayer}
}

\numberwithin{equation}{section}
\usepackage{relsize}

\makeatletter
\NAT@longnamesfalse
\makeatother
\pgfplotsset{compat=1.18}
\begin{document}
\title{A penalized distributed lag non-linear Lee-Carter framework for regional weekly mortality forecasting}
\author[1,*]{Jens Robben}
\author[2]{Karim Barigou}
\affil[1]{Research Centre for Longevity Risk (RCLR), Faculty of Economics and Business, University of Amsterdam, Amsterdam, the Netherlands.}
\affil[2]{Institute of Statistics, Biostatistics and Actuarial Science (ISBA), Louvain Institute of Data Analysis and Modeling (LIDAM), UCLouvain, Louvain-la-Neuve, Belgium }

\affil[*]{Corresponding author: \href{mailto:j.robben@uva.nl}{j.robben@uva.nl}}

\maketitle
\begin{abstract}
Accurate forecasts of weekly mortality are essential for public health and the insurance industry. We develop a forecasting framework that extends the Lee–Carter model with age- and region-specific seasonal effects and penalized distributed lag non-linear components that capture the delayed and non-linear effects of heat, cold, and influenza on mortality. The model accommodates overdispersed mortality rates via a negative binomial distribution. We model the temporal dynamics of the latent factors in the model using SARIMA processes and capture cross-regional dependencies through a copula-based approach. Using regional French mortality data (1990–2019), we demonstrate that the proposed framework yields well-calibrated forecast distributions and improves predictive accuracy relative to benchmark models. The results further show substantial heterogeneity in temperature- and influenza-related relative risks between ages and regions. These findings underscore the importance of incorporating exogenous drivers and dependence structures into a weekly mortality forecasting framework.
\end{abstract}

\textbf{Keywords}: stochastic mortality modeling; seasonal mortality; distributed lag non-linear models; excess mortality

\section{Introduction}
Life insurers, pension funds, and health care providers generally rely on mortality tables constructed from national vital statistics aggregated at an annual level \citep{wilmoth2007methods}. However, since the emergence of the COVID-19 pandemic, more granular mortality data have become available worldwide \citep{wang2022estimating}. These new data sources allow for extensions of traditional actuarial mortality models to capture finer temporal and geographical variation. Mortality rates show, for example, a seasonal pattern within a year, with higher rates in winter than in summer \citep{madaniyazi2022seasonal, pavia2022estimation}. Deviations from these seasonal patterns can occur due to various exogenous events such as heat waves, cold spells, or epidemic outbreaks \citep{gasparrini2011impact, gasparrini2015mortality}. Therefore, more fine-grained models are valuable not only for insurers to better price short-term life-contingent insurance products, but also for public health authorities, who can use them to assess and monitor excess mortality in near real time \citep{karlinsky2021tracking, fernandez2015seasonal}. 

Building on the well-known Lee–Carter framework for annual, country-level data \citep{lee1992modeling}, we introduce a stochastic mortality projection model that operates at a much finer spatiotemporal resolution: weekly, regional, and age-specific. Our proposed model captures both seasonal mortality patterns and short-term deviations from it, driven by extreme temperatures and influenza outbreaks. To achieve this, we embed distributed lag non-linear models (DLNMs) within the framework \citep{gasparrini2010distributed}, which enables the estimation of age-specific, delayed, and non-linear effects of these shocks on seasonal mortality. By integrating DLNMs into our model, our framework provides a flexible and interpretable tool to disentangle seasonal baselines from weather- and epidemic-related excess mortality across regions and age groups. Furthermore, the proposed model retains the ability to produce stochastic mortality projections comparable to those of traditional stochastic mortality models.

The study of seasonal mortality and the influence of exogenous factors such as heatwaves, cold spells, and influenza outbreaks has recently gained increasing attention in the actuarial literature. To our knowledge, \citet{seklecka2017mortality} is among the first to incorporate a temperature-related factor into the Lee–Carter mortality model. Their study, based on grouped monthly and annual data from the United Kingdom, shows improved predictive accuracy and applies the model to life insurance pricing. However, the use of aggregated data prevents the model from capturing regional heterogeneity or fine-scale temporal patterns. Next, \citet{li2022joint} apply concepts from the extreme value theory to investigate the dependence between monthly temperature extremes and mortality at the regional level in the United States. They find a particularly strong dependence between cold-weather extremes and mortality at older ages. However, their focus lies on describing extremal dependence rather than developing a stochastic mortality forecasting framework. Therefore, more recently, \citet{guibert2025impact} extend the Li–Lee multi-population model \citep{li2005coherent} by integrating a distributed lag non-linear model to capture non-linear and delayed temperature effects. They also provide mortality projections under alternative representative concentration pathway (RCP) scenarios. Although this constitutes one of the first actuarial attempts to link mortality forecasts to climate scenarios, their projections remain at the annual and country-specific level. Building upon this line of research, \citet{min2025assessing} combine single- and multi-population stochastic mortality models with a DLNM to disentangle long-term mortality trends from non-linear and lagged climate effects. Parallel to these developments, \citet{beginmodelling} propose an extension of the age–period–cohort (APC) model with a seasonal component based on periodic cubic splines. This allows them to generate mortality forecasts on the daily scale while accounting for seasonal fluctuations. Although this represents a step toward fully stochastic seasonal mortality projections, their framework does not incorporate exogenous factors or a multi-population structure.

The epidemiological and environmental health literature has repeatedly demonstrated that DLNMs offer a flexible and interpretable way to model short-term, non-linear relationships between temperature and mortality. For example, \cite{gasparrini2015mortality} estimate that 7.7$\%$ of deaths can be attributed to suboptimal temperatures, based on 74 million deaths in 384 locations in 13 countries. \cite{guo2017heat} also show that heat waves significantly increased mortality at higher temperature thresholds in more than 400 regions in 18 countries. In addition to temperature, DLNMs have also been applied to model the association between infectious diseases and mortality: \cite{li2023influenza} showed that influenza caused significant excess mortality in Guangzhou, mainly due to respiratory and cardiovascular deaths among the elderly. Although DLNMs capture associations with a single risk factor, they are rarely integrated into a broader mortality modeling framework that can jointly account for demographic structure, temporal trends, and spatial dependence. Moreover, most applications in the epidemiological literature do not use forward-looking stochastic mortality forecasts and are often estimated using age-aggregated data, which is not suitable for actuarial applications.

\added{We contribute to the literature in several ways. First, unlike existing temperature-only DLNM-based stochastic mortality models \citep{guibert2025impact, min2025assessing}, we explicitly incorporate influenza as a second exogenous driver of short-term mortality fluctuations, alongside temperature. As we demonstrate in our case study, this leads to substantial improvements in predictive performance, both in-sample and out-of-sample. Second, in the case study, we collect and analyze a rich data set of individual death counts data from the National Institute of Statistics and Economic Studies (INSEE), that cover all metropolitan regions of France from 1990 to 2019. Previous studies typically rely on shorter observation periods (e.g., 2015–2019 in \cite{min2025assessing}) or on a single region (e.g.,
Quebec in \cite{beginmodelling}). Because public health authorities generally require regional-level assessments of excess mortality for the entire nation, our collected data set, methodology, and application provide a framework for doing so. Third, our baseline mortality model specification differs fundamentally from existing models: we model a separate, explicit weekly seasonal effect with age-specific sensitivity rather than allowing the latent mortality index to absorb within-year variation. Our proposed model is therefore more parsimonious and avoids estimating a separate parameter for each (week, year), which would introduce hundreds of additional parameters for our case study. Fourth, we make a methodological contribution by creating similarity in the estimated temperature and influenza effects between adjacent regions. Specifically, we integrate a Laplacian spatial penalty matrix into the log-likelihood of our model, which enforces the smoothing of effects across regions that share borders. This spatial regularization step makes
the model more robust by drawing knowledge from geographically close regions. Fifth, from a more technical point of view, we fit our model simultaneously across all ages and regions within a single, coherent negative binomial Lee–Carter framework, which improves upon existing approaches that estimate DLNM effects separately for each age and often do not account for overdispersion. Lastly, we quantify parameter uncertainty through a parametric bootstrap procedure and dempose out-of-sample predictive uncertainty by separating the contribution of simulated time-series variation from negative binomial sampling variability. This provides insight into how much forecast uncertainty is driven by the latent mortality index and, when covariates are forecasted, by temperature and influenza dynamics.}

The structure of this paper is as follows. Section~\ref{sec:data} sets out the notation and provides a detailed description of the data sources, regarding mortality, population, temperature, and influenza surveillance, accompanied by an initial exploratory analysis. Section~\ref{sec:modelspecification} develops the proposed mortality modeling framework, which is specified at the weekly, regional, and age-specific level, and is designed to capture seasonal variation, as well as the impact of heat waves, cold spells, and influenza outbreaks. Section~\ref{sec:calibrationandforecasting} details the calibration strategy based on the maximum likelihood estimation, the procedure to generate forecasts, and the evaluation of the uncertainty of the parameters. Section~\ref{sec:casestudy} presents a comprehensive case study to illustrate the application of the methodology to the data introduced. Section~\ref{sec:conclusion} concludes with a summary of the main findings.

\section{Notations and data} \label{sec:data}
\subsection{Notations} \label{subsec:noations}
Let $d_{x,t,w,r}$ denote the weekly death count in region $r$ for age group $x$ at week $w$ of year $t$. The time points $(t,w)$ follow the ISO 8601 standard, where each ISO week is exactly 7 days, and ISO week 1 starts on the Monday of the week that contains the first Thursday of the year. We define the following sets: \( \mathcal{R} \) is the set of all regions, \( \mathcal{X} \) the set of age groups, and \( \mathcal{T} = \{t_{\min}, t_{\min +1}, \dots, t_{\max}\} \) the range of all years under consideration. \added{For each year $t$, the set of weeks is given by \( \mathcal{W}_t = \{1, 2, \dots, W_t\} \), where $W_t \in \{52,53\}$. Under the ISO 8601 standard, a year $t$ contains 53 weeks if and only if it either (i) starts on a Thursday, or (ii) is a leap year that starts on a Wednesday; otherwise, it contains 52 weeks.}

Exposure to risk \( E_{x,t,w,r} \) represents the total number of person years lived by people aged \( [x, x + 1) \) during week \( w \) of year \( t \) in region $r$. The observed weekly death rate is then given by:
\begin{equation} \label{eq:deathrate}
    m_{x,t,w,r} = \frac{d_{x,t,w,r}}{E_{x,t,w,r}}.
\end{equation}
We refer to $\mu_{x,t,w,r}$ as the weekly force of mortality for people aged $x$ in week $w$ of year $t$ in region $r$. This quantity represents the instantaneous risk of death at $(x,t,w,r)$. Under the assumption of constant force of mortality within a week interval, $\mu_{x,t,w,r}$ can be approximated by the observed weekly death rate $m_{x,t,w,r}$. For ease of notation, gender is omitted from the notation, but all methods can be applied to both genders.

\subsection{Mortality data} \label{subsec:mortalitydata}
\paragraph{Death counts.} We use individual-level French death records from the National Institute of Statistics and Economic Studies (INSEE).\footnote{The database is available at \url{https://www.insee.fr/fr/information/4769950}.} Each record contains information on the name, gender, commune of birth, commune of death, date of birth, date of death, and death certificate number. We extract death records spanning from 1970 to the first quarter of 2025, resulting in approximately 25.5 million records.

We perform the following data preprocessing steps: (1) remove duplicate entries, (2) exclude records with missing birth or death dates, (3) exclude records where the date of death precedes the date of birth, and (4) impute missing or misspecified months or days of death by sampling from the empirical distribution of deaths by month and day. \added{The fraction of imputed death records is very small (376 out of approximately 20 million observations, i.e., $<0.002\%$), and will therefore only have a negligible impact on the seasonality}. Next, we aggregate the cleaned death records by 5-year age group, ISO year and week, administrative region of death, and gender. To do so, we first map the commune of death to its corresponding department by extracting the first two digits of the reported postal code. We then map departments to administrative regions using the matching table provided by INSEE.\footnote{The matching table is available at \url{https://www.insee.fr/fr/information/7766585}.}

\begin{figure}[ht!]
\centering
\begin{subfigure}{0.95\textwidth}
\centering
\includegraphics[width = \textwidth]{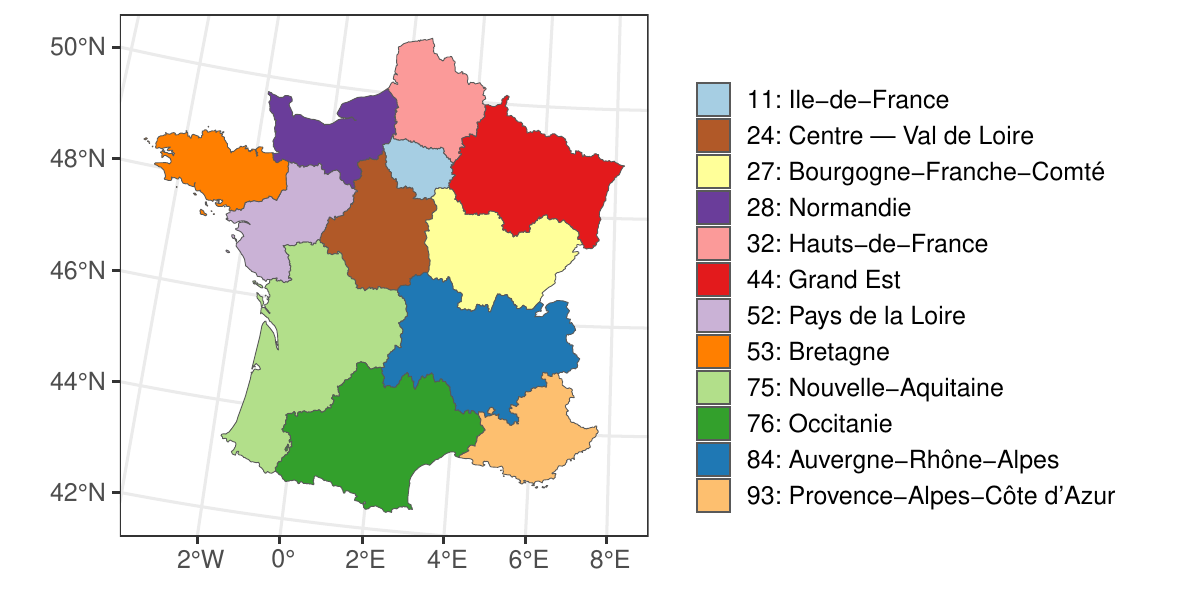}
\end{subfigure}
\hspace{0.1cm}
\caption{The 12 administrative regions in Metropolitan France. \label{fig:regions}}
\end{figure}

\begin{figure}[!htb]
    \centering
    \includegraphics[width=0.95\linewidth]{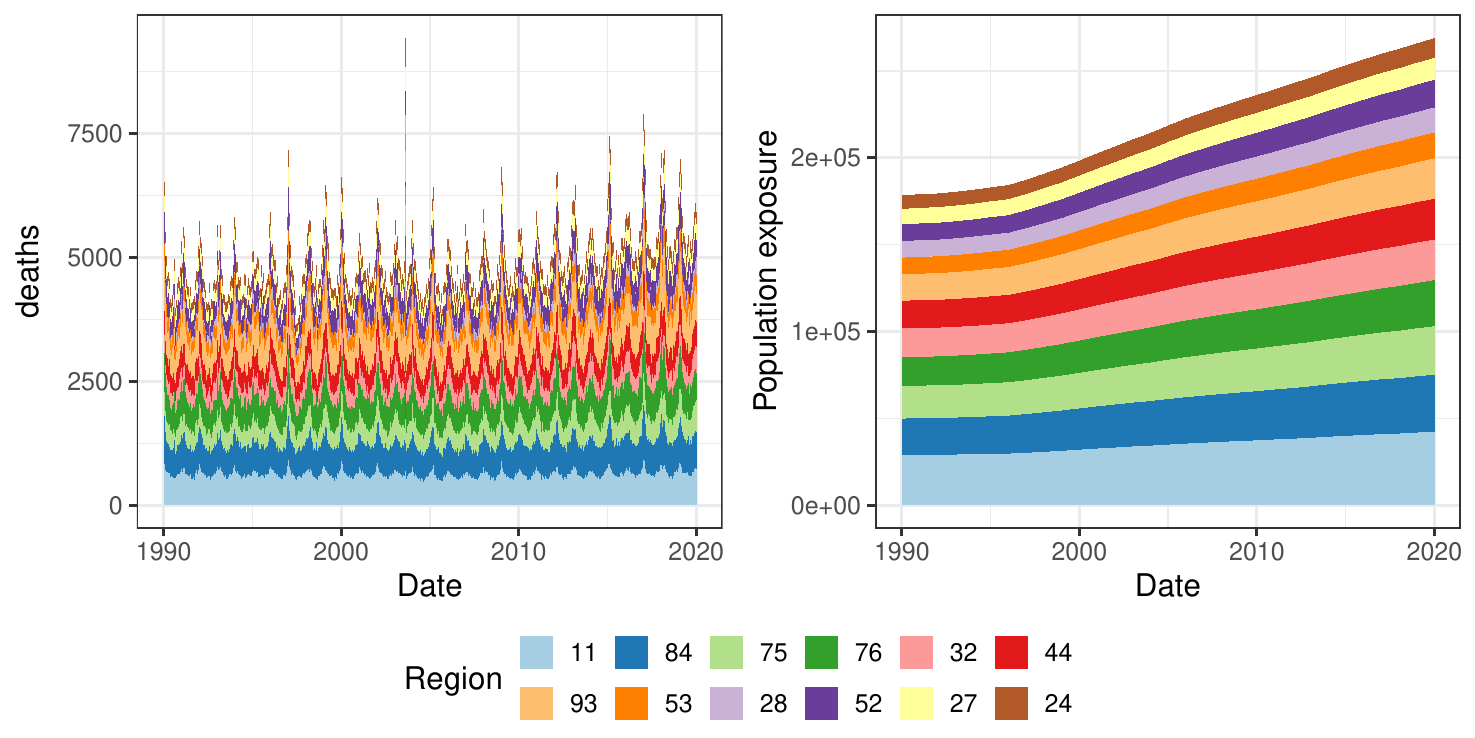}
    \caption{\added{Stacked female death counts (left panel) and exposures (right panel) per week for the twelve French administrative regions over the years 1990-2019. The weekly death counts and exposures per region are aggregated across the different age groups. \label{fig:overview}}}
\end{figure}

In this paper, we examine the 30-year period from 1990 to 2019, and focus on five-year age groups beginning at age 50 and including an open-ended 95+ age category. Our analysis covers the twelve administrative regions of Metropolitan France, excluding Corsica. Figure~\ref{fig:regions} displays these twelve regions, along with their respective geographic INSEE codes and names. The left panel of Figure~\ref{fig:overview} shows the stacked female death counts for the 12 administrative regions, aggregated among the different age groups considered.

Next, we analyze the overdispersion in our data set of weekly death counts across different regions and age groups. Compared to annual country-level death statistics typically used in actuarial mortality models, our data exhibit significantly greater overdispersion. Table~\ref{tab:dispersion} presents the variance-to-mean ratio, also called the dispersion index, for each region and age group in this study. The results show that overdispersion is consistently present (index $>$ 1) in all strata. Although younger age groups exhibit relatively mild overdispersion, the dispersion index increases substantially for older age groups. Furthermore, the dispersion index varies considerably by region, with southern regions generally exhibiting higher overdispersion compared to northern ones, except for Île-de-France. This regional variability may be related to the increased climate variability in the southern French regions. %\ali{Table 1 is really nice and informative. Regarding overdispersion, there is the interesting paper "Bayesian mortality forecasting with overdispersion" from Wong et al published in IME. They compare the Poisson and NB Lee-Carter model, and prove some statistical tests based on Pearson residuals. It could be interesting to compare weekly lee-carter model with and without overdispersion, and provide similar test statistics.}

\begin{table}[!ht]
\centering
\resizebox{\textwidth}{!}{%
\begin{tabular}{lcccccccccccc}
  \toprule
\footnotesize{\textbf{Age} $\backslash$ \textbf{Region}} & \textbf{11} & \textbf{24} & \textbf{27} & \textbf{28} & \textbf{32} & \textbf{44} & \textbf{52} & \textbf{53} & \textbf{75} & \textbf{76} & \textbf{84} & \textbf{93} \\ 
  \midrule
\textbf{50-54} & 1.28 & 1.15 & 1.09 & 1.22 & 1.30 & 1.15 & 1.10 & 1.11 & 1.29 & 1.21 & 1.13 & 1.11 \\ 
\textbf{55-59} & 1.32 & 1.09 & 1.16 & 1.17 & 1.40 & 1.23 & 1.12 & 1.18 & 1.36 & 1.46 & 1.28 & 1.18 \\ 
\textbf{60-64} & 1.41 & 1.15 & 1.11 & 1.24 & 1.53 & 1.41 & 1.29 & 1.26 & 1.40 & 1.42 & 1.33 & 1.22 \\ 
\textbf{65-69} & 1.86 & 1.31 & 1.37 & 1.33 & 2.03 & 1.80 & 1.38 & 1.37 & 1.70 & 1.56 & 1.68 & 1.50 \\ 
\textbf{70-74} & 2.55 & 1.48 & 1.68 & 1.55 & 2.85 & 2.31 & 1.40 & 1.83 & 2.06 & 1.92 & 2.31 & 1.87 \\ 
\textbf{75-79} & 3.84 & 1.90 & 2.08 & 1.87 & 3.77 & 3.00 & 1.91 & 2.34 & 2.92 & 2.66 & 3.17 & 2.69 \\ 
\textbf{80-84} & 5.64 & 2.17 & 2.47 & 2.18 & 3.90 & 3.55 & 2.14 & 2.46 & 3.88 & 3.25 & 4.01 & 2.83 \\ 
\textbf{85-89} & 6.35 & 2.95 & 3.45 & 3.08 & 5.62 & 5.45 & 3.69 & 3.74 & 5.42 & 4.96 & 6.26 & 4.32 \\ 
\textbf{90-94} & 8.50 & 4.36 & 4.80 & 5.08 & 7.55 & 8.19 & 6.15 & 7.22 & 9.19 & 9.06 & 10.32 & 7.32 \\ 
\textbf{95+} & 8.41 & 4.34 & 4.78 & 5.17 & 5.73 & 6.58 & 6.15 & 6.71 & 9.83 & 9.29 & 9.83 & 8.49 \\ 
  \bottomrule
\end{tabular}}
\caption{\added{Dispersion index of the weekly female death counts per considered age group and administrative region in Metropolitan France.}}
\label{tab:dispersion}
\end{table}

\paragraph{Population exposure.} To compute weekly population exposures by age, sex, and region, we first obtain annual population counts $P_{x,t,r}$ from INSEE.\footnote{We obtain the annual population counts from the following database: \url{https://www.insee.fr/fr/statistiques/1893198}.} These counts are recorded on January 1st of each year $t$, for each age $x$ and administrative region $r$, and we omit the dependence on sex to ease notation. To estimate population counts at a weekly time scale, we perform linear interpolation between subsequent years as follows:
\begin{equation*}
\tilde{P}_{x,t,w,r} = P_{x,t,r} + \frac{\text{days}(\text{date}_t, \text{date}_{t,w})}{\text{days}(\text{date}_t, \text{date}_{t+1})} \left( P_{x,t+1, r} - P_{x, t, r} \right),
\end{equation*}
where $\text{date}_t$ denotes January 1st of year $t$, $\text{date}_{t,w}$ denotes the start date of the $w$-th ISO week in year $t$, and $\text{days}(\cdot, \cdot)$ is the function that calculates the number of days between two given dates. We then compute weekly exposures by taking the midpoint:
\begin{equation*}
E_{x,t,w, r} = \frac{1}{2\cdot 52.18} \left( \tilde{P}_{x,t,w,r} + \tilde{P}_{x,t,w+1,r} \right).
\end{equation*}
where the time point $(t,w+1)$ corresponds to $(t+1,1)$ if the former time point does not exist. The division by the average number of weeks in a year ensures that the weekly exposure measure is expressed in units of person-years. The right panel of Figure~\ref{fig:overview} shows the stacked female exposures for the 12 administrative regions of metropolitan France over the years 1990-2019, aggregated across the different age groups considered. Due to linear interpolation, the weekly exposure estimates vary smoothly over the different weeks. This approach improves existing methods, which, for example, often assume that weekly exposures remain constant throughout the year \citep{stmfnote}. %\ali{I would be nice to add one or two references regarding this interpolation technique. I think you already used in some of your previous papers.}

\begin{figure}[!htb]
    \centering
    \includegraphics[width=0.95\linewidth]{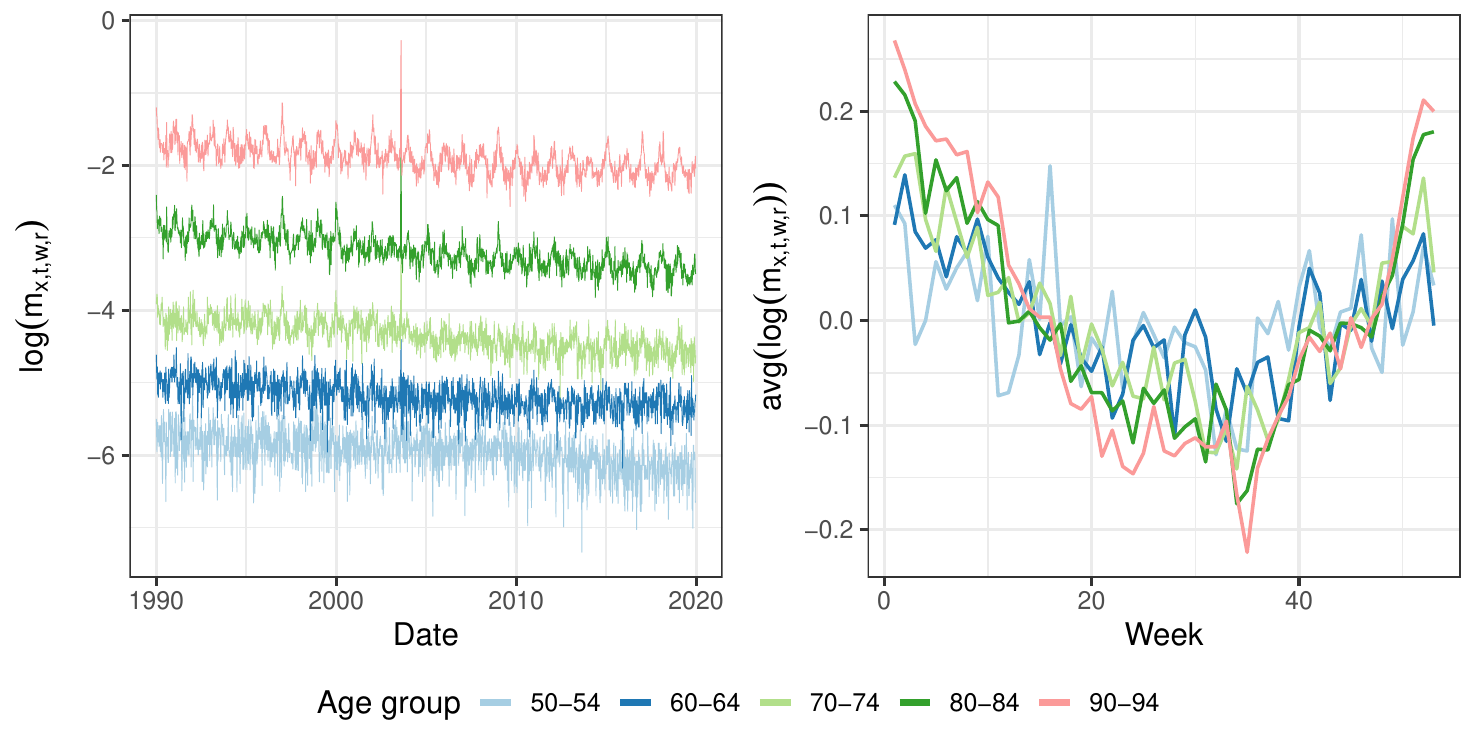}
    \caption{\added{Left panel: the logarithm of the weekly death rates for females in the administrative region of Ile-de-France for 5 different age groups, 50-54, 60-64, 70-74, 80-84, and 90-94, across the years 1990-2019. Right panel: the average of the log death rates per week across the years 1990-2019, centered around zero for comparability reasons. \label{fig:deathrate}}
    \label{fig:enter-label}}
\end{figure}

\paragraph{Death rates.} We compute the death rates following the formula in Eq.~\eqref{eq:deathrate}. Figure~\ref{fig:enter-label} illustrates the weekly death rates, on the logarithmic scale, for women in the administrative region of Île-de-France during the period 1990-2019. The left panel displays the log death rates for five selected age groups (50-54, 60-64, 70-74, 80-84, and 90-94) over time and reveals clear differences in the mortality level between ages. The severe excess mortality in the summer of 2003 corresponds to the 2003 European heatwave that particularly hit France. The right panel shows the average log death rate per week, aggregated over all years and centered around zero. This panel highlights the seasonal pattern with higher mortality during the winter weeks and a trough in the summer. Here, it is also clear that the seasonal pattern is more pronounced for the older age groups. 

\subsection{Temperature data} \label{subsec:temperaturedata}
\begin{figure}[!htb]
    \centering
    \includegraphics[width=0.95\linewidth]{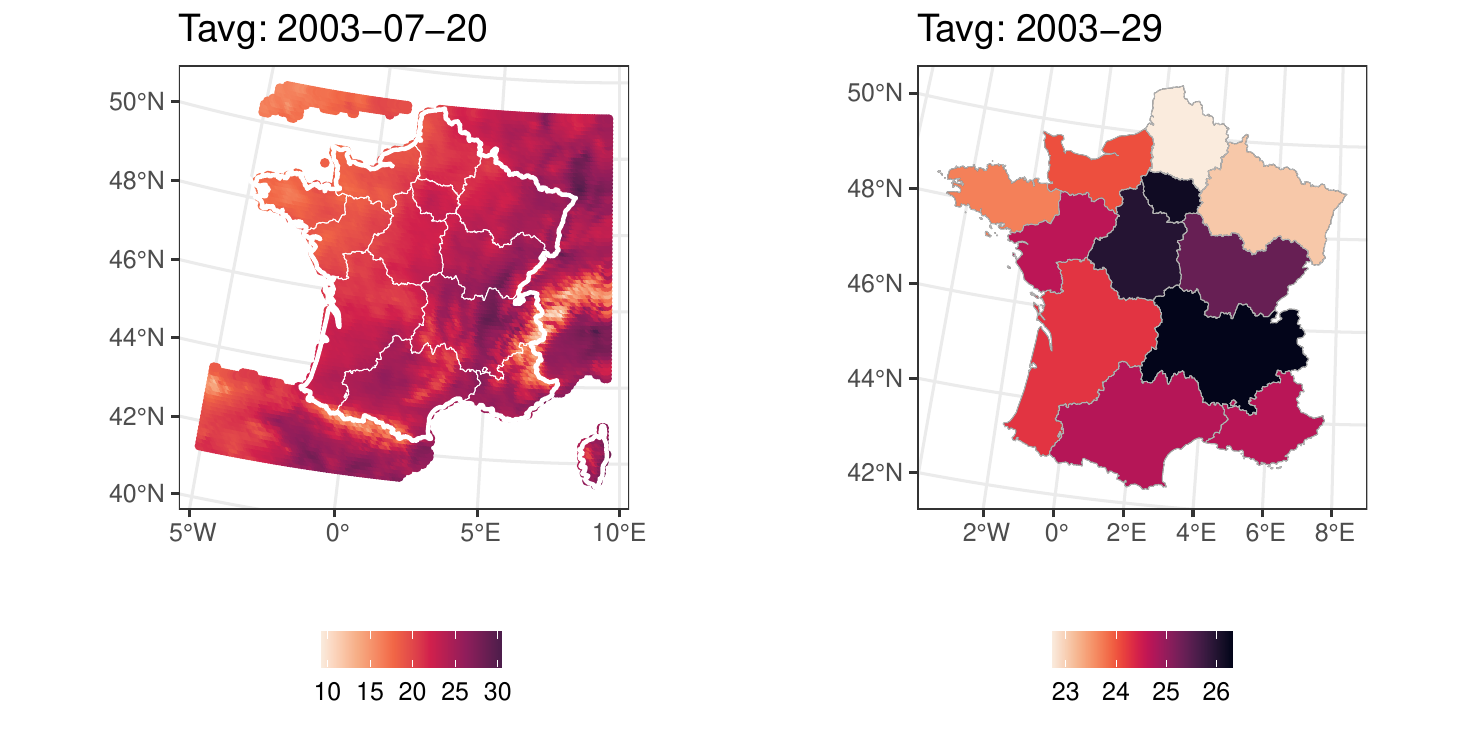}
    \caption{Left panel: the daily average temperature on a spatial grid covering metropolitan France on July 20, 2003. Right panel: the weekly average of the daily average temperature values in the 29th ISO week of the year 2003 per administrative region. \label{fig:tempdata}}
\end{figure}

We obtain temperature data from the E-OBS dataset, a land-only, gridded meteorological dataset available on the Copernicus Climate Data Store \citepalias{cds}. This data set provides daily weather variables for Europe with high spatial resolution (0.1° × 0.1°, ~10 km), and is derived from interpolated observations from a large network of European meteorological stations. Following the approach of \cite{gasparrini2015mortality}, we focus on the daily average temperature as our exposure metric. \added{Alternative measures (e.g., maximum or minimum temperature) have not been consistently shown to improve the prediction of mean mortality \citep{barnett2010measure}.} Henceforth, we extract daily average temperatures across a spatial grid that encompasses metropolitan France from 1990 to 2019. For reference, the left panel of Figure~\ref{fig:tempdata} illustrates the spatial distribution of temperatures on a randomly selected date during this period.

Since the mortality data in Section~\ref{subsec:mortalitydata} is defined at the level of administrative regions, we aggregate the daily temperature data into population-weighted regional averages using gridded population data \citepalias{nasadataset}. This weighting principle is described in \cite{robben2025short} in more detail and ensures that temperature values better represent areas with a higher population density within each region. We then calculate the weekly average of these daily temperature values to match the temporal resolution of the mortality data. \added{In Suppl.~Mat.~\ref{app:temp_robustness}, we also investigate alternative aggregations beyond the weekly average, such as the weekly maximum, weekly minimum, or maximum three-day moving average, to assess whether these aggregations might better capture heat wave or cold spell effects. For the case study in Section~\ref{sec:casestudy}, we find that the weekly average temperature is sufficiently flexible to capture the non-linear and delayed effects associated with temperature extremes.} The right panel of Figure~\ref{fig:tempdata} displays the resulting processed data for a randomly selected ISO week and year.

\subsection{Influenza-like illness surveillance data}  \label{subsec:ILIdata}
We use data on influenza-like illness (ILI) from the French Sentinelles Network \citep{valleron1986computer}. This is a nationwide epidemiological surveillance system established in 1984, coordinated by the Pierre Louis Institute of Epidemiology and Public Health under Inserm and Sorbonne University. It monitors communicable diseases, including ILI, through a network of volunteer general practitioners in mainland France. In 2024, the network consisted of 135 general practitioners who report weekly data on ILI cases.

\begin{figure}[!htb]
    \centering
    \includegraphics[width=0.8\linewidth]{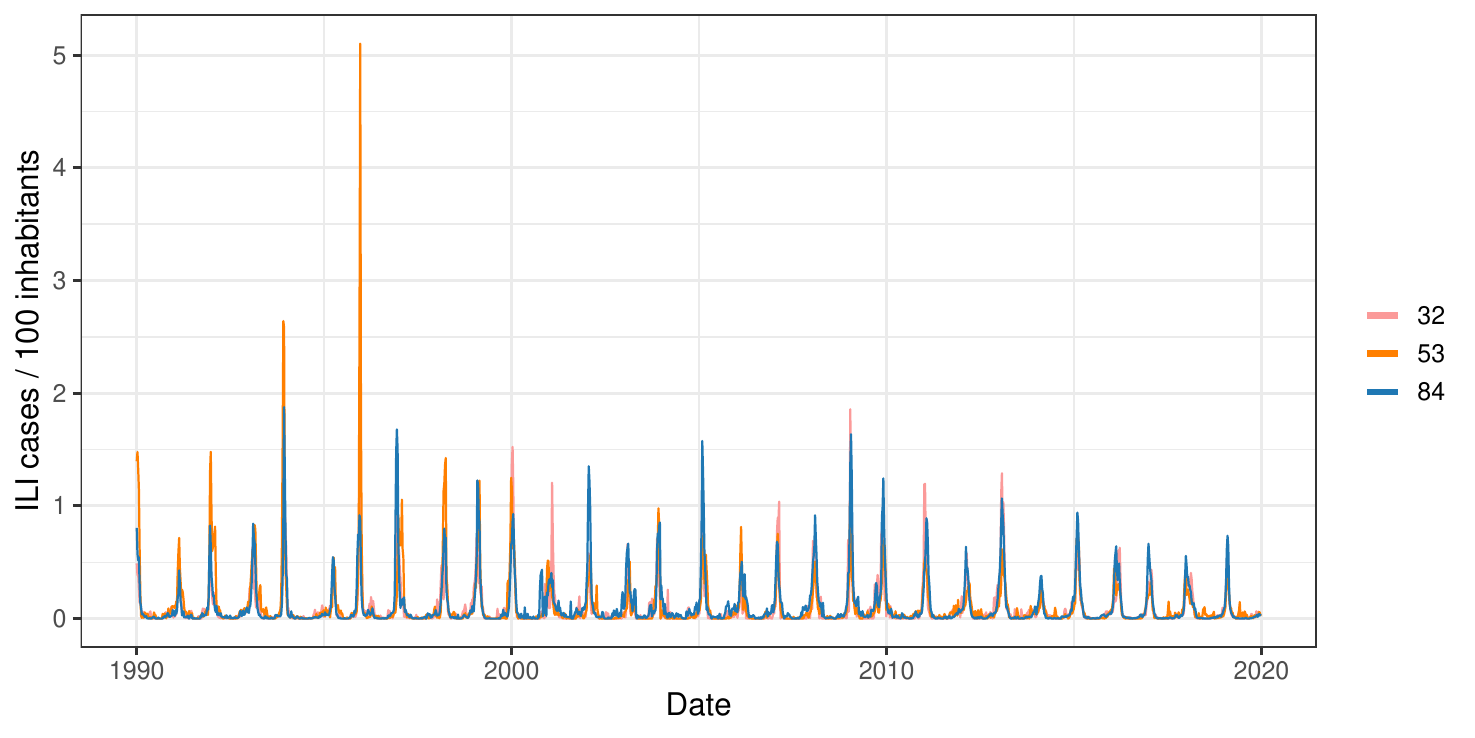}
    \caption{The weekly incidence rates of influenza-like illness in three administrative regions in France: Hauts-de-France (32), Bretagne (53), and Auvergne-Rhône-Alpes (84) from 1990-2019.  \label{fig:infldata}}
\end{figure}

We obtain the weekly incidence rate of influenza-like illness (ILI) per administrative region in France from the French Sentinelles network. These incidence rates represent the number of new cases of ILI per unit of population. Figure~\ref{fig:infldata} displays the trends in the incidence of ILI in three French administrative regions. Although the overall patterns are broadly similar, the intensity of influenza outbreaks varies significantly between regions.

\section{Model specification} \label{sec:modelspecification}
We develop a weekly, region, and age-specific mortality model designed to (1) capture seasonal variations in weekly mortality rates and (2) quantify the effects of heat waves, cold spells, and influenza outbreaks on deviations from these seasonal mortality patterns. Our proposed model extends the traditional Lee-Carter framework \citep{lee1992modeling}, typically applied to country-level annual mortality data, to a finer spatiotemporal resolution. 

\subsection{Model structure} \label{subsec:modelstructure}
To account for the overdispersion in death counts (see Table~\ref{tab:dispersion} in Section~\ref{subsec:mortalitydata}), we model the number of deaths, $D_{x,t,w,r}$, for the age group $x \in \mathcal{X}$, year $t \in \mathcal{T}$, week $w \in \mathcal{W}$, and region $r \in \mathcal{R}$, using a Negative Binomial (NB) distribution:
\begin{align} \label{eq:modeldist}
D_{x,t,w,r} \sim \text{NB}(E_{x,t,w,r} \cdot \mu_{x,t,w,r}, \phi_{x,r}),
\end{align}
where $E_{x,t,w,r}$ is the exposure to risk, $\mu_{x,t,w,r}$ is the weekly force of mortality, and $\phi_{x,r}$ is the age- and region-specific dispersion parameter that captures extra-Poisson variability within the death counts. The NB model was considered in a frequentist setting by \cite{delwarde2007negative} and in a Bayesian setting by \cite{wong2018bayesian} and \cite{barigou2023bayesian}, among others.

We express the logarithm of the weekly force of mortality as the sum of two additive components. The first, the seasonal baseline component, represents the expected seasonal mortality pattern in the absence of exogenous shocks. The second captures short-term deviations from this baseline due to extreme temperatures and influenza outbreaks. Formally, we write:
\begin{align} \label{eq:modelstructure}
\log \mu_{x,t,w,r} = \underbrace{\alpha_{x,r} + \beta_x \kappa_{t,r} + \gamma_x \lambda_{w,r}}_{\text{Seasonal baseline}} + \underbrace{\delta_x f_r^{(1)}(\text{Tavg}_{t,w,r}) + \epsilon_x f_r^{(2)}(\text{ILI}_{t,w,r})}_{\text{Deviations}}, 
\end{align}
where:
\begin{itemize}
  \item[-] $\alpha_{x,r}$ reflects the region-specific average log mortality for the age group $x$. We make this parameter region dependent to allow for overall differences in mortality between regions, e.g., due to socio-economic differences,
  \item[-] $\kappa_{t,r}$ is a region-specific mortality index that captures the overall annual mortality decline, and $\beta_x$ measures the age-specific sensitivity to this decline,
  \item[-] $\lambda_{w,r}$ is a region-specific week effect capturing periodic week-to-week variations in mortality (higher mortality in winter compared to summer), and $\gamma_x$ accounts for differences in this seasonal intensity between different age groups,
  \item[-] $f_r^{(1)}(\text{Tavg}_{t,w,r})$ is a region-specific distributed lag non-linear model (DLNM) that describes the non-linear and delayed association between the weekly average temperature $\text{Tavg}_{t,w,r}$ and mortality, and $\delta_x$ models the \added{time-independent} age-specific effect of it,
  \item[-] $f_r^{(2)}(\text{ILI}_{t,w,r})$ is a region-specific DLNM that models the non-linear and delayed association between weekly average influenza-like illness (ILI) and mortality, and $\epsilon_x$ models the \added{time-independent} age-specific impact of it.
\end{itemize}
We model the dispersion parameter on the logarithmic scale as an additive combination of age- and region-specific components, i.e.,
\begin{align} \label{eq:dispparam}
    \phi_{x,r} = \exp\left(\phi_x + \phi_r\right).
\end{align}
This decomposition enhances both the model parsimony and interpretability. We ensure identifiability in the additive composition of the dispersion parameter by imposing $\phi_{x_{\min}} = 0$, with $x_{\min} \in \mathcal{X}$ the minimum age. \added{We also explored an extension allowing for week-specific variation in the dispersion parameter, i.e., $\phi_{w,x,r} = \exp(\phi_w + \phi_x + \phi_r)$. This led to a modest increase in the log-likelihood, but the BIC deteriorated, which indicates that the additional complexity is not supported by the data.}

In summary, our proposed model extends the Lee-Carter modeling framework by adding the following:
\begin{itemize}
    \item[1.] age-specific weekly effects to capture seasonal mortality patterns,
    \item[2.] DLNMs to account for the age-specific, non-linear, and lagged effects of temperature and influenza on mortality.
\end{itemize}
Furthermore, our model is defined in a multi-population setting where we use a common age effect across different regions, as proposed in \cite{kleinow2015common}. 

\paragraph{\added{Remark on the seasonal specification.}}
\added{The proposed decomposition into separate year ($\kappa_{t,r}$) and week ($\lambda_{w,r}$) effects is considerably more parsimonious compared to a single, latent weekly mortality index (which would require 18\,780 parameters for our case study). More importantly, this structure ensures a clean separation between long-term mortality trends and recurring seasonal patterns. It also prevents $\kappa_{t,r}$ from absorbing short-term fluctuations due to extreme temperatures or influenza. As we will show in panels B and D of Figure~\ref{fig:param1} (Section~\ref{subsec:modelestimationresults}), the age-specific annual and weekly sensitivities also exhibit fundamentally distinct patterns, which further justifies this separation.}

\subsection{Identifiability constraints} 
To ensure identifiability in the model structure of Eq.~\eqref{eq:modelstructure}, we impose the following constraints on the respective parameters:
\begin{equation}\label{eq:constraints}
\begin{aligned} 
   &\beta_{x_{\min}} = 1, \hspace{0.1cm} \gamma_{x_{\min}} = 1, \hspace{0.1cm} \delta_{x_{\min}} = 1, \hspace{0.1cm} \epsilon_{x_{\min}} = 1 \\
   &\kappa_{t_{\min}, r} = 0, \hspace{0.1cm} \added{\lambda_{1,r} = 0}, \hspace{0.25cm} \forall r \in \mathcal{R},
\end{aligned}
\end{equation}
where $x_{\min}$ and $t_{\min}$ are the minimum age and year values in the sets $\mathcal{X}$ and $\mathcal{T}$, respectively. Since $\kappa_{t,r}$ and $\lambda_{w,r}$ are defined on different time scales (years and weeks), no additional constraints (e.g., orthogonality) are needed between them or their associated age effects. In total, we impose $2\cdot|\mathcal{R}| + 4$ constraints. The constraints in Eq.~\eqref{eq:constraints} differ from the standard Lee-Carter constraints but allow for an easier computation of standard errors for the model parameters, as discussed in Section~\ref{subsec:calibration}. However, all methods and estimation procedures can be performed using straightforward extensions of the typical Lee-Carter constraints.

\subsection{Distributed lag non-linear models} \label{subsec:dlnm}
We model the non-linear and delayed effects of temperature and influenza on mortality in Eq.~\eqref{eq:modelstructure} using a distributed lag non-linear model (DLNM). This approach is well-established in the epidemiological literature for assessing covariate-response associations with delayed effects, such as temperature-mortality associations. Mathematically, we can express the DLNM for a specific region $r$ as:
\begin{align} \label{eq:dlnm1}
f_r(\xi_{t,w,r}) = \sum_{\ell=0}^{L} \sum_{j=1}^{v_\xi} \sum_{k=1}^{v_l}  \eta_{jk,r} \cdot b_j(\xi_{t,w-\ell,r}) \cdot c_k(\ell),
\end{align}
where $\xi_{t,w,r}$ refers to the covariate value (temperature or ILI) at the time point $(t,w)$ in the region $r$, $\ell= 0, 1, \dots, L$ denote the lag values and $L$ the maximum lag considered. In addition, $\eta_{jk,r}$ are the parameter coefficients for the region $r$, $b_j(\xi_{t,w-\ell,r})$ refers to the $\ell$-week lagged covariate value transformed through its $j$th basis function, and $c_k(\ell)$ is the $k$th basis function for the lag dimension evaluated at lag value $\ell$. Basis functions are smooth functions that capture non-linear relationships. \added{Note that when $w-\ell < 1$, i.e., the lag extends to the previous year, the covariate value $\xi_{t,w-\ell,r}$ should be pulled at week $W_{t-1}-(\ell-w)$ of year $t-1$, where $W_{t-1} \in \{52,53\}$ denotes the number of ISO weeks in year $t-1$.} Lastly, we refer to $v_\xi$ and $v_l$ as the number of basis functions for the covariate and lag dimensions. Importantly, the basis dimensions $v_\xi$ and $v_l$, which determine the flexibility of the covariate and lag effects, are the same for all regions.

To simplify the notation, we introduce a cross-basis representation. Hereto, we denote, for every time point $(t,w)$ and region $r$, $\boldsymbol{Z}_{t,w,r} \in \mathbb{R}^{v_\xi \cdot v_l}$ as the vector with entries:
\begin{align} \label{eq:Ztwjkr}
  Z_{t,w,jk,r} := \displaystyle \sum_{\ell = 0}^L   b_j(\xi_{t,w-\ell,r}) \cdot c_k(\ell),
\end{align}
for $j \in \{1,2,...,v_\xi\}$ and $k\in\{1,2,..,v_l\}$. Furthermore, let $\boldsymbol{\eta}_r := (\eta_{jk,r})_{j,k} \in \mathbb{R}^{v_\xi \cdot v_l}$ be the parameter vector for region $r$. We can then write the DLNM from Eq.~\eqref{eq:dlnm1} as:
\begin{align} \label{eq:dlnm2}
    f_r(\xi_{t,w,r}) = \boldsymbol{Z}_{t,w,r}' \boldsymbol{\eta}_r,
\end{align}
where $'$ refers to the transpose operator. The cross-basis matrix $\boldsymbol{Z}_{r} \in \mathbb{R}^{T\times (v_\xi \cdot v_l)}$ stacks all vectors $\boldsymbol{Z}_{t,w,r}$ as rows, where $T := |\mathcal{T}|$ is the total number of time points $(t,w)$ considered.

In the context of Equation~\eqref{eq:modelstructure}, we define the cross-basis matrices for the weekly average temperature ($\text{Tavg}$) and the weekly incidence rates of influenza-like illnesses ($\text{ILI}$) as $\boldsymbol{Z}_{1,r}$ and $\boldsymbol{Z}_{2,r}$, respectively, for each region $r\in \mathcal{R}$. We denote the corresponding parameter vectors by $\boldsymbol{\eta}_{1,r}$ and $\boldsymbol{\eta}_{2,r}$. Note that the cross-basis matrices $\boldsymbol{Z}_{1,r}$ and $\boldsymbol{Z}_{2,r}$ may differ in dimension, since we allow for a different maximum lag and number and type of basis functions in both the covariate and lag dimensions for temperature and ILI. %\ali{I have seen in the paper of Min et al. they propose to use bootstrapping to account for the uncertainty of the DLNM parameters; maybe good to consider this as well.}

\subsection{Relative risk} \label{subsec:rr}
The concept of relative risk (RR) is a useful way to quantify the effect of the DLNM components in Eq.~\eqref{eq:modelstructure}. At a given covariate level $\xi$, i.e., at a specific weekly average temperature or influenza-like illness incidence, the RR quantifies the factor by which the baseline force of mortality increases (if RR $>$ 1) or decreases (if RR $<$ 1) relative to a particular reference value. This RR captures both non-linear and delayed effects modeled by the DLNM. 

For temperature and influenza incidence, we compute the RR curves for the region $r \in \mathcal{R}$ and age group $x\in \mathcal{X}$ as:
\begin{equation} \label{eq:RRcurve}
\begin{aligned}
    \text{RR}_{\text{Tavg},x,r}(\xi) &= \exp\left(\delta_x \, \left[f_r^{(1)}(\xi) - f_r^{(1)}(\xi_{\text{ref}})\right]\right),  && \xi \in \Xi_{\text{Tavg}, r}, \\
    \text{RR}_{\text{ILI},x,r}(\xi) &= \exp\left(\epsilon_x \, \left[f_r^{(2)}(\xi) - f_r^{(2)}(\xi_{\text{ref}})\right]\right),  && \xi \in \Xi_{\text{ILI},r}, 
\end{aligned}
\end{equation}
where \(\Xi_{\text{Tavg},r}\) and \(\Xi_{\text{ILI},r}\) denote grids of covariate values for temperature and influenza incidence in region \(r\), respectively. The parameters \(\delta_x\) and \(\epsilon_x\), for \(x \in \mathcal{X}\), are the corresponding age-specific effects in Eq.~\eqref{eq:modelstructure}, and $\xi_{\text{ref}}$ denotes a reference value for the weekly average temperature or influenza-like illness incidence. From Eq.~\eqref{eq:dlnm2}, we obtain
\begin{equation*}
    f_r^{(i)}(\xi) = \boldsymbol{Z}'_{i,r}(\xi) \boldsymbol{\eta}_{i,r},
\end{equation*}
where $\boldsymbol{Z}_{i,r}(\xi)$ is defined through Eq.~\eqref{eq:Ztwjkr} by fixing the covariate value to $\xi$ over the different lags.

\section{Model calibration and forecasting} \label{sec:calibrationandforecasting}

\subsection{Calibration} \label{subsec:calibration}

\paragraph{Log-likelihood.} We calibrate the model parameters in Eq.~\eqref{eq:modelstructure} using maximum likelihood estimation. Hereto, we set up the negative binomial log-likelihood based on the distributional assumption in Eq.~\eqref{eq:modeldist}. From the mean parametrization, we obtain the following log-likelihood:
\begin{equation}\label{eq:unpenloglik}
\begin{aligned} 
    l_{\text{nb}}(\boldsymbol{\theta}) = \displaystyle \sum_{x,t,w,r} \Big[ \log\:  &\Gamma(d_{x,t,w,r} + \phi_{x,r}) - \log \Gamma(\phi_{x,r}) - \log \Gamma(d_{x,t,w,r} + 1) \:+  \\
    &d_{x,t,w,r} \cdot \log(E_{x,t,w,r} \cdot \mu_{x,t,w,r}) + \phi_{x,r} \cdot \log \phi_{x,r}\: - \\
    &(d_{x,t,w,r} + \phi_{x,r}) \cdot \log (E_{x,t,w,r} \cdot \mu_{x,t,w,r} + \phi_{x,r})\Big],
\end{aligned}
\end{equation}
where $\mu_{x,t,w,r}$ is the model structure for the weekly force of mortality in Eq.~\eqref{eq:modelstructure}, $\phi_{x,r}$ is the parameterized dispersion parameter in Eq.~\eqref{eq:dispparam}, and $\boldsymbol{\theta}$ denotes the vector of all parameters to be estimated.

To reduce model complexity and enforce spatial coherence in the DLNM components, we add a penalty term to the log-likelihood that smoothens the DLNM parameters across neighboring regions, i.e., regions that share a border. Hereto, we introduce the Laplacian matrix $\boldsymbol{L}$, which is a matrix representation of a graph that encodes the spatial adjacency (neighborhood) structure of the 12 French administrative regions. The entries $L_{ij}$ of the Laplacian $\boldsymbol{L}$ are defined as:
\begin{equation}\label{eq:laplacian}
\begin{aligned} 
    L_{ij} = \begin{cases}
    \text{deg}(\text{Region}_i) & \text{if } i = j, \\
    -1 & \text{if } i \neq j \:\: \text{and}\:\: \text{region }  i \text{ is neighbour of region } j \\
    0 & \text{elsewhere,}
    \end{cases}
\end{aligned}
\end{equation}
where $\text{deg}(\text{Region}_i)$ refers to the number of neighbors of region $i$. 

The penalized log-likelihood then results in:
\begin{align} \label{eq:penloglik}
    l_p(\boldsymbol{\theta}; \psi_1, \psi_2) = l_{\text{nb}}(\boldsymbol{\theta}) - \frac{\psi_1}{2} \displaystyle \sum_{q=1}^{Q_1} \boldsymbol{\eta}_{1,q}' \boldsymbol{L} \boldsymbol{\eta}_{1,q} - \frac{\psi_2}{2} \displaystyle \sum_{q=1}^{Q_2} \boldsymbol{\eta}_{2,q}' \boldsymbol{L} \boldsymbol{\eta}_{2,q},
\end{align}
where $l_{\text{nb}}(\boldsymbol{\theta})$ is the non-penalized negative binomial log-likelihood defined in Eq.~\eqref{eq:unpenloglik}, $\boldsymbol{L}$ is the Laplacian matrix in Eq.~\eqref{eq:laplacian}, $Q_i$ represents the number of columns in the cross-basis matrices $\boldsymbol{Z}_{i,r}$ for $i=1,2$, $\boldsymbol{\eta}_{1,q} = (\eta_{1,q,r})_r \in \mathbb{R}^R$, with $R = |\mathcal{R}|$ the number of regions in the set $\mathcal{R}$, and $\psi_i$, for $i=1,2$, are the smoothing parameters that reflect the degree of spatial smoothness imposed on the DLNM parameters.

\paragraph{Estimation.} To maximize the penalized log-likelihood in Eq.~\eqref{eq:penloglik}, we use a Newton-type optimization strategy. Specifically, we iteratively update subsets of the full parameter vector $\boldsymbol{\theta}$ while keeping the remaining parameters fixed. This blockwise approach is computationally efficient and is well-suited for models with high-dimensional parameter spaces. A similar approach has also been used in \cite{pitacco2009modelling} to estimate the parameters in a Lee-Carter model under a Poisson distributional assumption. 

In our modeling framework defined in Eq.~\eqref{eq:modelstructure}, we consider 11 parameter subsets: 
\begin{align*}
    (\alpha_{x,r})_{x\in\mathcal{X},r\in\mathcal{R}}, \hspace{0.5cm}(\beta_x)_{x\in\mathcal{X}}, \hspace{0.5cm}(\kappa_{t,r})_{t\in\mathcal{T},r\in\mathcal{R}}, \hspace{0.5cm}(\gamma_x)_{x\in\mathcal{X}}, \hspace{0.5cm}(\lambda_{w,r})_{w\in\mathcal{W},r\in\mathcal{R}}, \\(\delta_x)_{x\in\mathcal{X}}, \hspace{0.5cm}(\boldsymbol{\eta}_{1,r})_{r\in\mathcal{R}}, \hspace{0.5cm}(\epsilon_x)_{x\in\mathcal{X}}, \hspace{0.5cm}(\boldsymbol{\eta}_{2,r})_{r\in\mathcal{R}}, \hspace{0.5cm}(\phi_x)_{x\in\mathcal{X}}, \hspace{0.5cm}(\phi_r)_{r\in\mathcal{R}},
\end{align*}
which constitute the complete vector of parameters $\boldsymbol{\theta}$.
Let $\boldsymbol{\zeta} \subset \boldsymbol{\theta}$ be any set of parameters to be updated (e.g., the set of $\alpha_{x,r}$'s for all ages $x$ and regions $r$). In the iteration step $\nu+1$, the update for $\boldsymbol{\zeta}$ is given by:
\begin{align*} 
    \hat{\boldsymbol{\zeta}}^{(\nu+1)} = \hat{\boldsymbol{\zeta}}^{(\nu)} - \left[ \boldsymbol{H}_{\boldsymbol{\zeta}}\left( \hat{\boldsymbol{\zeta}}^{(\nu)}\right)\right]^{-1} 
    \boldsymbol{J}_{\boldsymbol{\zeta}}\left( \hat{\boldsymbol{\zeta}}^{(\nu)}\right),
\end{align*}
where $\boldsymbol{J}_{\boldsymbol{\zeta}}(\hat{\boldsymbol{\zeta}}^{(\nu)})$ and $\boldsymbol{H}_{\boldsymbol{\zeta}}(\hat{\boldsymbol{\zeta}}^{(\nu)})$ denote, respectively, the gradient vector and the Hessian matrix of the penalized negative binomial log-likelihood (see Eq.~\eqref{eq:penloglik}) with respect to the parameter vector $\boldsymbol{\zeta}$, evaluated in the previous estimate $\hat{\boldsymbol{\zeta}}^{(\nu)}$. We repeat this procedure sequentially across all parameter subsets until convergence of the full log-likelihood.

In the case study in Section~\ref{sec:casestudy}, we adopt a two-step calibration procedure. First, we estimate the parameters in the weekly Lee-Carter model, i.e., $(\alpha_{x,r}, \beta_x, \kappa_{t,r}, \gamma_x, \lambda_{w,r}, \phi_x, \phi_r)$, using the iterative procedure described above. In the second step, we estimate the parameters in the DLNM component, i.e., $(\delta_x, \boldsymbol{\eta}_{1,r}, \epsilon_x, \boldsymbol{\eta}_{2,r})$, while keeping the first-stage parameters fixed. \added{This two-step approach has two main advantages: (1) it avoids identification issues that may arise from confounding between seasonality and environmental effects; and (2) it facilitates a clearer interpretation of the covariate-response relationships as deviations from baseline seasonal mortality patterns.} \added{To further reduce sensitivity to the initial separation between baseline and DLNM effects, we repeat this two-step procedure, allowing first the baseline parameters and subsequently the DLNM parameters to be re-estimated conditional on the updated counterpart. In practice, this additional cycle leads to a reallocation of short-term covariate effects from the baseline model to the DLNM component, while preserving the overall fit of the model. Further iterations are in principle possible, but were not considered as they did not lead to meaningful changes in the resulting likelihood, while increasing computational time substantially. We therefore restrict our attention to only two rounds in the case study of Section~\ref{sec:casestudy}.}

\paragraph{Model selection.} We select the smoothing parameters $\psi_1$ and $\psi_2$ of the penalized DLNM (see Eq.~\eqref{eq:penloglik}) by minimizing the \added{Bayesian Information Criterion (BIC)}, where the complexity of the model is measured by the effective degrees of freedom (EDF) rather than the raw number of parameters \citep{wood2016smoothing,wood2017generalized}. This approach accounts for the shrinkage induced by the penalization.

Hereto, we define grids $\Psi_1$ and $\Psi_2$ of candidate values for $\psi_1$ and $\psi_2$, respectively. For each pair $(\psi_1, \psi_2) \in \Psi_1 \times \Psi_2$, we estimate the parameter vector
\begin{align*}
\hat{\boldsymbol{\theta}}_{\psi_1, \psi_2} = \arg \max_{\boldsymbol{\theta}} l_p(\boldsymbol{\theta}; \psi_1, \psi_2),
\end{align*}
where $l_p(\cdot)$ is the penalized log-likelihood defined in Eq.~\eqref{eq:penloglik}, using the iterative Newton method described above. We then compute the effective degrees of freedom as \citep{wood2016smoothing}:
\begin{align} \label{eq:edf}
\mathrm{edf}(\psi_1, \psi_2) = \operatorname{trace}\left[
\widetilde{\boldsymbol{H}}_{\boldsymbol{\theta}}\left(\hat{\boldsymbol{\theta}}_{\psi_1, \psi_2}\right)
\cdot
\boldsymbol{H}_{\boldsymbol{\theta}}\left(\hat{\boldsymbol{\theta}}_{\psi_1, \psi_2}\right)^{-1}
\right],
\end{align}
where $\widetilde{\boldsymbol{H}}_{\boldsymbol{\theta}}(\cdot)$ denotes the Hessian of the unpenalized negative binomial log-likelihood (see Eq.~\eqref{eq:unpenloglik}), and \(\boldsymbol{H}_{\boldsymbol{\theta}}(\cdot)\) denotes the Hessian of the penalized log-likelihood (Eq.~\eqref{eq:penloglik}). Note that when $\psi_1 = \psi_2 = 0$ (no penalization), the effective degrees of freedom are equal to the total number of parameters in the model minus the number of identifiability constraints.

Subsequently, we calculate the \added{BIC} for each pair of smoothing parameters:
\added{\begin{align} \label{eq:edfAIC}
\mathrm{BIC}(\psi_1, \psi_2) = -2 \, l_p\left(\hat{\boldsymbol{\theta}}_{\psi_1, \psi_2}\right) + \log n_{\text{obs}} \cdot \mathrm{edf}(\psi_1, \psi_2),
\end{align}
where $n_{\text{obs}}$ is the number of observations in the dataset.} We then select the optimal smoothing parameters $(\hat{\psi}_1, \hat{\psi}_2)$ by minimizing the \added{BIC} on the candidate grids.

\paragraph{Parameter uncertainty.} \added{Following \cite{brouhns2005bootstrapping}, we construct confidence intervals for the estimated parameter vector $\hat{\boldsymbol{\theta}}$ using a parametric bootstrap procedure. We generate $B$ bootstrap samples $(E_{x,t,w,r}, d_{x,t,w,r}^{(b)})$, where the bootstrapped death counts $d_{x,t,w,r}^{(b)}$ are drawn from a negative binomial distribution with mean:
\begin{align*}
    \hat{\mu}_{x,t,w,r} = \exp\left( \hat{\alpha}_{x,r} + \hat{\beta}_x \hat{\kappa}_{t,r} + \hat{\gamma}_x \hat{\lambda}_{w,r} + \hat{\delta}_x \hat{f}_{r}^{(1)} (\text{Tavg}_{t,w,r}) + \hat{\epsilon}_x \hat{f}_{r}^{(2)} (\text{ILI}_{t,w,r})\right),
\end{align*}
and dispersion parameter $\hat{\phi}_{x,r} = \exp(\hat{\phi}_x + \hat{\phi}_r)$. This procedure introduces observation-level variability around the fitted mean structure while preserving the estimated overdispersion.}

\added{For each bootstrap sample $(E_{x,t,w,r}, d_{x,t,w,r}^{(b)})$, we re-estimate the full weekly Lee–Carter model with DLNM specification described in Section~\ref{subsec:modelstructure}, yielding a set of bootstrap parameter estimates
\begin{align*}
    (\hat{\alpha}^{(b)}_{x,r})_{x,r}, \quad (\hat{\beta}^{(b)}_x)_x, \quad (\hat{\kappa}^{(b)}_{t,r})_{t,r}, \quad (\hat{\gamma}^{(b)}_x)_x, \quad (\hat{\lambda}^{(b)}_{w,r})_{w,r}, \quad (\hat{\delta}^{(b)}_x)_x, \quad (\hat{\boldsymbol{\eta}}^{(b)}_{1,r})_r, \quad (\hat{\epsilon}^{(b)}_x)_x, \quad (\hat{\boldsymbol{\eta}}^{(b)}_{2,r})_r .
\end{align*}
Pointwise confidence intervals for each parameter are then obtained from the empirical quantiles of the corresponding bootstrap distributions.}

% Following \cite{brouhns2002measuring}, we derive confidence intervals for the estimated parameter vector by computing the covariance matrix as the inverse of the observed Fisher information matrix. The observed Fisher information matrix is defined as the negative of the Hessian of the penalized log-likelihood function evaluated at $\hat{\boldsymbol{\theta}}_{\hat{\psi}_1, \hat{\psi}_2}$:
% \begin{equation} \label{eq:fisherinfo}
% \widehat{\mathrm{Var}}(\hat{\boldsymbol{\theta}}_{\hat{\psi}_1, \hat{\psi}_2}) = \left[ -\boldsymbol{H}_{\boldsymbol{\theta}}\left( \hat{\boldsymbol{\theta}}_{\hat{\psi}_1, \hat{\psi}_2} \right) \right]^{-1},
% \end{equation}
% where $\boldsymbol{H}_{\boldsymbol{\theta}}(\cdot)$ denotes the Hessian matrix of the penalized log-likelihood in Eq.~\eqref{eq:penloglik} with respect to $\boldsymbol{\theta}$. We compute these second-order derivatives analytically to ensure computational efficiency. Other approaches to quantify parameter uncertainty in the literature include the semiparametric bootstrap proposed by \cite{brouhns2005bootstrapping} and the residual bootstrap approach introduced by \cite{koissi2006evaluating}.

\subsection{Forecasting} \label{subsec:forecasting}
To forecast weekly mortality, we adopt a two-step procedure: (1) forecast the latent mortality index $\kappa_{t,r}$ and the exogenous predictors (temperature and ILI), and (2) calculate the predicted weekly death rates using the estimated model from Section~\ref{subsec:modelstructure} and the distributed lag structures from Section~\ref{subsec:dlnm}.

\paragraph{Step 1: Copula-based time series forecasting with covariates.}  
The latent mortality index $\kappa_{t,r}$, the weekly average temperature $\text{Tavg}_{t,w,r}$, and the influenza-like illness rates $\text{ILI}_{t,w,r}$ evolve over time and jointly shape the mortality dynamics. In addition, strong spatial dependencies are present: for instance, high temperatures in one region are likely to coincide with high temperatures in neighboring regions. Our forecasting framework should therefore account for the regional dependencies within each process.

Therefore, we jointly model the regional dependencies within each process using a copula-based framework, while capturing the temporal evolution in each region through a Seasonal Autoregressive Integrated Moving Average process (SARIMA). We also test for the inclusion of exogenous drives in Suppl.~Mat.~\ref{app:compareSARIMAX}. A SARIMA$(p,d,q)(P,D,Q)_s$ model for a time series $y_{t}$ is specified as \citep{hyndman2008automatic}:
\begin{equation}\label{eq:sarimax}
\begin{aligned} 
\Phi(B^s)\,\phi(B)\,(1-B^s)^D(1-B)^d \,y_t &= \added{d} + \Theta(B^s)\,\theta(B)\,\added{\omega_t},
\end{aligned}
\end{equation}
where
\begin{itemize}
    \item[-] $p,d,q$ and $P,D,Q$ are the non-seasonal and seasonal AR, differencing, and MA orders, with $s$ the seasonal period,
    \item[-] $\phi(B), \theta(B)$ and $\Phi(B^s), \Theta(B^s)$ are the corresponding non-seasonal and seasonal AR and MA polynomials in the backshift operators $B$ and $B^s$, respectively,
    \item[-] $(1-B)^d$ and $(1-B^s)^D$ denote the ordinary and seasonal differencing operators,
    \item[-] \added{$d$} is the drift parameter,
    \item[-] \added{$\omega_t$} is white noise with mean zero and variance $\sigma^2$.
\end{itemize}
We proceed by applying the following modeling steps:
\begin{enumerate}
    \item \textbf{Temperature:} For each region $r \in R$, we fit a SARIMA$(p_1, d_1, q_1)(P_1, D_1, Q_1)_{52}$ model to the weekly average temperature $\text{Tavg}_{t, w, r}$. The seasonal and non-seasonal orders are kept identical across all regions. Second, to capture the regional dependencies between the SARIMA processes, we fit a $|R|$-dimensional $t$-copula on the standardized innovations $\hat{\varepsilon}_{t, w} = (\hat{\varepsilon}_{t, w, r})_{r \in R}$ from the individual SARIMA processes. The $t$-copula is chosen to better model dependencies in the tails of the distribution compared to a Gaussian copula. 
    \item \textbf{Influenza:} We fit a \added{SARIMA$(p_2, d_2, q_2)(P_2, D_2, Q_2)_{52}$} process to the logarithmic transformation of weekly incidence rates of ILI, $\log(\text{ILI}_{t, w, r} + 0.01)$. The logarithmic transformation improves the seasonal structure and reduces the impact of extreme peaks, while the small constant $0.01$ avoids taking the logarithm of zero. As before, a $|R|$-dimensional $t$-copula is then fitted to the standardized innovations to account for regional dependencies between the SARIMA processes.
    \item \textbf{Mortality index:} For each region $r \in R$, we fit an \added{ARIMA$(p_3, d_3, q_3)$} process on the estimated latent mortality index $\hat{\kappa}_{t, r}$, i.e., a SARIMA process with $P = D = Q = 0$. Since this index is defined on an annual basis, no seasonal component is included. We fit a Gaussian copula to the standardized residuals to account for regional dependencies.
\end{enumerate}

\paragraph{Step 2: Generating mortality forecasts.}  
Once the model is fitted to historical data, we produce forecasts for $h = 1, \dots, H$ weeks ahead, where $H$ denotes the desired forecast horizon.

\begin{itemize}
    \item The forecasted $\widehat{\kappa}_{t,w+h,r}$ is plugged into the model in Eq.~\eqref{eq:modelstructure}.
    \item The forecasted $\widehat{\text{Tavg}}_{t,w+h,r}$ and $\widehat{\text{ILI}}_{t,w+h,r}$ are transformed into their respective cross-basis matrices $\boldsymbol{Z}_{1,r}^{\text{new}}$ and $\boldsymbol{Z}_{2,r}^{\text{new}}$ using the same DLNM basis functions and lag structures used in the estimation.
    \item Since DLNMs require lagged values, the forecasts are initialized with observed covariate values and updated recursively using previous predictions for lags beyond the observed horizon.
\end{itemize}

The final forecasted weekly force of mortality is computed as follows:
\begin{align}\label{eq:logmortality}
\log \widehat{\mu}_{x,t,w+h,r} = \hat{\alpha}_{x,r} + \widehat{\beta}_x \widehat{\kappa}_{t,r} + \widehat{\gamma}_x \widehat{\lambda}_{w+h,r} + \widehat{\delta}_x \left(\boldsymbol{Z}_{1,r}^{\text{new}} \widehat{\boldsymbol{\eta}}_{1,r} \right) + \widehat{\epsilon}_x \left(\boldsymbol{Z}_{2,r}^{\text{new}} \widehat{\boldsymbol{\eta}}_{2,r} \right).
\end{align}

The forecasted number of deaths is then obtained by:
\begin{align} \label{eq:sampleNB}
\widehat{D}_{x,t,w+h,r} \sim \text{NB}(E_{x,t,w+h,r} \cdot \widehat{\mu}_{x,t,w+h,r}, \phi_{x,r}).
\end{align}

\paragraph{Uncertainty quantification.}  
To account for uncertainty, we propose a simulation-based approach:
\begin{itemize}
    \item Generate multiple paths for the mortality index, temperature, and ILI forecasts using the estimated (S)ARIMA models and innovations sampled from the corresponding estimated copulas.
    \item For each simulated path, propagate through the mortality model to yield a corresponding forecast distribution of deaths using Eq.~\eqref{eq:logmortality} and \eqref{eq:sampleNB}.
    \item Construct prediction intervals by computing empirical quantiles of the simulated forecast distribution.
\end{itemize}
\added{Note that all parameters in the mortality and copula-(S)ARIMA models are treated as fixed in the forecasting procedure. The uncertainty quantification reflects stochastic variation arising from the simulated time-series innovations and the negative binomial sampling distribution.}

\section{Case study} \label{sec:casestudy}

\subsection{Specification and assessment of DLNM-based covariate effects} \label{subsec:specdlnmcase}
\paragraph{DLNM specifications.} We model the effects of temperature and influenza-like illness (ILI) on mortality using distributed lag non-linear models (see Section~\ref{subsec:dlnm}). For temperature, the covariate–response relationship is modeled with cubic B-spline basis functions. \added{We use a common basis specification for all regions, with two internal knots placed at the 10th and 90th percentiles of the pooled temperature distribution over the historical period 1990–2019.} The spline and knot choices are based on common practices in DLNM studies that examine the temperature–mortality relationship \citep{gasparrini2015mortality, martinez2021projections}. \added{Boundary knots are defined based on the global temperature range, with the lower boundary set slightly below the observed minimum (by 5$^\circ$C) to obtain sufficient flexibility in the lower tail. This avoids overly restrictive behavior of the spline near the boundary and ensures that extreme cold effects can be better captured. This common specification ensures that the DLNM parameter vectors have the same interpretation across regions, and that the spatial smoothing penalty operates on comparable quantities.}

\added{For influenza, we work with exceedance anomalies, defined as $\text{ILI}_{t,w,r} - q_{90}(r,w)$,
where $q_{90}(r,w)$ is the region- and week-specific 90th percentile of the raw ILI incidence rate over the historical period 1990--2019. 
We assume a linear covariate–response relationship for these anomalies 
\citep{li2023influenza}. The lag period is set at \added{3 weeks for temperature and 7 weeks for influenza}, reflecting the typically longer persistence of influenza effects \citep{chaves2023global}. Lag–response relationships are modeled with natural cubic splines, including an intercept, with two internal knots placed between lags of 0–1 and 1–2 weeks. In the remainder of this paper, whenever we write $\text{ILI}_{t,w,r}$ in the mortality model or DLNM specification (Eq.~\eqref{eq:modelstructure}), we refer to these 
exceedance anomalies unless otherwise stated.}

\paragraph{\added{Robustness of the DLNM specification.}} \added{In Suppl.~Mat.~\ref{app:sensitivitydlnm}, we evaluate the robustness of the DLNM specification through a series of sensitivity analyses. We vary the maximum lag lengths for both temperature and influenza-like illness (ILI), consider alternative specifications of the covariate–response relationships (including different spline bases, polynomial forms, and knot placements), and examine alternative parameterizations of the lag structure. We compare the different specifications through the Bayesian Information Criterion (BIC), which balances model fit and parsimony. Across all the considered specifications, our proposed specification for the DLNMs, as described above, consistently performs best in terms of BIC.}

\paragraph{\added{Comparison of alternative model specifications.}} \added{In Suppl.~Mat.~\ref{app:modelcomparison}, we compare five alternative specifications of the model in Eq.~\eqref{eq:modelstructure} which differ in the inclusion of DLNMs for temperature and influenza. One of these specifications specifically looks at the inclusion of a DLNM for modeling the interaction between them. Starting from a weekly Lee–Carter baseline, we sequentially incorporate DLNM effects for temperature, influenza, both covariates jointly, and finally their interaction. The results indicate a substantial improvement in fit when including environmental and epidemic covariates. In particular, the specification with both temperature and influenza DLNMs achieves the lowest BIC, suggesting that these effects capture important variation in weekly mortality. Although the interaction model leads to a slightly higher log-likelihood, it is penalised more heavily for its increased complexity and does not improve the BIC. This indicates that the added complexity of the interaction DLNM between temperature and influenza does not outweigh the extra explanatory power it brings. We therefore continue with the model specification as outlined in Eq.~\eqref{eq:modelstructure} in Section~\ref{sec:modelspecification}}

\paragraph{\added{Model selection.}} \added{Following the model selection in the calibration setup in Section~\ref{subsec:calibration}, we set up a tuning grid for the smoothing parameters $\psi_1$ and $\psi_2$, which control the degree of spatial smoothness imposed on the DLNM parameters. Since the $\psi_i$-values vary by many orders of magnitude, we define the parameter grid formed by all combinations of two sequences whose values are equally spaced on a logarithmic scale between $10^4$ and $10^{10}$:
\begin{align*}
    \psi_i \in \{10^4, 10^{4.5}, 10^5, ..., 10^{10}\},
\end{align*}
for $i=1,2$. By selecting the pair $(\psi_1, \psi_2)$ that leads to the lowest value for the edf-based BIC (see Eq.~\eqref{eq:edfAIC}), we obtain the estimated values $\hat{\psi}_1=10^{4.5}$ and $\hat{\psi}_2 = 10^{9.5}$, see Suppl.~Mat.~\ref{app:smoothingpsi}.}

\subsection{Model estimation results and goodness-of-fit assessment} \label{subsec:modelestimationresults}

\begin{figure}[!ht]
    \centering
    \includegraphics[width=\linewidth]{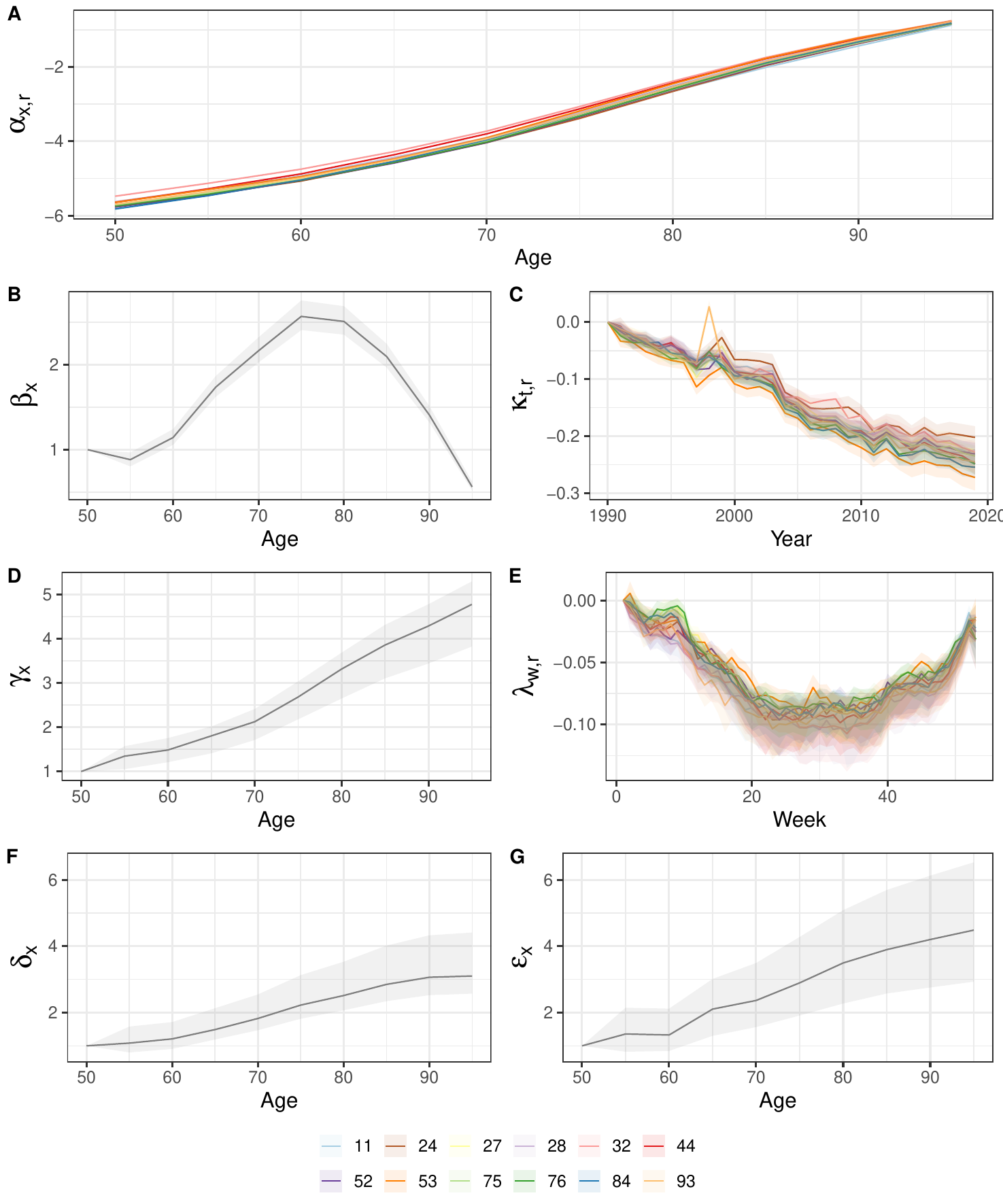}
    \caption{\added{Estimated parameters from the model: $\hat{\alpha}_{x,r}$ (A), $\hat{\beta}_x$ (B), $\hat{\kappa}_{t,r}$ (C), $\hat{\gamma}_x$ (D), $\hat{\lambda}_{w,r}$ (E), $\hat{\delta}_x$ (F), and $\hat{\epsilon}_x$ (G). Female data, French administrative regions, 5-year age groups $50-54$, $55-59$, ..., $95+$, and years $1990-2019$. We present $95\%$ confidence intervals based on the parametric bootstrap procedure. The age-specific parameters, with the exception of $\hat{\alpha}_{x,r}$, do not vary by region $r$ and are shown in black.
    \label{fig:param1}}}
\end{figure}

\paragraph{Parameter estimates.} \added{Using the optimally selected DLNM specifications, covariate effects, and smoothing parameters $\hat{\psi}_i$ for $i=1,2$, we present the parameter estimates in Figure~\ref{fig:param1}. We add $95\%$ confidence bounds based on the parametric bootstrap procedure with 5000 replications, as described in Section~\ref{subsec:calibration}.} \added{As a computationally lighter alternative, Suppl.~Mat.~\ref{App:parameter.uncertainty.Fisher} outlines how to construct confidence intervals based on the inverse observed Fisher information matrix, which conditions on the stage-one estimates of the seasonal baseline model and thus does not fully propagate estimation uncertainty.}

The trends and patterns in panels A, B, and C closely resemble a traditional Lee-Carter model, as they depend only on region, age, and year. From panel C, we see that the mortality decline is strongest in region 53 (Bretagne) and lowest in region 24 (Centre-Val de Loire). In panel B,  $\hat{\beta}_x$ shows that the mortality decline first increases with age, reaches a maximum for the 75-79 and 80-84 age groups, and then declines sharply. This means that the decline in mortality rates between 1990 and 2019 is strongest for the middle-aged age groups of 75-79 and 80-84, but the decline slows down significantly for older age groups. This is in line with traditional applications of the Lee-Carter model on annual mortality data in France, see, e.g., \citep{delwarde2007negative}.

Moreover, for the weekly effect parameter $\hat{\lambda}_{w,r}$ in panel E, we see a clear U- or V-shaped pattern, which reflects the seasonality. This means that mortality rates are highest during the winter weeks and lowest during the summer weeks. To a lesser extent, we can also observe some regional differences. For example, the seasonal variation, measured as the variance in the estimated $\hat{\lambda}_{w,r}$ for every region $r$, is the lowest in the administrative regions in the \added{west} of France, i.e., regions 52 (Pays de la Loire) and 53 (Bretagne). Regarding the estimated age effect $\hat{\gamma}_x$ in panel D, we see an increasing pattern. This means that seasonal variation is stronger for the oldest age groups. Moreover, the uncertainty surrounding the estimated age effect also increases with age.

Lastly, panels F and G present the age effects associated with the distributed lag non-linear components of temperature and influenza, respectively. Both panels display an upward trend, indicating that the impacts of temperature and influenza increase steadily with age. On the other hand, we also observe an increasing uncertainty with age. We evaluate in more depth the association of temperature and influenza with mortality deviations from the seasonal baseline model by means of the relative risk in Section~\ref{subsec:case.study.relative.risk}.

% \paragraph{Dispersion parameter.} We examine the estimated dispersion parameter $\hat{\phi}_{x,r} = \exp(\hat{\phi}_x + \hat{\phi}_r)$ to spot the region- and age group-specific differences in Table~\ref{tab:est.dispersion}. We clearly see a decreasing trend with age in the age-specific component $\hat{\phi}_x$, indicating a higher estimated variance-to-mean ratio for the oldest age groups since
% \begin{align*}
%     \widehat{\text{Var}}[D_{x,t,w,r}] = E_{x,t,w,r} \, \hat{\mu}_{x,t,w,r} + \dfrac{(E_{x,t,w,r} \, \hat{\mu}_{x,t,w,r})^2}{\hat{\phi}_{x,r}}.
% \end{align*}
% From $\hat{\phi}_r$, the largest estimated variance-to-mean ratios are obtained for the regions Ile-de-France (11), Centre-Val de Loire (24), and Bourgogne-Franche-Comté (27).

% \begin{table}[!ht]
% \centering
% \begin{tabular}{ccccc}
% \toprule
% Age & $\hat{\phi}_x$ & & Age & $\hat{\phi}_x$ \\
% \midrule
% 50-54 & 0.18 & & 75-79 & -0.07 \\
% 55-59 & 0.50 & & 80-84 & -0.16 \\
% 60-64 & 0.72 & & 85-89 & -0.38 \\
% 65-69 & 0.18 & & 90-94 & -0.58 \\
% 70-74 & 0.19 & & 95+   & -0.58 \\
%       &      &       &       \\  
% \bottomrule
% \end{tabular} \:\:\:\:\:\:\:\:\:
% \begin{tabular}{ccccc}
% \toprule
% Region & $\hat{\phi}_r$ & & Region & $\hat{\phi}_r$ \\
% \midrule
% 11 & 4.93 & & 52 & 5.24 \\
% 24 & 4.91 & & 53 & 5.43 \\
% 27 & 4.88 & & 75 & 5.58 \\
% 28 & 5.75 & & 76 & 5.58 \\
% 32 & 5.52 & & 84 & 5.45 \\
% 44 & 5.37 & & 93 & 5.52 \\
% \bottomrule
% \end{tabular}
% \caption{Estimated values of the components $\hat{\phi}_x$ and $\hat{\phi}_r$ in the dispersion parameter $\hat{\phi}_{x,r}$. \label{tab:est.dispersion}}
% \end{table}

\paragraph{In-sample fit of death rates.} Using the estimated parameters $\hat{\alpha}_{x,r}$, $\hat{\beta}_x$, $\hat{\kappa}_{t,r}$, $\hat{\gamma}_x$, $\hat{\lambda}_{w,r}$, $\hat{\delta}_x$, $\hat{\boldsymbol{\eta}}_{1,r}$, $\hat{\epsilon}_x$, and $\hat{\boldsymbol{\eta}}_{2,r}$, for every $x\in \mathcal{X}$, $t \in \mathcal{T}$, $w \in \mathcal{W}$, and $r \in \mathcal{R}$, as visualized in Figure~\ref{fig:param1}, we construct an in-sample estimate of the force of mortality $\hat{\mu}_{x,t,w,r}$ using Eq.~\eqref{eq:modelstructure}. This estimated force of mortality $\hat{\mu}_{x,t,w,r}$ serves as an estimate for the observed weekly death rates $m_{x,t,w,r}$, as discussed in Section~\ref{subsec:noations}. This gives us a visual idea of how well our in-sample predictions match the observations. 

To disentangle the contribution of the DLNM component from the baseline structure in the model of Eq.~\eqref{eq:modelstructure}, we calculate the estimated force of mortality as the multiplication of the baseline force of mortality and the DLNM component:
\begin{equation} \label{eq:estfom}
    \begin{aligned}
        \hat{\mu}_{x,t,w,r}^{(\text{base})} &= \exp\left(\hat{\alpha}_{x,r} + \hat{\beta}_x \,\hat{\kappa}_{t,r} + \hat{\gamma}_x \,\hat{\lambda}_{w,r}\right) \\
        \hat{\mu}_{x,t,w,r} &= \hat{\mu}_{x,t,w,r}^{(\text{base})} \cdot  \exp\left(\hat{\delta}_x \, \hat{f}_r^{(1)}(\text{Tavg}_{t,w,r}) + \hat{\epsilon}_x\, \hat{f}_r^{(2)}(\text{ILI}_{t,w,r}) \right),
    \end{aligned}
\end{equation}
where $\hat{f}_r^{(1)}(\text{Tavg}_{t,w,r})$ and $\hat{f}_r^{(2)}(\text{ILI}_{t,w,r})$ are based on the estimated parameter vectors $\hat{\boldsymbol{\eta}}_{1,r}$ and $\hat{\boldsymbol{\eta}}_{2,r}$, respectively (see Eq.~\eqref{eq:dlnm1}).

\begin{figure}[!ht]
    \centering
    \includegraphics[width=\linewidth]{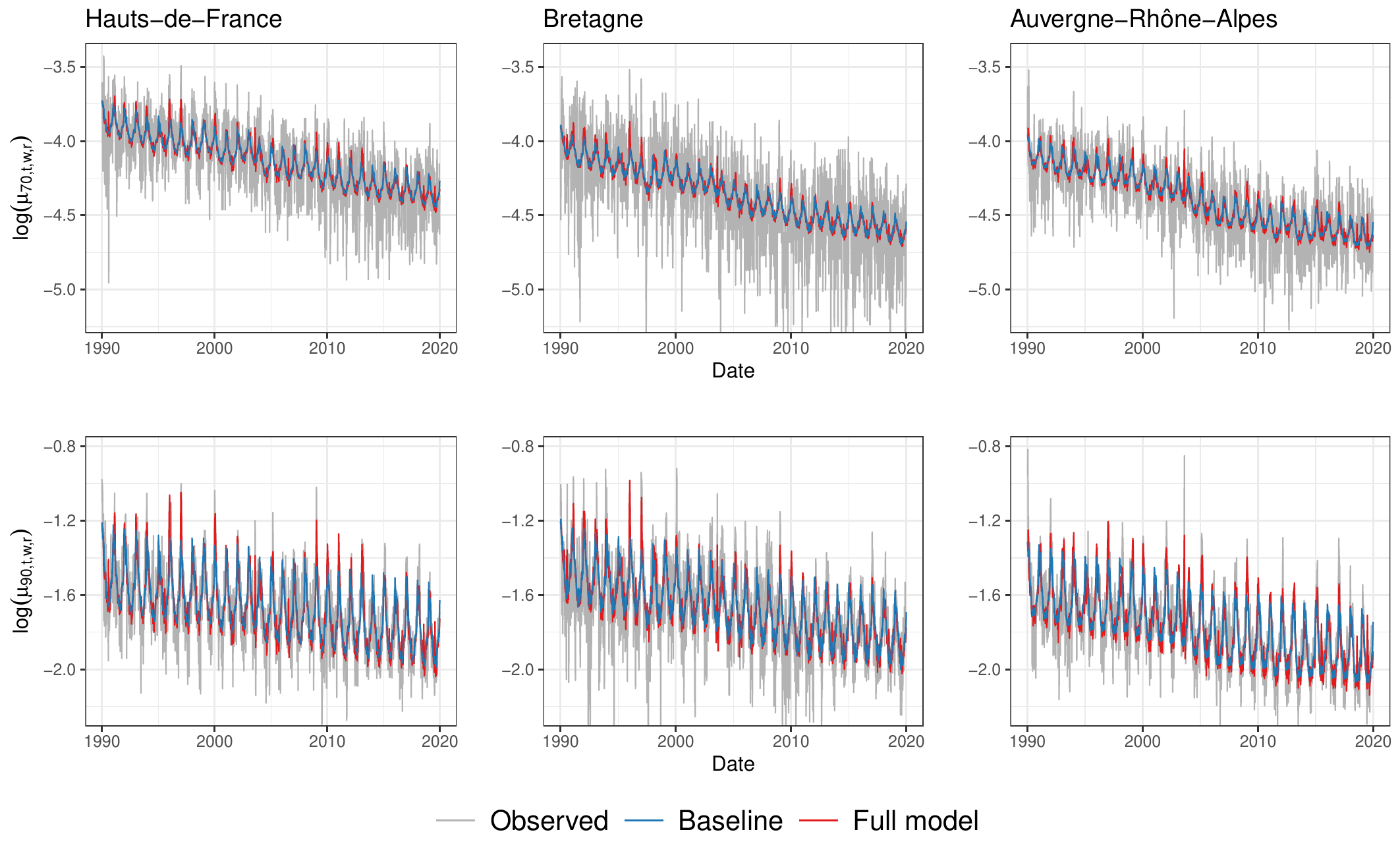}
    \caption{\added{Estimated female weekly death rates from 1990-2019 for the age groups 70-74 (top) and 90-94 (bottom) in Hauts-de-France (left), Bretagne (middle), and Auvergne-Rh\^one-Alpes (right). The gray line represents the logarithm of the observed death rates, while the blue and red lines reflect, respectively, the logarithm of the death rates as estimated by the baseline model and the full model including the DLNM component. \label{fig:insamplefit}}}
\end{figure}

Figure~\ref{fig:insamplefit} compares the observed and estimated weekly death rates for women between 1990 and 2019 for the age groups 70-74 and 90-94, and for three administrative regions in France: Hauts-de-France, Bretagne, and Auvergne-Rhône-Alpes. In addition to the empirical rates, we plot the fit of the baseline model and the full model that incorporates the DLNM component, capturing the effects of extreme temperatures and influenza activity. The results show that seasonal mortality patterns are well reproduced, with a stronger seasonal signal in the 90–94 age group compared to the 70–74 age group. Furthermore, the integration of the DLNM component is particularly valuable for older age groups, who are more vulnerable to temperature extremes and influenza outbreaks. The \textit{Online Appendix} contains the in-sample fits for the remaining regions and age groups.

\paragraph{Pearson residuals.} To evaluate the goodness-of-fit of the model in a quantitative way, we calculate the squared Pearson residuals for each region $r\in\mathcal{R}$:
\begin{equation*}
\begin{aligned}
\rho_{x,t,w,r}^2 &= \dfrac{\left(d_{x,t,w,r} - \mathbb{E}[D_{x,t,w,r}]\right)^2}{\text{Var}[D_{x,t,w,r}]}\Big|_{\mu_{x,t,w,r} = \hat{\mu}_{x,t,w,r}} \\
&=\dfrac{\left(d_{x,t,w,r} - E_{x,t,w,r} \, \hat{\mu}_{x,t,w,r}\right)^2}{E_{x,t,w,r} \, \hat{\mu}_{x,t,w,r} + (E_{x,t,w,r} \, \hat{\mu}_{x,t,w,r})^2 / \hat{\phi}_{x,r}},
\end{aligned}
\end{equation*}
where $d_{x,t,w,r}$ is the observed number of deaths, $\hat{\mu}_{x,t,w,r}$ the estimated force of mortality, $\hat{\mu}_{x,t,w,r} \, E_{x,t,w,r}$ the estimated expected number of deaths, and $\hat{\phi}_{x,r} = \exp(\hat{\phi}_x + \hat{\phi}_r)$ the estimated dispersion parameter. 

\begin{figure}[!ht]
    \centering
    \includegraphics[width=0.95\linewidth]{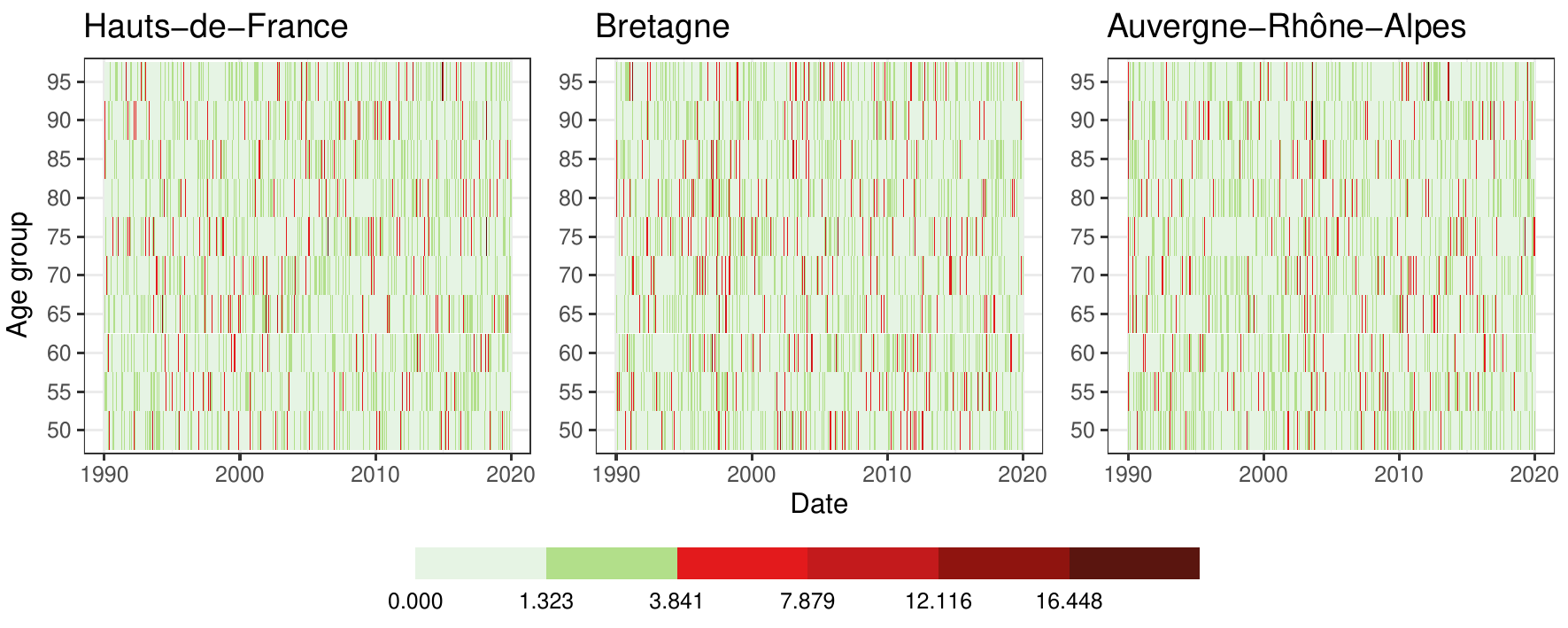}
    \caption{\added{Heat map of the squared Pearson residuals $\rho_{x,t,w,r}^2$ across age and time for Hauts-de-France (left), Bretagne (middle), and Auvergne-Rhône-Alpes (right). Green cells indicate areas with a good fit, while red and black cells correspond to areas with a poor fit ($\rho_{x,t,w,r}^2 > \chi_1^2$).}}
    \label{fig:pearson}
\end{figure}

Figure~\ref{fig:pearson} visualizes the squared Pearson residuals for Hauts-de-France, Bretagne, and Auvergne-Rhône-Alpes. Under some mild regularity conditions and assuming that the model specified in Eq.\eqref{eq:modeldist} and~\eqref{eq:modelstructure} is correctly specified \citep{wong2018bayesian}, each $\rho_{x,t,w,r}^2$ is asymptotically distributed as a chi-square random variable with one degree of freedom ($\chi^2_1$). Consequently, for each region, around $5\%$ of the cells in Figure~\ref{fig:pearson} are expected to exceed the 95th percentile of the $\chi^2_1$-distribution, i.e., $\rho_{x,t,w,r}^2 > 3.841$. For the three regions shown, the observed proportions are \added{$4.74\%$, $5.11\%$, and $4.68\%$}, respectively. We find comparable percentages for the remaining French regions, all between \added{$4.64\%$ and $5.11\%$}. Moreover, no particular age groups or time points stand out as systematic sources of poor fit.

Lastly, we assess the overall model's goodness-of-fit by aggregating the squared Pearson residuals across all age groups, years, weeks, and regions \citep{wong2018bayesian}:
\begin{align*}
    \rho^2 = \displaystyle \sum_{x,t,w,r} \rho_{x,t,w,r}^2,
\end{align*}
which leads to a total value of \added{187\,025.8}. Under the same set of assumptions as for the cell-specific tests, we now compare this statistic with the 95th percentile of a Chi-square distribution with degrees of freedom equal to the total number of observations. In addition, we use the effective degrees of the model to account for the regularization of the parameters in the DLNM component of the model, see Equation~\eqref{eq:edf}. The corresponding 95th percentile is 187\,466.6. As such, we consider the negative binomial assumption as adequate for our analysis, as it explains the majority of the extra-Poisson variability within the data. In Suppl.~Mat.~\ref{app:gofPoisson}, we repeat the analysis under the assumption that the weekly death counts follow a Poisson distribution, i.e., without accounting for overdispersion. In that case, more than 5$\%$ of the cells in the heat map of the squared Pearson residuals exceed the critical, and the aggregated chi-square statistic also exceeds its reference value. These results confirm that the negative binomial specification is more appropriate for our application.

\begin{figure}[!ht]
    \centering
    \includegraphics[width=0.95\linewidth]{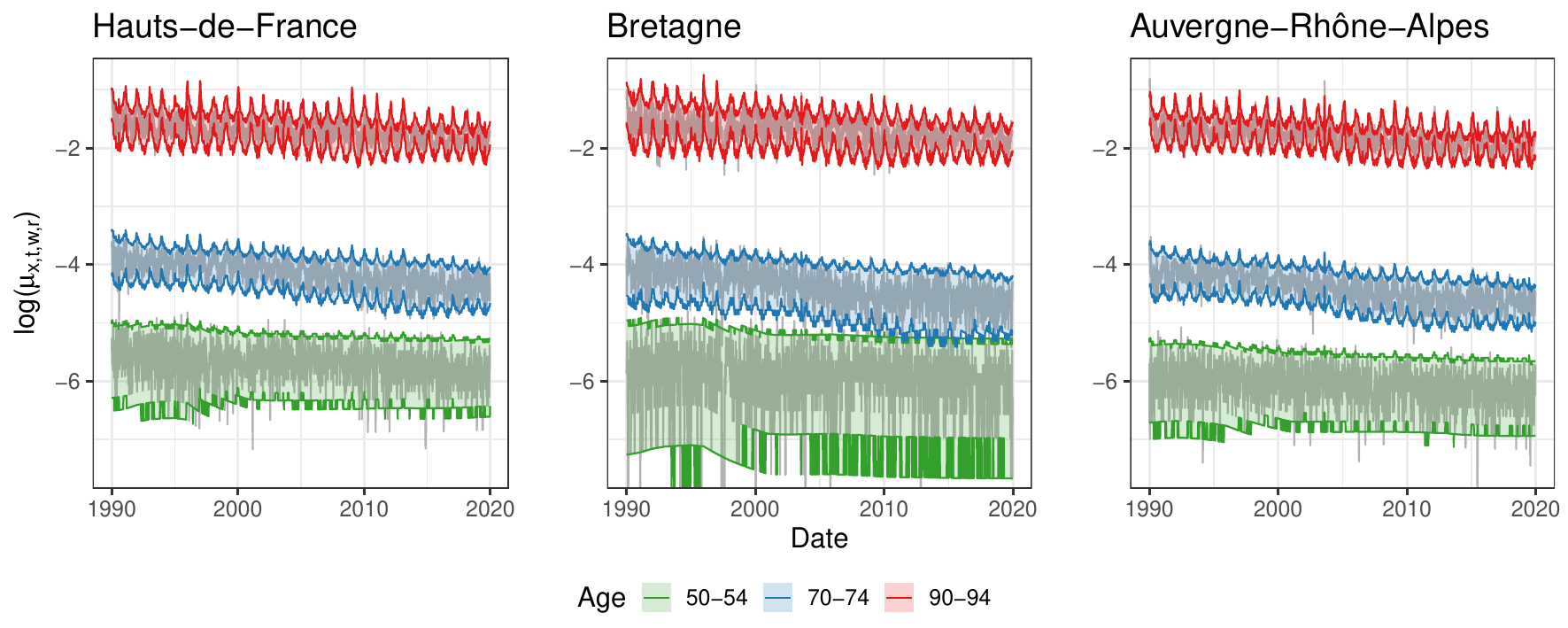}
    \caption{\added{$95\%$ uncertainty bounds around the logarithm of the estimated female weekly death rates from 1990-2019 for the age groups 50-54, 70-74, and 90-94 in Hauts-de-France (left), Bretagne (middle), and Auvergne-Rh\^one-Alpes (right). Wiggly intervals for the 50-54 age group in Bretagne reflect sampling variability due to low death counts. \label{fig:uncertainty}}}
\end{figure}

\paragraph{Uncertainty bounds.} To complement the in-sample estimates, we also quantify uncertainty around the estimated weekly death rates. We incorporate two types of uncertainty. First, we account for parameter uncertainty \added{using a parametric bootstrap. More specifically, we generate bootstrap parameter vectors by resampling from the estimated sampling distribution of the fitted parameter vector, following Section~\ref{subsec:calibration}.} For each draw of the parameter set, we reconstruct the force of mortality according to Eq.~\eqref{eq:modelstructure}. Second, we account for aleatoric or observation uncertainty by simulating deaths from the negative binomial distribution with mean equal to the estimated force of mortality times the population exposure and variance determined by the estimated dispersion. We repeat this process 5\ 000 times and combine the two sources of uncertainty to construct $95\%$ uncertainty bounds for the weekly death rates. These bounds provide a measure of the goodness-of-fit of our model. Figure~\ref{fig:uncertainty} shows the results for the age groups 50–54, 70–74, and 90–94 in Hauts-de-France, Bretagne, and Auvergne-Rhône-Alpes. The \textit{Online Appendix} contains the results for the other regions. The observed death rates fall mainly within the constructed $95\%$ uncertainty bounds, even during extreme events such as severe influenza outbreaks and heat waves. 

\subsection{Relative risk}  \label{subsec:case.study.relative.risk}
Using the estimated parameter vectors \(\hat{\boldsymbol{\eta}}_{1,r}\) and \(\hat{\boldsymbol{\eta}}_{2,r}\) from Eq.~\eqref{eq:modelstructure}, we visualize the effect of temperature and influenza on mortality, in excess of the seasonal baseline, through the relative risk (RR), as discussed in Section~\ref{subsec:rr}.

\begin{figure}[!ht]
    \centering
    \includegraphics[width=0.95\linewidth]{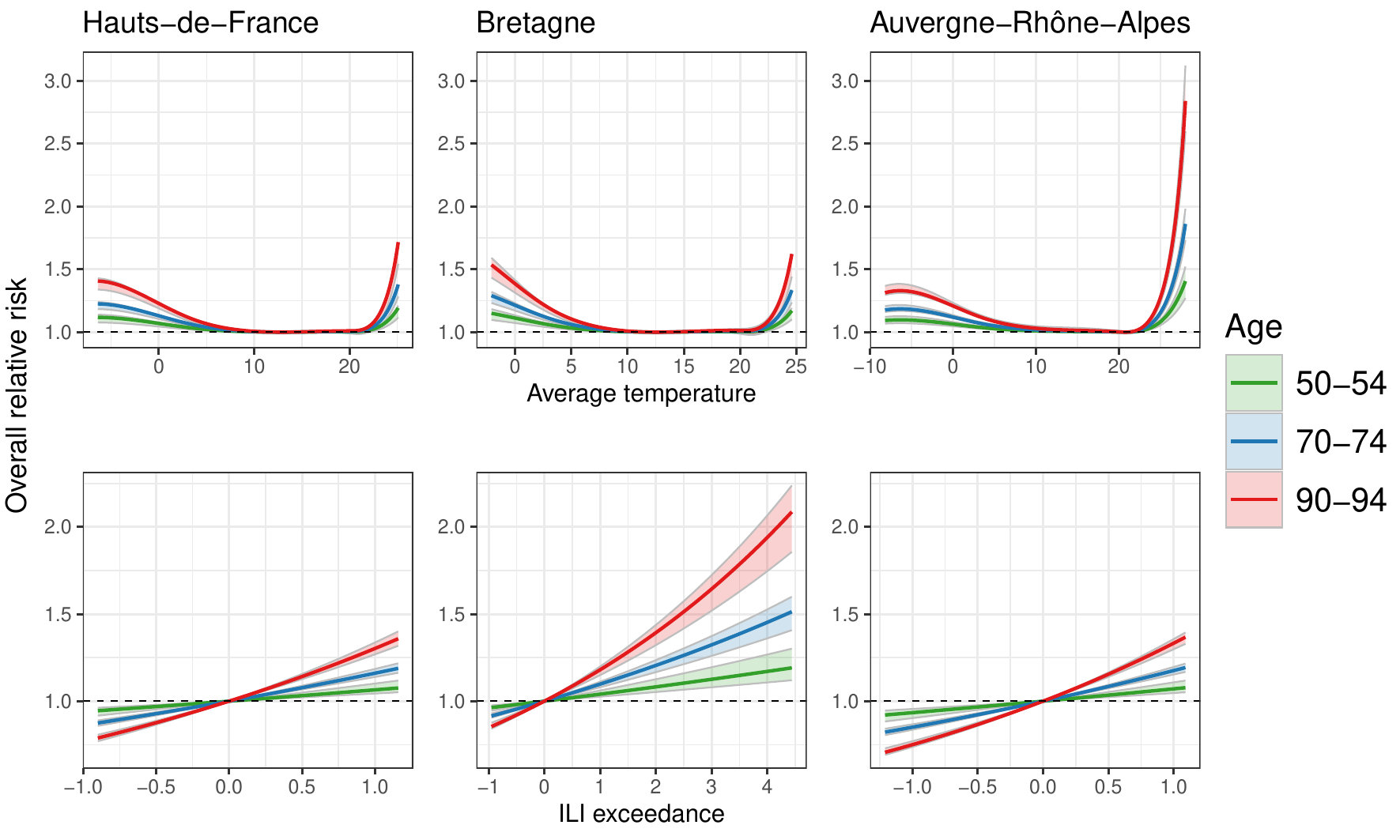}
    \caption{Estimated overall relative risk for temperature (top row) and influenza (bottom row) in Hauts-de-France (left), Bretagne (middle), and Provence-Alpes-Côte d’Azur (right) for the age groups 50–54, 70–74, and 90–94. Shaded areas represent 95$\%$ point-wise confidence intervals. \label{fig:RRoverall}}
\end{figure}

\paragraph{Overall relative risk.} Figure~\ref{fig:RRoverall} visualizes the overall relative risk, as defined in Eq.~\eqref{eq:RRcurve}, for three administrative regions in France: Hauts-de-France, Bretagne, and Provence-Alpes-Côte d’Azur. These regions represent a region in the north, west, and south of France, respectively. We provide the plots for the remaining regions in the \textit{Online Appendix}. For each region, we compute the overall relative risk for three age groups: 50–54, 70–74, and 90–94 years. \added{Furthermore, we express all relative risks with respect to a region-specific reference temperature. For each region, we determine the minimum mortality temperature (MMT) as the temperature at which the overall relative risk curve is minimized, and use this value as the reference point. This choice ensures that the relative risks are interpreted relative to the locally optimal temperature.} As the reference value for the relative risk of the ILI exceedances, we take 0.

\added{In Figure~\eqref{fig:RRoverall}, we also compute point-wise $95\%$ confidence intervals around the relative risk curves using the parametric bootstrap procedure with 5000 replications, as described in Section~\ref{subsec:calibration}. For each bootstrap replication of the parameter vector, we evaluate Eq.~\eqref{eq:RRcurve} over the range of temperatures and ILI exceedances while holding the reference value fixed. Point-wise confidence intervals are then obtained from the empirical 2.5th and 97.5th percentiles of the bootstrap distribution at each temperature level and ILI exceedance.}

Among the three regions, the effect of temperature on mortality increases rapidly at high temperatures and is most pronounced in Auvergne-Rhône-Alpes, followed by Bretagne and Hauts-de-France. For influenza-like illness (ILI), we clearly observe the highest relative risk in Bretagne. Furthermore, we find that the relative risk of both temperature and ILI is substantially higher for the oldest age group (90–94) compared to the youngest (50–54), although the uncertainty also increases with age. \added{Notable, the cold effect is most pronounced in Bretagne, and is relatively minor in Auvergne-Rhône-Alpes compared to its heat effect.} %Notably, the cold effect is minor. This is likely because most excess mortality in winter is captured by ILI rather than by low temperatures.

Table~\ref{tab:RRageregion} reports the relative risks (RR) by age group and region at the 99.5th percentile of the region-specific weekly average temperatures during the period 1990–2019. Although the highest 99.5th temperature percentile is observed in the southern region Provence-Alpes-Côte d’Azur, we observe the highest relative risks in Ile-de-France. \added{The lowest relative risks at the 99.5th percentile are observed in the north-western French regions, particularly Normandie (code 28) and Bretagne (code 53).} Additionally, the relative risk increases consistently with age across all regions.
\begin{table}[!ht]
\centering
\resizebox{\textwidth}{!}{%
\begin{tabular}{lcccccccccccc}
  \toprule
\footnotesize{\textbf{Age} $\backslash$ \textbf{Region}} & \textbf{11} & \textbf{24} & \textbf{27} & \textbf{28} & \textbf{32} & \textbf{44} & \textbf{52} & \textbf{53} & \textbf{75} & \textbf{76} & \textbf{84} & \textbf{93} \\ 

 & (25.3) & (24.5) & (24.2) & (21.3) & (22.4) & (24.4) & (23.6) & (21.1) & (24.4) & (25.8) & (24.4) & (26.1)\\ \hline
 
\textbf{50-54} & 1.16 & 1.07 & 1.06 & 1.01 & 1.03 & 1.08 & 1.04 & 1.01 & 1.06 & 1.07 & 1.06 & 1.06 \\  
\textbf{55-59} & 1.18 & 1.08 & 1.07 & 1.01 & 1.03 & 1.09 & 1.05 & 1.01 & 1.06 & 1.08 & 1.06 & 1.07\\ 
\textbf{60-64} & 1.20 & 1.09 & 1.08 & 1.01 & 1.03 & 1.10 & 1.05 & 1.01 & 1.07 & 1.09 & 1.07 & 1.08 \\ 
\textbf{65-69} & 1.25 & 1.11 & 1.10 & 1.01 & 1.04 & 1.13 & 1.07 & 1.01 & 1.08 & 1.11 & 1.09 & 1.10 \\
\textbf{70-74} & 1.32 & 1.14 & 1.12 & 1.01 & 1.05 & 1.16 & 1.08 & 1.01 & 1.10 & 1.13 & 1.11 & 1.12 \\ 
\textbf{75-79} & 1.40 & 1.17 & 1.15 & 1.01 & 1.06 & 1.20 & 1.10 & 1.01 & 1.13 & 1.16 & 1.14 & 1.15 \\
\textbf{80-84} & 1.46 & 1.19 & 1.17 & 1.02 & 1.07 & 1.23 & 1.12 & 1.01 & 1.15 & 1.19 & 1.15 & 1.17 \\ 
\textbf{85-89} & 1.54 & 1.22 & 1.19 & 1.02 & 1.08 & 1.26 & 1.13 & 1.02 & 1.17 & 1.21 & 1.18 & 1.19 \\  
\textbf{90-94} & 1.59 & 1.24 & 1.21 & 1.02 & 1.09 & 1.28 & 1.14 & 1.02 & 1.18 & 1.23 & 1.19 & 1.21 \\
\textbf{95+} & 1.60 & 1.25 & 1.21 & 1.02 & 1.09 & 1.29 & 1.15 & 1.02 & 1.18 & 1.23 & 1.19 & 1.21 \\  
\hline
\end{tabular}
}
\caption{\added{The overall relative risk per age group at the 99.5th percentile of the historical temperature distribution. The value of the percentile is given below the region's INSEE code.}}
\label{tab:RRageregion}
\end{table}

\paragraph{Relative risk by lag.} The overall RR summarizes the cumulative effect across all lags, as discussed in Section~\ref{subsec:rr}. To gain further insight, we also examine the effect of a fixed temperature or ILI exceedance value at each individual lag, i.e., at lags 0–4 weeks for temperature and lags 0–6 weeks for ILI. This allows us to evaluate whether the estimated effects occur immediately or persist over time.

\begin{figure}[!ht]
    \centering
    \includegraphics[width=0.95\linewidth]{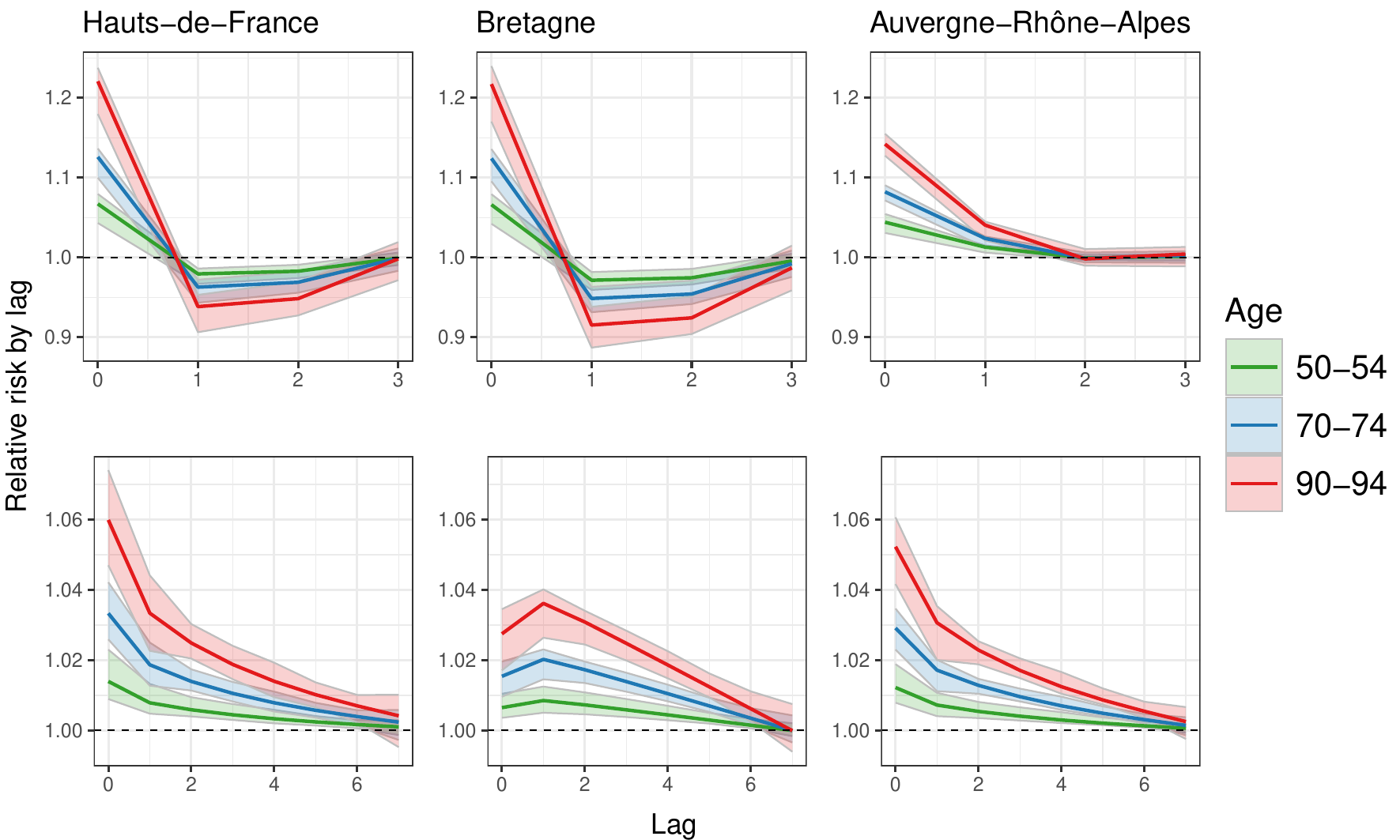}
    \caption{Estimated relative risk per lag at the 99.5th percentile of the region-specific weekly average temperatures (top row) and ILI exceedance rates (bottom row) in Hauts-de-France (left), Bretagne (middle), and Auverge-Rhône-Alpes (right) for the age groups 50–54, 70–74, and 90–94. Shaded areas represent 95$\%$ point-wise confidence intervals. \label{fig:RRlag}}
\end{figure}

Figure~\ref{fig:RRlag} presents the estimated lag-specific RRs at the 99.5th percentile of the region-specific weekly average temperatures (top row) and ILI exceedance rates (bottom row), for the same set of regions and age groups as in Figure~\ref{fig:RRoverall}. The effect of heat appears to be largely immediate, with elevated RRs \added{(RR $>$ 1)} at lag 0, followed by a harvesting effect reflected in \added{relative risks below 1 in subsequent lags}. In contrast, the effect of ILI shows no evidence of harvesting: it peaks at lag 0 and gradually diminishes across subsequent lags. We refer to the \textit{Online Appendix} for visualisations of the lag-specific RRs of the other administrative regions in France.

\paragraph{Relative risk surface.} Lastly, to provide a comprehensive overview of how lagged temperature and ILI exceedances are associated with mortality, we present the relative risk (RR) surface plot. This plot depicts the estimated associations across the entire grid of predictor and lag values.

\begin{figure}[!ht]
    \centering
    \includegraphics[width=0.95\linewidth]{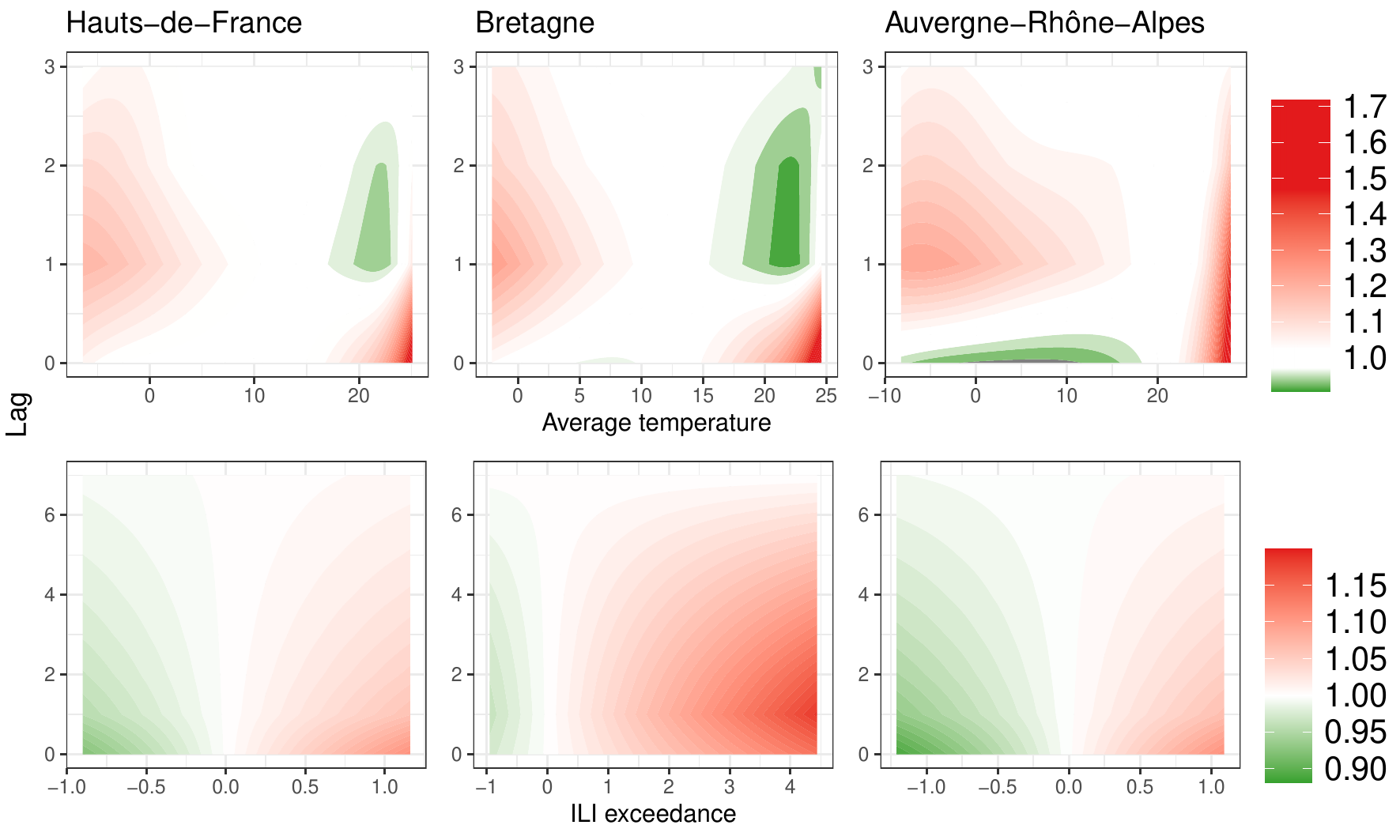}
    \caption{\added{Estimated relative risk (RR) surfaces of the lag-specific temperature– and ILI–mortality associations in Hauts-de-France (left), Bretagne (middle), and Auvergne-Rhône-Alpes (right) for the age group 90–94. Red areas indicate an increased RR, while green areas indicate a decreased RR.\label{fig:RRsurface}}}
\end{figure}

Figure~\ref{fig:RRsurface} displays the RR surface plots for Hauts-de-France, Bretagne, and Auvergne-Rhône-Alpes in the age group 90–94. Red areas correspond to regions where RR $>$ 1 (increased mortality risk), and green areas to regions where RR $<$ 1 (decreased mortality risk). The surfaces interpolate smoothly between weekly lags. In the top row, which shows the RR surfaces for temperature, we observe clear red areas at the bottom right of each panel, indicating an immediate and strong increase in mortality risk during periods of extreme heat. \added{Directly above these red zones, green zones emerge in Hauts-de-France and Bretagne, reflecting the harvesting effect: mortality falls below expected levels after a hot week. This is less the case for Auvergne-Rhône-Alpes. On the left-hand side of each panel, the cold effect is visible. This effect appears to be longer-lasting, but smaller than that of heat, and no harvesting is observed.} In the bottom row, the RR surface plots for ILI exceedances reveal an effect that grows with the level of exceedance and gradually decreases with subsequent lags. In the \textit{Online Appendix} we provide the RR surface plots for age group 90-94 for all the French administrative regions under consideration.

\subsection{Predictive performance} \label{subsec:performance}
To ensure a fair evaluation of the predictive performance of our proposed mortality model (Eqs.~\eqref{eq:modeldist} and \eqref{eq:modelstructure}), we restrict our attention to pre-pandemic data and avoid any distortions induced by the COVID-19 pandemic. Therefore, we recalibrate the model on data from 1990 to 2014. Using the recalibrated model, we then forecast death counts for 2015–2019 following the procedure outlined in Section~\ref{subsec:forecasting}. The forecasted counts are then compared with observed deaths using multiple metrics and scoring rules, along with several benchmark models. 

\paragraph{Proposed model vs.~benchmarks.} We compare the mortality forecasts with those from several benchmark models, each assuming a negative binomial distribution for the death counts:
\begin{align*}
    D_{x,t,w,r} \sim \text{NB}\!\left(E_{x,t,w,r} \cdot \mu_{x,t,w,r}, \, \phi_{x,r}\right).
\end{align*}

\added{We consider three benchmark specifications that do not include exogenous covariates:}
\begin{itemize}
    \item[-] \added{\textbf{Model 1}: $\log \mu_{x,t,w,r} = \alpha_{x,r} + \beta_x \kappa_{t,r}$ \hfill (annual, regional).}
    \item[-] \added{\textbf{Model 2}: $\log \mu_{x,t,w,r} = \alpha_{x} + \beta_x \kappa_t + \gamma_x \lambda_w$ \hfill (weekly, national).}
    \item[-] \added{\textbf{Model 3}: $\log \mu_{x,t,w,r} = \alpha_{x,r} + \beta_x \kappa_{t,r} + \gamma_x \lambda_{w,r}$ \hfill (weekly, regional).}
\end{itemize}

\added{We then extend Model~3 by incorporating temperature and influenza incidence through a DLNM structure, to assess the added value of exogenous drivers in terms of out-of-sample predictive performance. We consider three covariate-augmented model specifications:}
\begin{itemize}
    \item[-] \added{\textbf{Model 4}: $\log \mu_{x,t,w,r} = \text{Model 3} + \delta_x f_r^{(1)}(\text{Tavg}_{t,w,r})$ \hfill (weekly, regional).}
    \item[-] \added{\textbf{Model 5}: $\log \mu_{x,t,w,r} = \text{Model 3} + \epsilon_x f_r^{(2)}(\text{ILI}_{t,w,r})$ \hfill (weekly, regional).}
    \item[-] \added{\textbf{Model 6}: $\log \mu_{x,t,w,r} = \text{Model 3} + \delta_x f_r^{(1)}(\text{Tavg}_{t,w,r}) + \epsilon_x f_r^{(2)}(\text{ILI}_{t,w,r})$ \hfill (weekly, regional).}
\end{itemize}

\added{Model 6 corresponds to our proposed full model specification from Section~\ref{sec:modelspecification} (Eqs.~\eqref{eq:modeldist} and \eqref{eq:modelstructure}). Models 4 and 5 represent special cases of this framework. Model 4 isolates the effect of temperature, while Model 5 isolates the effect of influenza. This allows us to quantify the contribution of each covariate to the predictive performance.}

\added{For Models 1--6, we capture the dynamics of the latent mortality index $\hat{\kappa}_{t,r}$ or $\hat{\kappa}_t$ with an ARIMA(0,1,1) model, fitted separately for each region if the model is regional.}\footnote{We compared AIC and BIC values across a range of candidate ARIMA specifications over all regions jointly.} \added{We account for regional dependence across the fitted processes via a Gaussian copula applied to the vector of standardized innovations, see Section~\ref{subsec:forecasting}. In Models 4--6, we further distinguish between two forecasting scenarios for the exogenous covariates over 2015--2019. For the temperature ($\text{Tavg}_{t,w,r}$) and ILI ($\text{ILI}_{t,w,r}$) covariates, we consider the following two cases:}
\begin{enumerate}
    \item \added{\textbf{Observed covariates.} We treat future covariate values as known and directly substitute the observed values of $\text{Tavg}_{t,w,r}$ and $\text{ILI}_{t,w,r}$ into the model. This setting isolates the contribution of the mortality model conditional on exogenous inputs. This scenario is hypothetical but becomes more realistic in short-term forecasting settings, where meteorological forecasts, in particular temperature, are typically accurate at short horizons.}

    \item \added{\textbf{Simulated covariates.} We generate out-of-sample forecasts for temperature and ILI using stochastic time-series models. Specifically, for each region, we fit a SARIMA$(1,0,1)(1,1,1)_{52}$ model for temperature and ILI. We again used AIC/BIC to select the orders jointly across regions. Regional dependence is accounted for using the copula approach in Section~\ref{subsec:forecasting}.}
\end{enumerate}

\added{In the simulated covariate setting, we emphasize that the SARIMA models are fitted to and forecast the raw ILI incidence rates (transformed by $\log(\cdot + 0.01)$ to stabilize variance). To obtain the exceedance anomalies required as input to the DLNM in Models 5 and 6, we subtract the historical region- and week-specific 90th percentiles $q_{90}(r,w)$ (computed over the period 1990-2014) from the forecasted raw rates. This ensures consistency between the estimation and forecasting stages.}

\added{Lastly, in Suppl.~Mat.~\ref{app:compareSARIMAX}, we explore whether exogenous variables should enter the dynamics of $\kappa_{t,r}$ and ILI, for instance by including temperature as an external regressor for ILI, and temperature and ILI as external regressors for $\kappa_{t,r}$. The motivation is that long-term mortality trends might be linked to gradual environmental or epidemiological changes, and that temperature may drive influenza dynamics. However, for ILI, the SARIMA without temperature yielded a lower BIC for all regions. For $\kappa_{t,r}$, the ARIMA without external regressors performed better for 11 out of 12 regions, with only negligible improvement for the remaining region. These extensions were therefore not retained.}

\paragraph{\added{Conditional predictive accuracy.}} \added{To evaluate how well the mortality model captures the association between mortality and its environmental drivers, we first consider a conditional evaluation framework in which temperature and ILI are treated as observed. This allows us to assess the model’s ability to reproduce observed mortality dynamics given the realised exogenous environment, rather than its full ex ante forecasting performance in which these drivers are unknown. In this setting, we compute point forecasts based on the predicted mean of the ARIMA(0,1,1) process for the latent mortality index, combined with observed values of the covariates where applicable. We evaluate predictive accuracy using the root mean squared error (RMSE) and the mean absolute error (MAE).} We calculate RMSE and MAE per age group by averaging across time points and regions. For age group $x$, we obtain:
\begin{align*}
    \text{RMSE}_{x} &= \sqrt{\dfrac{1}{n_x} \displaystyle \sum_{r \in \mathcal{R}} \sum_{t = 2015}^{2019} \sum_{w = 1}^{\added{W_t}} \left( d_{x,t,w,r} - E_{x,t,w,r} \, \hat{\mu}_{x,t,w,r}\right)^2}, \\
    \text{MAE}_{x} &= \dfrac{1}{n_x} \sum_{r \in \mathcal{R}}\sum_{t = 2015}^{2019} \sum_{w = 1}^{\added{W_t}} \left| d_{x,t,w,r} - E_{x,t,w,r} \, \hat{\mu}_{x,t,w,r}\right|,
\end{align*}
where $n_x$ is the number of observations in the data set per age group, $\mathcal{R}$ the set of regions, \added{$W_t$ the number of ISO weeks in year $t$}, and $\hat{\mu}_{x,t,w,r}$ is the forecasted death rate for the age group $x$ in the region $r$ at time point $(t,w)$. In addition, we calculate the overall RMSE and MAE by also including the sum over the age groups in the formulae.

Table~\ref{tab:forecast_errors} reports the overall RMSE and MAE as well as the corresponding values by age group. \added{Our proposed specification, incorporating both temperature and ILI (Model 6), consistently achieves the best performance across age groups, except for the 50–54 age group. The second-best performing model varies between Model 4 and Model 5. Overall, when exogenous conditions are accurately reflected, Models 4–6 systematically outperform the benchmark specifications (Models 1–3). Importantly, incorporating both temperature and ILI (Model 6) leads to a consistent improvement over specifications that include only temperature (Model 4) or only influenza (Model 5). This further indicates that both drivers contain complementary information to explain short-term mortality fluctuations. We also observe that differences in error metrics are relatively small for younger age groups, mainly due to the lower degree of seasonality and variability in mortality at these ages. Finally, the results clearly indicate that a weekly, regional specification provides substantial gains over both the national weekly model (Model 2) and the regional annual model (Model 1).}
\sisetup{detect-all = true}
\begin{table}[!ht]
\centering
\begin{adjustbox}{max width=\textwidth}
\begin{tabular}{l *{6}{S} *{6}{S}}
\toprule 
 & \multicolumn{6}{c}{\textbf{RMSE}} & \multicolumn{6}{c}{\textbf{MAE}} \\
\cmidrule(lr){2-7} \cmidrule(lr){8-13}
Age  
  & \text{Model 1} & \text{Model 2} & \text{Model 3} & \text{Model 4} & \text{Model 5} & \text{Model 6}
  & \text{Model 1} & \text{Model 2} & \text{Model 3} & \text{Model 4} & \text{Model 5} & \text{Model 6}\\
\midrule
50--54 & 3.07 & 3.19 & 3.06 & 3.06 & \bfseries 3.05 & 3.07 & 2.41 & 2.49 & 2.39 & 2.39 & \bfseries 2.39 & 2.40 \\ 
55--59 & 3.67 & 3.80 & 3.64 & 3.63 & 3.64 & \bfseries3.62 & 2.84 & 2.94 & 2.81 & 2.81 & 2.81 & \bfseries 2.80 \\ 
60--64 & 4.50 & 4.66 & 4.41 & 4.39 & 4.41 & \bfseries4.37 & 3.47 & 3.58 & 3.41 & 3.40 & 3.41 & \bfseries 3.39 \\ 
65--69 & 5.67 & 5.85 & 5.50 & 5.46 & 5.53 & \bfseries 5.41 & 4.39 & 4.48 & 4.28 & 4.25 & 4.30 & \bfseries 4.21 \\ 
70--74 & 5.93 & 5.95 & 5.67 & 5.60 & 5.69 & \bfseries 5.53 & 4.52 & 4.54 & 4.35 & 4.30 & 4.36 & \bfseries 4.24 \\ 
75--79 & 7.23 & 7.14 & 6.61 & 6.48 & 6.63 & \bfseries 6.36 & 5.48 & 5.42 & 5.07 & 5.00 & 5.08 & \bfseries 4.91 \\ 
80--84 & 11.65 & 10.66 & 9.50 & 9.31 & 9.52 & \bfseries 9.01 & 8.70 & 8.13 & 7.29 & 7.19 & 7.31 & \bfseries 6.97 \\ 
85--89 & 18.65 & 15.55 & 13.73 & 13.35 & 13.72 & \bfseries 12.62 & 13.34 & 11.62 & 10.17 & 9.97 & 10.19 & \bfseries 9.51 \\ 
90--94 & 21.92 & 16.82 & 15.43 & 15.18 & 15.33 &  \bfseries 14.18 & 15.40 & 12.41 & 11.14 & 11.09 & 11.10 & \bfseries 10.50 \\ 
95+ & 14.22 & 10.72 & 10.34 & 10.28 & 10.00 & \bfseries 9.87 & 10.83 & 8.10 & 7.92 & 7.76 & 7.69 & \bfseries 7.46 \\ 
\midrule
Overall & 11.53 & 9.59 & 8.78 & 8.64 & 8.73 & \bfseries 8.25 & 7.14 & 6.37 & 5.88 & 5.82 & 5.86 & \bfseries 5.64     \\
\bottomrule
\end{tabular}
\end{adjustbox}
\caption{\added{Out-of-sample predictive performance (RMSE and MAE) across age groups for the six competing mortality models, evaluated over the 2015–2019 test period. Models 4–6 use observed covariate values. The best-performing model within each age group is highlighted in bold. \label{tab:forecast_errors}}}
\end{table}

\paragraph{Evaluation of probabilistic forecasts.} To assess the accuracy of the probabilistic forecast distributions of the future death counts, we put focus on two commonly used scoring rules: the logarithmic score (LogS) and the continuous ranked probability score (CRPS):
\begin{align*}
    \text{LogS}_x &= - \frac{1}{n_x} \sum_{r\in \mathcal{R}} \sum_{t=2015}^{2019} \sum_{w = 1}^{\added{W_t}} \log \left[ f_{x,t,w,r} \left( d_{x,t,w,r} \right) \right], \\
    \text{CRPS}_x &= \frac{1}{n_x} \sum_{r\in \mathcal{R}} \sum_{t=2015}^{2019} \sum_{w = 1}^{\added{W_t}} \int_{\mathbb{R}} \left( F_{x,t,w,r}(s) - \mathbbm{1}\left\{d_{x,t,w,r} \le s \right\}\right) d s.
\end{align*}
Here, $n_x$ is the number of observations for age group $x$, $F_{x,t,w,r}$ is the predictive cumulative distribution function, and $f_{x,t,w,r}$ its density. Since the predictive distributions for the death counts are available through simulated forecast scenarios $E_{x,t,w,r} \cdot \hat{\mu}_{x,t,w,r}^{(i)}$, for $i=1,2,...,5\,000$, we approximate $F$ and $f$ by their empirical counterparts constructed from these simulations. The scores are evaluated using the \texttt{R} package \texttt{scoringRules} \citep{jordan2019evaluating}. These scoring rules have also been applied in a Bayesian mortality modeling approach by \cite{barigou2023bayesian}.

\added{Table~\ref{tab:probabilistic_errors} reports the CRPS and LogS by age group, together with overall values, for all model specifications. For Models 4–6, we show the results under both observed and simulated covariate scenarios. This allows us to distinguish between conditional predictive performance and fully out-of-sample forecasting performance.}

\begin{table}[!ht]
\centering
\begin{adjustbox}{max width=\textwidth}
\begin{tabular}{l *{3}{S} *{3}{S} *{3}{S}}
\toprule 
\multicolumn{10}{c}{\textbf{Continuous Ranked Probability Score (CRPS)}} \\
\toprule
 & & & & \multicolumn{3}{c}{\text{\small Observed covariates}} & \multicolumn{3}{c}{\text{\small Simulated covariates}}  \\
 \cmidrule(lr){5-7} \cmidrule(lr){8-10}
Age  
  & \text{Model 1} & \text{Model 2} & \text{Model 3} & \text{Model 4} & \text{Model 5} & \text{Model 6}
   & \text{Model 4} & \text{Model 5} & \text{Model 6} \\
\midrule
50--54 & 1.66 & 1.72 & 1.66 & 1.66 & \bfseries 1.65 & 1.66 & 1.66 & 1.66 & 1.66 \\ 
  55--59 & 2.01 & 2.09 & 1.99 & 1.99 & 1.99 & \bfseries 1.98 & 1.99 & 1.99 & 1.99 \\ 
  60--64 & 2.49 & 2.56 & 2.44 & 2.44 & 2.44 & 2.42 & 2.43 & 2.45 & \bfseries 2.41 \\ 
  65--69 & 3.15 & 3.24 & 3.06 & 3.04 & 3.08 & 3.01 & 3.04 & 3.08 &  \bfseries3.01 \\ 
  70--74 & 3.23 & 3.26 & 3.09 & 3.06 & 3.11 & \bfseries 3.02 & 3.06 & 3.11 & 3.02 \\ 
  75--79 & 3.93 & 3.91 & 3.62 & 3.56 & 3.64 & \bfseries 3.50 & 3.59 & 3.65 & 3.54 \\ 
  80--84 & 6.28 & 5.85 & 5.20 & 5.11 & 5.21 & \bfseries 4.95 & 5.15 & 5.25 & 5.06 \\ 
  85--89 & 9.78 & 8.42 & 7.35 & 7.16 & 7.35 & \bfseries 6.82 & 7.27 & 7.46 & 7.08 \\ 
  90--94 & 11.34 & 8.98 & 8.09 & 8.02 & 8.05 & \bfseries 7.55 & 8.12 & 8.22 & 7.89 \\ 
  95+ & 7.55 & 5.71 & 5.56 & 5.48 & 5.41 & \bfseries 5.27 & 5.55 & 5.52 & 5.52 \\ 
\midrule
Overall & 5.14 & 4.57 & 4.21 & 4.15 & 4.19 & \bfseries 4.02 & 4.19 & 4.24 & 4.12 \\
\bottomrule
\end{tabular}
\end{adjustbox}

\vspace{0.5cm}

\begin{adjustbox}{max width=\textwidth}
\begin{tabular}{l *{3}{S} *{3}{S} *{3}{S}}
\toprule 
\multicolumn{10}{c}{\textbf{Logarithmic Score (LogS)}} \\
\toprule 
 & & & & \multicolumn{3}{c}{\text{\small Observed covariates}} & \multicolumn{3}{c}{\text{\small Simulated covariates}}  \\
 \cmidrule(lr){5-7} \cmidrule(lr){8-10}
Age  
  & \text{Model 1} & \text{Model 2} & \text{Model 3} & \text{Model 4} & \text{Model 5} & \text{Model 6}
   & \text{Model 4} & \text{Model 5} & \text{Model 6} \\
\midrule
50--54 & \bfseries 2.41 & 2.46 & 2.42 & 2.42 & 2.42 & 2.42 & 2.42 & 2.42 & 2.42 \\ 
55--59 & 2.66 & 2.69 & 2.65 & 2.65 & 2.64 & \bfseries 2.64 & 2.65 & 2.65 & 2.64 \\ 
60--64 & 2.88 & 2.92 & 2.87 & 2.86 & 2.86 & 2.85 & 2.86 & 2.87 & \bfseries 2.85 \\ 
65--69 & 3.12 & 3.16 & 3.09 & 3.09 & 3.10 & 3.08 & 3.08 & 3.10 &  \bfseries 3.08 \\ 
70--74 & 3.13 & 3.16 & 3.09 & 3.08 & 3.09 & \bfseries 3.06 & 3.08 & 3.09 & 3.07 \\ 
75--79 & 3.34 & 3.40 & 3.26 & 3.25 & 3.26 & \bfseries 3.22 & 3.25 & 3.27 & 3.24 \\ 
80--84 & 3.80 & 3.76 & 3.61 & 3.59 & 3.61 & \bfseries 3.57 & 3.61 & 3.62 & 3.59 \\ 
85--89 & 4.25 & 4.15 & 3.95 & 3.93 & 3.96 & \bfseries 3.89 & 3.95 & 3.97 & 3.92 \\ 
90--94 & 4.40 & 4.20 & 4.05 & 4.05 & 4.05 & \bfseries 3.98 & 4.07 & 4.07 & 4.02 \\ 
95+    & 3.95 & 3.68 & 3.66 & 3.65 & 3.64 & \bfseries 3.61 & 3.66 & 3.66 & 3.65 \\ 
\midrule
Overall & 3.39 & 3.36 & 3.27 & 3.25 & 3.26 & \bfseries 3.23 & 3.26 & 3.27 & 3.25 \\ 
\bottomrule
\end{tabular}
\end{adjustbox}
\caption{\added{Out-of-sample probabilistic forecast performance (CRPS and LogS) across age groups for the six competing mortality models, evaluated over the 2015–2019 test period. Models 4–6 are reported under both observed and simulated covariate scenarios. The best-performing model within each age group is highlighted in bold. \label{tab:probabilistic_errors}}}
\end{table}

\added{When covariates are included and treated as observed, Model 6 generally achieves the best performance, particularly at older ages. In these age groups, combining temperature and influenza information leads to consistent improvements over both Model 3 and the single-covariate specifications (Models 4 and 5). This also clearly indicates that the two covariates provide complementary information for explaining short-term mortality fluctuations.} 

\added{Importantly, these improvements persist when moving to the fully out-of-sample setting with simulated covariates. Model 6 continues to outperform Model 3 across the older age groups under both CRPS and LogS. This thus demonstrates that the benefits of incorporating temperature and influenza are not limited to the conditional setting but also translate into improved probabilistic forecasts when these covariates are forecasted themselves.}

\paragraph{Calibration of prediction intervals.} We assess the calibration of the constructed prediction intervals using two measures: (1) the coverage ratio, which evaluates the proportion of observed death counts that fall within the $100(1-\alpha)\%$ prediction interval $[\hat d_{x,t,w,r}^{\text{low}}, \hat d_{x,t,w,r}^{\text{up}}]$, and (2) the interval score proposed by \cite{gneiting2007strictly}, which penalizes both the width of the interval and any deviations of the observed values outside the interval. We compute them as follows:
\begin{align*}
\text{Coverage Ratio}_x &= \frac{1}{n_x} \sum_{r\in \mathcal{R}} \sum_{t=2015}^{2019} \sum_{w = 1}^{\added{W_t}} \mathbbm{1}\left\{ d_{x,t,w,r} \in [\hat{d}_{x,t,w,r}^{\text{low}}, \hat{d}_{x,t,w,r}^{\text{up}}] \right\},  \\
\text{Interval Score}_x 
&= \frac{1}{n_x} \sum_{r\in \mathcal{R}} \sum_{t=2015}^{2019} \sum_{w = 1}^{\added{W_t}} \Bigg[ (\hat d_{x,t,w,r}^{\text{up}} - \hat d_{x,t,w,r}^{\text{low}}) 
\,+ \\ &\hspace{-1.5cm}\frac{2}{\alpha} (\hat d_{x,t,w,r}^{\text{low}} -  d_{x,t,w,r}) \mathbbm{1}\left\{ d_{x,t,w,r} < \hat d_{x,t,w,r}^{\text{low}} \right\} 
+  \frac{2}{\alpha} (d_{x,t,w,r} - \hat d_{x,t,w,r}^{\text{up}}) \mathbbm{1}\left\{ d_{x,t,w,r} > \hat d_{x,t,w,r}^{\text{up}} \right\}\Bigg].
\end{align*}
where $n_x$ is the number of observations for the age group $x$. We compute $95\%$ pointwise intervals $[\hat d_{x,t,w,r}^{\text{low}}, \hat d_{x,t,w,r}^{\text{up}}]$ based on 5\,000 simulated forecast scenarios $E_{x,t,w,r} \cdot \hat{\mu}^{(i)}_{x,t,w,r}$. Values for the coverage ratio close to the nominal level of 0.95 indicate well-calibrated intervals.  

\added{Table~\ref{tab:COV.IS} reports the coverage ratios and interval scores by age group for all six model specifications. We show the results of Models 4-6 under both observed and simulated covariates. In general, the coverage ratios are close to the nominal level across most of the models and age groups. This indicates that the predictive distributions are overall well-calibrated. We only observe minor differences, although Model 6 under the simulated covariate setting attains the highest coverage in several of the older age groups.}

\begin{table}[!ht]
\centering
\begin{adjustbox}{max width=\textwidth}
\begin{tabular}{l *{3}{S} *{3}{S} *{3}{S}}
\toprule 
\multicolumn{10}{c}{\textbf{Coverage ratio}} \\
\toprule
 & & & & \multicolumn{3}{c}{\text{\small Observed covariates}} & \multicolumn{3}{c}{\text{\small Simulated covariates}}  \\
 \cmidrule(lr){5-7} \cmidrule(lr){8-10}
Age  
  & \text{Model 1} & \text{Model 2} & \text{Model 3} & \text{Model 4} & \text{Model 5} & \text{Model 6}
   & \text{Model 4} & \text{Model 5} & \text{Model 6} \\
\midrule
50--54 & \bfseries 0.91 & 0.90 & 0.91 & 0.91 & 0.91 & 0.91 & 0.91 & 0.91 & 0.91 \\ 
55--59 & \bfseries 0.92 & 0.91 & 0.92 & 0.92 & 0.92 & 0.92 & 0.92 & 0.92 & 0.92 \\ 
60--64 & 0.91 & 0.90 & 0.91 & 0.91 & 0.91 & 0.91 & 0.91 & 0.91 & \bfseries 0.91 \\ 
65--69 & 0.91 & 0.88 & 0.90 & 0.91 & 0.90 & 0.91 & 0.90 & 0.90 & \bfseries 0.91 \\ 
70--74 & 0.93 & 0.90 & 0.93 & 0.92 & 0.93 & 0.93 & 0.93 & 0.93 & \bfseries 0.94 \\ 
75--79 & 0.93 & 0.90 & 0.94 & 0.93 & 0.93 & 0.93 & 0.94 & 0.93 & \bfseries 0.94 \\ 
80--84 & 0.94 & 0.91 & 0.95 & 0.94 & 0.95 & 0.95 & 0.95 & 0.95 & \bfseries 0.96 \\ 
85--89 & 0.94 & 0.90 & 0.94 & 0.94 & 0.94 & 0.93 & 0.94 & 0.94 & \bfseries 0.95 \\ 
90--94 & 0.93 & 0.90 & \bfseries 0.94 & 0.92 & 0.94 & 0.92 & 0.94 & 0.94 & 0.94 \\ 
95+ & 0.95 & 0.93 & 0.96 & 0.94 & \bfseries 0.96 & 0.94 & 0.95 & 0.96 & 0.95 \\ 
\midrule
Overall & 0.93 & 0.90 & 0.93 & 0.92 & 0.93 & 0.93 & 0.93 & 0.93 & \bfseries 0.93 \\ 
\bottomrule
\end{tabular}
\end{adjustbox}

\vspace{0.5cm}

\begin{adjustbox}{max width=\textwidth}
\begin{tabular}{l *{3}{S} *{3}{S} *{3}{S}}
\toprule 
\multicolumn{10}{c}{\textbf{Interval Score}} \\
\toprule 
 & & & & \multicolumn{3}{c}{\text{\small Observed covariates}} & \multicolumn{3}{c}{\text{\small Simulated covariates}}  \\
 \cmidrule(lr){5-7} \cmidrule(lr){8-10}
Age  
  & \text{Model 1} & \text{Model 2} & \text{Model 3} & \text{Model 4} & \text{Model 5} & \text{Model 6}
   & \text{Model 4} & \text{Model 5} & \text{Model 6} \\
\midrule
50--54 & 13.45 & 14.02 & \bfseries13.39 & 13.46 & 13.47 & 13.56 & 13.48 & 13.39 & 13.54 \\ 
55--59 & 16.98 & 17.63 & 16.77 & 16.64 & 16.76 & \bfseries 16.62 & 16.75 & 16.82 & 16.65 \\ 
60--64 & 20.87 & 22.19 & 20.38 & 20.13 & 20.27 & 20.11 & 20.17 & 20.30 & \bfseries20.01 \\ 
65--69 & 26.25 & 28.64 & 25.35 & 25.07 & 25.57 & 24.97 & 24.85 & 25.52 & \bfseries 24.69 \\ 
70--74 & 27.04 & 28.49 & 25.86 & 25.57 & 25.93 & \bfseries 25.04 & 25.60 & 26.00 & 25.19 \\ 
75--79 & 33.98 & 35.70 & 30.26 & 29.84 & 30.32 & \bfseries 29.44 & 30.34 & 30.55 & 30.00 \\ 
80--84 & 55.64 & 53.33 & 43.12 & 41.75 & 43.24 & \bfseries 41.01 & 43.22 & 43.65 & 42.54 \\ 
85--89 & 94.51 & 79.95 & 64.70 & 61.77 & 64.93 & \bfseries 58.73 & 65.67 & 66.37 & 63.21 \\ 
90--94 & 116.51 & 86.02 & 75.10 & 73.33 & 74.92 & \bfseries 68.45 & 77.77 & 77.65 & 74.71 \\ 
95+ & 64.76 & 46.76 & 45.82 & 45.93 & 44.76 & \bfseries 44.51 & 46.77 & 45.88 & 46.63 \\ 
\midrule
Overall & 47.00 & 41.27 & 36.07 & 35.35 & 36.02 & \bfseries 34.24 & 36.46 & 36.61 & 35.72 \\ 
\bottomrule
\end{tabular}
\end{adjustbox}
\caption{\added{Out-of-sample probabilistic forecast performance (coverage ratio and interval score) across age groups for the six competing mortality models, evaluated over the 2015–2019 test period. Models 4–6 are reported under both observed and simulated covariate scenarios. The best-performing model within each age group is highlighted in bold. \label{tab:COV.IS}}}
\end{table}

\added{We observe clearer differences in the interval score. Moving from the simpler benchmark models (Models 1–2) to the weekly, regional specification (Model 3) leads to a substantial reduction in interval scores among all age groups. Moreover, if we add exogenous covariates, the performance further improves, particularly at older ages.}

\added{Model 6 consistently achieves the lowest interval scores in the older age groups and under the observed covariate setting. We conclude that combining temperature and influenza information leads to sharper and better-calibrated predictive intervals. Importantly, these gains largely persist under the simulated covariate setting: Model 6 continues to outperform Model 3 for the older age groups. This again demonstrates the benefits of incorporating temperature and influenza both in terms of conditional predictive performance and fully out-of-sample forecasting performance. Similar to previous results, the differences between models remain small at younger ages.}

\paragraph{\added{Uncertainty decomposition.}} 
\added{Table~\ref{tab:variance_decomp_all} reports the share of the total 95\% predictive interval width that is attributable to the simulated time-series components. For each age group, this share is computed as the ratio between the summed width of the 95\% interval obtained from simulations without negative binomial sampling and the summed width of the corresponding 95\% interval obtained after adding negative binomial sampling variation. The remaining share is therefore attributed to negative binomial sampling variability.}

\begin{table}[!ht]
\centering
\begin{adjustbox}{max width=\textwidth}
\begin{tabular}{l *{3}{c} *{3}{c} *{3}{c}}
\toprule 
 & & &  
 & \multicolumn{3}{c}{\textbf{Observed covariates}} 
 & \multicolumn{3}{c}{\textbf{Simulated covariates}} \\
 \cmidrule(lr){5-7} \cmidrule(lr){8-10}
Age 
& Model 1 & Model 2 & Model 3 
& Model 4 & Model 5 & Model 6
& Model 4 & Model 5 & Model 6 \\
\midrule
50--54 & 0.04 & 0.03 & 0.04 & 0.04 & 0.04 & 0.04 & 0.07 & 0.04 & 0.07 \\ 
55--59 & 0.05 & 0.04 & 0.05 & 0.05 & 0.05 & 0.05 & 0.08 & 0.05 & 0.08 \\ 
60--64 & 0.08 & 0.06 & 0.08 & 0.08 & 0.08 & 0.08 & 0.12 & 0.08 & 0.13 \\ 
65--69 & 0.13 & 0.11 & 0.13 & 0.14 & 0.14 & 0.14 & 0.18 & 0.14 & 0.19 \\ 
70--74 & 0.15 & 0.13 & 0.17 & 0.17 & 0.17 & 0.17 & 0.23 & 0.17 & 0.24 \\ 
75--79 & 0.19 & 0.17 & 0.22 & 0.23 & 0.22 & 0.22 & 0.30 & 0.22 & 0.32 \\ 
80--84 & 0.21 & 0.21 & 0.26 & 0.28 & 0.26 & 0.28 & 0.39 & 0.27 & 0.42 \\ 
85--89 & 0.18 & 0.20 & 0.25 & 0.27 & 0.25 & 0.28 & 0.44 & 0.27 & 0.49 \\ 
90--94 & 0.12 & 0.14 & 0.17 & 0.20 & 0.17 & 0.20 & 0.41 & 0.20 & 0.48 \\ 
95+    & 0.03 & 0.04 & 0.05 & 0.05 & 0.04 & 0.05 & 0.29 & 0.08 & 0.36 \\ 
\bottomrule
\end{tabular}
\end{adjustbox}
\caption{\added{Proportion of uncertainty attributed to stochastic components across age groups for all model specifications, evaluated over the 2015–2019 test period. Stochastic uncertainty reflects variation arising from the latent mortality index and, where applicable, from forecasted exogenous covariates. Models 4--6 are shown under both observed and simulated covariate scenarios.}
\label{tab:variance_decomp_all}}
\end{table}

\added{For Models 1--3, the stochastic component reflects uncertainty arising from the latent mortality index $\kappa_t$ or $\kappa_{t,r}$. Here, all short-term fluctuations not captured by this latent index are absorbed by the negative binomial sampling variability. As a result, the proportion of uncertainty attributed to the stochastic components of the model remains relatively small and stable across age groups.}

\added{For Models 4--6 under the observed covariate setting, observed temperature and influenza values enter the model deterministically through the DLNM specification. They therefore do not introduce additional stochastic uncertainty. The decomposition remains broadly similar to Model 3.}

\added{We observe more pronounced changes under the simulated covariate setting. In this case, we generate future temperature and influenza trajectories from the fitted SARIMA models, and the resulting covariate forecast uncertainty is propagated through the DLNM into the mortality forecasts. This set-up increases the share of total predictive uncertainty attributed to the stochastic components in the model, particularly at older ages where exogenous drivers play a more important role in mortality fluctuations.}

\added{Overall, the decomposition shows that Model 6 under the simulated covariate setting shifts a substantial portion of predictive uncertainty away from the negative binomial sampling component towards structured variation driven by temperature and influenza, especially in the simulated setting. Relative to Model 3, this implies that a considerably larger share of the uncertainty in mortality forecasts can be explained by identifiable exogenous drivers rather than being treated as unstructured noise.}

\begin{figure}[!ht]
    \centering
    \includegraphics[width=0.95\linewidth]{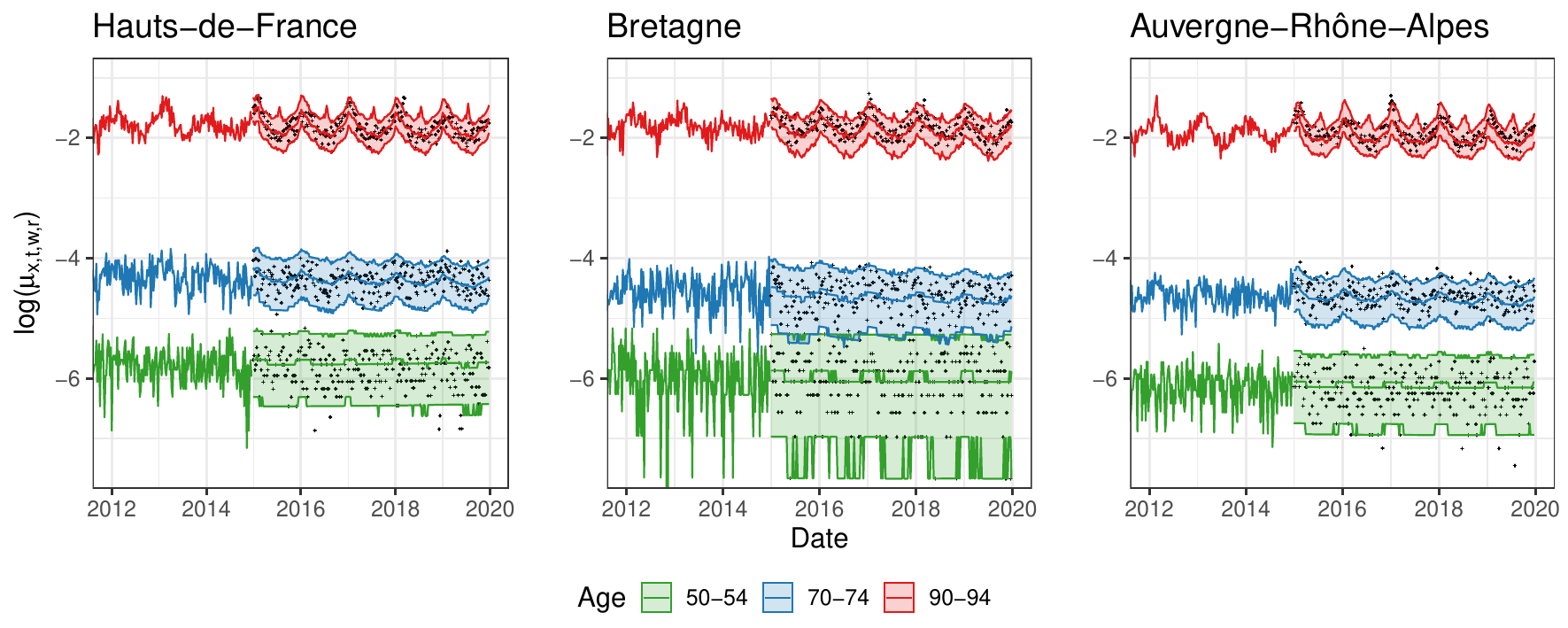}
    \caption{\added{Forecasted female weekly death rates on log-scale according to Model 6 (simulated covariates) in Hauts-de-France (left), Bretagne (middle), and Auvergne-Rhône-Alpes (right) over the period 2015–2019 using simulated exogenous factors. We show the $95\%$ prediction intervals based on 5\,000 simulated trajectories of temperature, ILI incidence rates, and the latent mortality index for the age groups 50-54, 70-74, and 90-94. The observed death rates from 2015-2019 are visualized in black with a '+'. Wiggly intervals for the 50-54 age group in Bretagne reflect sampling variability due to low death counts.}}
    \label{fig:predperform}
\end{figure}

\paragraph{Visualisation.} We evaluate the predictive performance of our proposed model visually in Figure~\ref{fig:predperform} by plotting the observed death rates against the corresponding 95$\%$ prediction intervals for three administrative regions in France and for the age groups 50-54, 70-74, and 90-94. Specifically, we focus on \added{Model 6 under the simulated covariate setting}, which incorporates simulated trajectories of the latent mortality index, temperature, and ILI incidence rates in the forecasting procedure. We find that the observed death counts fall mostly within the constructed intervals, even during periods with increased mortality in summer or winter.

\subsection{Assessing post-COVID excess mortality} \label{subsec:casestudy.forecasting}
Lastly, we forecast death rates for the 2020–2024 period using our proposed mortality model calibrated on 1990–2019 data. We focus on \added{Model 6} and condition the forecasts on the observed exogenous characteristics (temperature and influenza activity). By doing so, we can isolate the effect of COVID-19, since the model already accounts for the observed temperature conditions and the markedly lower influenza activity during the pandemic. This leads, as such, to a more accurate assessment of excess mortality.

Hereto, we first collect daily gridded average temperatures for the 12 administrative regions of France from the E-OBS data set (Copernicus Climate Data Store) for the period January 1, 2020, to December 31, 2024, and construct population-weighted regional, weekly temperature series using the approach described in Section~\ref{subsec:temperaturedata}. Second, we take weekly incidence rates of influenza-like illnesses (ILI) from the French Sentinelles network for the same period, as covered in Section~\ref{subsec:ILIdata}. We then transform both observed series into cross-basis matrices using the DLNM specifications introduced in Section~\ref{subsec:dlnm}. 

\begin{figure}[!ht]
    \centering
    \includegraphics[width=0.95\linewidth]{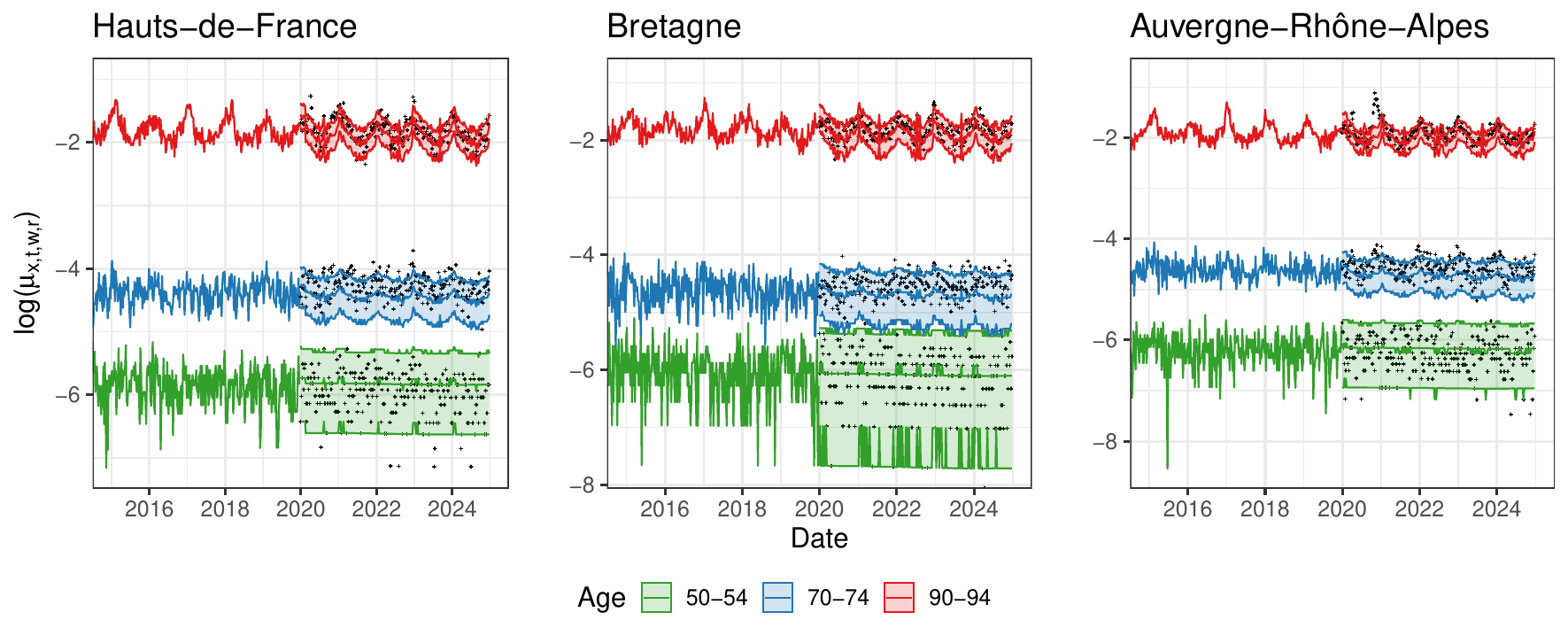}
    \caption{\added{Forecasted weekly death rates for women in Hauts-de-France (left), Bretagne (middle), and Auvergne-Rhône-Alpes (right) over the period 2020–2024 using the realized exogenous factors. We show the $95\%$ prediction intervals based on 10 000 simulated trajectories of the latent mortality index for the age groups 50-54, 70-74, and 90-94. The observed death rates from 2020-2024 are visualized in black with a '+'.}}
    \label{fig:forecast_obs}
\end{figure}

By following the forecasting approach for \added{Model 6 under the observed covariate setting}, Figure~\ref{fig:forecast_obs} shows the 95$\%$ prediction intervals of the forecasted weekly death rates for women in three administrative French regions for age groups 50-54, 70-74, and 90-94. We observe that the observed death rates clearly fall outside the 95$\%$ uncertainty bound during the severe COVID-19 waves. However, after 2022, death rates still fall outside the bounds for the 70-74 age group, which indicates that there is still post-COVID excess mortality for certain age groups.

\section{Conclusion} \label{sec:conclusion}
This paper develops and implements a new high-resolution stochastic mortality projection model that extends the classic Lee-Carter framework to a weekly, regional, and age-specific context. Our model is specifically designed to disentangle the baseline seasonal mortality patterns from the acute impacts of heat waves, cold spells, and influenza outbreaks. The methodological innovation lies in the integration of distributed lag non-linear models (DLNMs) within a multi-population Negative Binomial Lee-Carter structure. This allows for an in-depth quantification of the non-linear and delayed effects of temperature and influenza on mortality, while accounting for overdispersion. The application of spatial penalization via a Laplacian smoothness prior ensures a coherent estimation of these effects across neighboring regions.

Through a case study on French data from 1990 to 2019, we calibrate our proposed framework and validate the model's performance. First, we obtain a good in-sample fit and capture seasonal patterns and extreme mortality events well. Pearson residuals confirm a good overall fit. Second, the mortality impacts of both heat and influenza show a strong, increasing effect with age, with the oldest age groups (90+) showing the highest relative risk. Third, the strength of the temperature-mortality and influenza-mortality associations varies significantly between regions in France. The effects of heat are immediate, with a clear harvesting effect (mortality displacement), whereas the effects of influenza peak immediately and gradually decrease over several weeks. Third, our model shows superior predictive performance compared to benchmark specifications, with particularly strong gains at older ages. \added{While improvements in mortality forecasts are more limited when exogenous variables must be forecasted, the proposed framework leads to clear gains in terms of predictive distributions, as reflected by improved CRPS and interval scores.}

The model in its current form is designed mainly for short- to medium-term mortality forecasts. The proposed framework is therefore particularly valuable for insurers to price short- to medium-term life-contingent insurance products or for public health to assess excess mortality across different administrative regions and age groups within a country. However, the model is less suitable for long-term mortality projections, as the seasonal (weekly) component is currently assumed to be fixed over time, whereas empirical evidence indicates that seasonal mortality patterns may change over longer horizons \citep{madaniyazi2022seasonal}. Future extensions of this work could therefore deal with incorporating a time-varying seasonal component in the model, which could capture potential shifts in seasonal mortality patterns.

\section*{Supplementary information}
The Supplementary Material (Suppl.~Mat.) is available on pages 38-47 of this manuscript.\\

The \textit{Online Appendix} is available at: \url{https://jensrobben.github.io/articles/wLC-DLNM/Online_Appendix.pdf}.\\

The R code used to implement and analyze the case study presented in this paper will be made available on the GitHub repository: \url{https://github.com/jensrobben/wLC-DLNM}.

\section*{Data and code availability statement}
The data sets used in this paper are publicly available. We consult the ten-year files of all individual people who have died since 1970 and the population estimates by department, sex, and five-year age group from INSEE (\url{https://www.insee.fr/fr/information/4769950} and \url{https://www.insee.fr/fr/statistiques/8331297}). We retrieve the daily average temperature from the E-OBS dataset in the Copernicus Climate Data Store (\url{https://cds.climate.copernicus.eu/datasets/insitu-gridded-observations-europe}) and the incidence rates of influenza activity from the French Sentinelles network (\url{https://www.sentiweb.fr/france/en/?page=table&maladie=25}). 

\section*{Funding statement}
This study is part of the research programme at the Research Centre for Longevity Risk, a joint initiative of NN Group and the University of Amsterdam, with additional funding from the Dutch government’s Public-Private Partnership (PPP) programme. 

\section*{Conflict of interest disclosure}
The authors declare no conflict of interest.

{\bibliography{References}}

@article{delwarde2007negative,
  title={Negative binomial version of the {L}ee--{C}arter model for mortality forecasting},
  author={Delwarde, Antoine and Denuit, Michel and Partrat, Christian},
  journal={Applied Stochastic Models in Business and Industry},
  volume={23},
  number={5},
  pages={385--401},
  year={2007},
  publisher={Wiley Online Library},
  doi={10.1002/asmb.679}
}

@article{barigou2023bayesian,
  title={Bayesian model averaging for mortality forecasting using leave-future-out validation},
  author={Barigou, Karim and Goffard, Pierre-Olivier and Loisel, St{\'e}phane and Salhi, Yahia},
  journal={International Journal of Forecasting},
  volume={39},
  number={2},
  pages={674--690},
  year={2023},
  publisher={Elsevier},
  doi={10.1016/j.ijforecast.2022.01.011}
}

@misc{cds2,
  author       = {{Copernicus Climate Change Service}},
  title        = {Temperature and precipitation climate impact indicators from 1970 to 2100 derived from {E}uropean climate projections},
  year         = {2021},
  publisher    = {Copernicus Climate Change Service (C3S) Climate Data Store (CDS)},
  howpublished = {\url{https://cds.climate.copernicus.eu}},
  note         = {DOI: \href{10.24381/cds.9eed87d5}{https://doi.org/10.24381/cds.9eed87d5}}
}

@article{valleron1986computer,
  title={A computer network for the surveillance of communicable diseases: the {F}rench experiment},
  author={Valleron, ALAIN-JACQUES and Bouvet, Elisabeth and Garnerin, Philippe and M{\'e}nares, JUAN and Heard, ISABELLE and Letrait, SYLVIA and Lefaucheux, JACQUES},
  journal={American Journal of Public Health},
  volume={76},
  number={11},
  pages={1289--1292},
  year={1986},
  publisher={American Public Health Association},
  doi={10.2105/ajph.76.11.1289}
}

@article{pavia2022estimation,
  title={Estimation of the combined effects of ageing and seasonality on mortality risk: An application to {S}pain},
  author={Pav{\'\i}a, Jose M and Lled{\'o}, Josep},
  journal={Journal of the Royal Statistical Society Series A: Statistics in Society},
  volume={185},
  number={2},
  pages={471--497},
  year={2022},
  publisher={Oxford University Press},
  doi={doi.org/10.1111/rssa.12769}
}

@article{lee1992modeling,
  title={Modeling and forecasting {US} mortality},
  author={Lee, Ronald D and Carter, Lawrence R},
  journal={Journal of the American Statistical Association},
  volume={87},
  number={419},
  pages={659--671},
  year={1992},
  publisher={Taylor \& Francis},
  doi={10.1080/01621459.1992.10475265}
}

@article{kleinow2015common,
  title={A common age effect model for the mortality of multiple populations},
  author={Kleinow, Torsten},
  journal={Insurance: Mathematics and Economics},
  volume={63},
  pages={147--152},
  year={2015},
  publisher={Elsevier},
  doi={10.1016/j.insmatheco.2015.03.023}
}

@article{barnett2010measure,
  title={What measure of temperature is the best predictor of mortality?},
  author={Barnett, A.},
  journal={Environmental Research},
  volume={110},
  number={6},
  pages={604--611},
  year={2010},
  publisher={Elsevier},
  doi = {10.1097/01.ede.0000362214.90491.53}
}

@article{wong2018bayesian,
  title={Bayesian mortality forecasting with overdispersion},
  author={Wong, Jackie ST and Forster, Jonathan J and Smith, Peter WF},
  journal={Insurance: Mathematics and Economics},
  volume={83},
  pages={206--221},
  year={2018},
  publisher={Elsevier},
  doi={10.1016/j.ijforecast.2022.01.011}
}

@misc{brouhns2002measuring,
  title={Measuring the longevity risk in mortality projections},
  author={Brouhns, Natacha and Denuit, Michel and Vermunt, Jeroen K},
  journal={Bulletin of the Swiss Association of Actuaries},
  number={2},
  pages={105--130},
  year={2002},
  publisher={Staempfli Verlag AG},
  note={Preprint at \url{https://research.tilburguniversity.edu/files/510433/brouhns2002.PDF}}
}

@article{guibert2025impact,
  title={Impact of climate change on mortality: An extrapolation of temperature effects based on time series data in {F}rance},
  author={Guibert, Quentin and Pincemin, Ga{\"e}lle and Planchet, Fr{\'e}d{\'e}ric},
  journal={International Journal of Forecasting},
  year={2025},
  publisher={Elsevier},
  doi={10.1016/j.ijforecast.2025.07.004}
}

@article{li2005coherent,
  title={Coherent mortality forecasts for a group of populations: An extension of the {L}ee-{C}arter method},
  author={Li, Nan and Lee, Ronald Demos},
  journal={Demography},
  volume={42},
  number={3},
  pages={575--594},
  year={2005},
  publisher={Population Association of America},
  doi={10.1353/dem.2005.0021}
}

@misc{cds,
  author = "{{Copernicus Climate Change Service, Climate Data Store}}",
  title = {{E-OBS} daily gridded meteorological data for {E}urope from 1950 to present derived from in-situ observations},
  publisher = {Copernicus Climate Change Service (C3S) Climate Data Store (CDS)},
  year = {2020},
  doi = {10.24381/cds.151d3ec6},
  note = {Accessed 01-09-2023}
}

@misc{nasadataset,
  title={Gridded Population of the World, Version 4 ({GPWv4}): Population Count, Revision 11},
  author = {{Center for International Earth Science Information Network -- CIESIN -- Columbia University}},
  publisher = {Palisades, New York: NASA Socioeconomic Data and Applications Center (SEDAC)},
  year = {2018},
  doi = {10.7927/H4JW8BX5},
  note = {Accessed 01-08-2023}
}

@article{wang2022estimating,
  title={Estimating excess mortality due to the {COVID}-19 pandemic: a systematic analysis of {COVID}-19-related mortality, 2020--21},
  author={Wang, Haidong and Paulson, Katherine R and Pease, Spencer A and Watson, Stefanie and Comfort, Haley and Zheng, Peng and Aravkin, Aleksandr Y and Bisignano, Catherine and Barber, Ryan M and Alam, Tahiya and others},
  journal={The Lancet},
  volume={399},
  number={10334},
  pages={1513--1536},
  year={2022},
  publisher={Elsevier},
  doi={10.1016/S0140-6736(21)02796-3 }
}

@misc{wilmoth2007methods,
  title={Methods protocol for the human mortality database},
  author={Wilmoth, John R and Andreev, Kirill and Jdanov, Dmitri and Glei, Dana A and Boe, C and Bubenheim, M and Philipov, D and Shkolnikov, V and Vachon, P},
  note={University of California, Berkeley, and Max Planck Institute for Demographic Research, Rostock. URL: \url{http://mortality. org [version 31/05/2007]}},
  volume={9},
  pages={10--11},
  year={2007}
}

@article{gasparrini2011impact,
  title={The impact of heat waves on mortality},
  author={Gasparrini, Antonio and Armstrong, Ben},
  journal={Epidemiology},
  volume={22},
  number={1},
  pages={68},
  year={2011},
  publisher={Europe PMC Funders},
  doi = {10.1097/EDE.0b013e3181fdcd99}
}

@article{gasparrini2010distributed,
  title={Distributed lag non-linear models},
  author={Gasparrini, Antonio and Armstrong, Ben and Kenward, M. G.},
  journal={Statistics in Medicine},
  volume={29},
  number={21},
  pages={2224--2234},
  year={2010},
  publisher={Wiley Online Library},
  doi = {10.1002/sim.3940}
}

@article{martinez2021projections,
  title={Projections of temperature-attributable mortality in {E}urope: a time series analysis of 147 contiguous regions in 16 countries},
  author={Mart{\'\i}nez-Solanas, {\`E}rica and Quijal-Zamorano, Marcos and Achebak, Hicham and Petrova, Desislava and Robine, Jean-Marie and Herrmann, Fran{\c{c}}ois R and Rod{\'o}, Xavier and Ballester, Joan},
  journal={The Lancet Planetary Health},
  volume={5},
  number={7},
  pages={e446--e454},
  year={2021},
  publisher={Elsevier},
  doi={10.1016/S2542-5196(21)00150-9}
}

@article{gasparrini2015mortality,
  title={Mortality risk attributable to high and low ambient temperature: A multicountry observational study},
  author={Gasparrini, Antonio and Guo, Yuming and Hashizume, Masahiro and Lavigne, Eric and Zanobetti, Antonella and Schwartz, Joel and Tobias, Aurelio and Tong, Shilu and Rockl{\"o}v, Joacim and Forsberg, Bertil and Leone, Michela and De Sario, Manuela and Bell, Michelle L. and Guo, Yue-Liang Leon and Wu, Chang-fu and Kan, Haidong and Yi, Seung-Muk and de Sousa Zanotti Stagliorio Coelho, Micheline and Saldiva, Paulo Hilario Nascimento and Honda, Yasushi and Kim, Ho and Armstrong, Ben},
  journal={The Lancet},
  volume={386},
  number={9991},
  pages={369--375},
  year={2015},
  publisher={Elsevier},
  doi = {10.1016/S0140-6736(14)62114-0}
}

@article{lowen2014roles,
  title={Roles of humidity and temperature in shaping influenza seasonality},
  author={Lowen, Anice C and Steel, John},
  journal={Journal of Virology},
  volume={88},
  number={14},
  pages={7692--7695},
  year={2014},
  publisher={American Society for Microbiology 1752 N St., NW, Washington, DC},
  doi={10.1128/jvi.03544-13}
}

@article{hyndman2008automatic,
  title={Automatic time series forecasting: the forecast package for {R}},
  author={Hyndman, Rob J and Khandakar, Yeasmin},
  journal={Journal of Statistical Software},
  volume={27},
  pages={1--22},
  year={2008},
  doi={10.18637/jss.v027.i03}
}

@article{lowen2007influenza,
  title={Influenza virus transmission is dependent on relative humidity and temperature},
  author={Lowen, Anice C and Mubareka, Samira and Steel, John and Palese, Peter},
  journal={PLoS Pathogens},
  volume={3},
  number={10},
  pages={e151},
  year={2007},
  publisher={Public Library of Science San Francisco, USA},
  doi={10.1371/journal.ppat.0030151}
}

@article{li2023influenza,
  title={Influenza-associated excess mortality by age, sex, and subtype/lineage: population-based time-series study with a distributed-lag nonlinear model},
  author={Li, Li and Yan, Ze-Lin and Luo, Lei and Liu, Wenhui and Yang, Zhou and Shi, Chen and Ming, Bo-Wen and Yang, Jun and Cao, Peihua and Ou, Chun-Quan},
  journal={JMIR Public Health and Surveillance},
  volume={9},
  number={1},
  pages={e42530},
  year={2023},
  publisher={JMIR Publications Inc., Toronto, Canada},
  doi={10.2196/42530}
}

@article{brouhns2005bootstrapping,
  title={Bootstrapping the {P}oisson log-bilinear model for mortality forecasting},
  author={Brouhns, Natacha and Denuit, Michel and Van Keilegom, Ingrid},
  journal={Scandinavian Actuarial Journal},
  volume={2005},
  number={3},
  pages={212--224},
  year={2005},
  publisher={Taylor \& Francis},
  doi={10.1080/03461230510009754}
}

@article{chaves2023global,
  title={Global, regional and national estimates of influenza-attributable ischemic heart disease mortality},
  author={Chaves, Sandra S and Nealon, Joshua and Burkart, Katrin G and Modin, Daniel and Biering-S{\o}rensen, Tor and Ortiz, Justin R and Vilchis-Tella, Victor M and Wallace, Lindsey E and Roth, Gregory and Mahe, Cedric and others},
  journal={EClinicalMedicine},
  volume={55},
  year={2023},
  publisher={Elsevier},
  doi={10.1016/j.eclinm.2022.101740}
}

@article{madaniyazi2022seasonal,
  title={Seasonal variation in mortality and the role of temperature: a multi-country multi-city study},
  author={Madaniyazi, Lina and Armstrong, Ben and Chung, Yeonseung and Ng, Chris Fook Sheng and Seposo, Xerxes and Kim, Yoonhee and Tobias, Aurelio and Guo, Yuming and Sera, Francesco and Honda, Yasushi and others},
  journal={International Journal of Epidemiology},
  volume={51},
  number={1},
  pages={122--133},
  year={2022},
  publisher={Oxford University Press},
  doi={10.1093/ije/dyab143}
}

@article{fernandez2015seasonal,
  title={Seasonal mortality for fractional ages in short term life insurance},
  author={Fern{\'a}ndez-Dur{\'a}n, Juan Jos{\'e} and Gregorio-Dom{\'\i}nguez, Mar{\'\i}a Mercedes},
  journal={Scandinavian Actuarial Journal},
  volume={2015},
  number={3},
  pages={266--277},
  year={2015},
  publisher={Taylor \& Francis},
  doi={10.1080/03461238.2013.819028}
}

@article{karlinsky2021tracking,
  title={Tracking excess mortality across countries during the {COVID}-19 pandemic with the {World Mortality Dataset}},
  author={Karlinsky, Ariel and Kobak, Dmitry},
  journal={elife},
  volume={10},
  pages={e69336},
  year={2021},
  publisher={eLife Sciences Publications, Ltd},
  doi={10.7554/eLife.69336}
}

@article{guo2017heat,
  title={Heat wave and mortality: A multicountry, multicommunity study},
  author={Guo, Yuming and Gasparrini, Antonio and Armstrong, Ben G. and Tawatsupa, Benjawan and Tobias, Aurelio and Lavigne, Eric and Coelho, Micheline de Sousa Zanotti Stagliorio and Pan, Xiaochuan and Kim, Ho and Hashizume, Masahiro and ... and Tong, Shilu},
  journal={Environmental Health Perspectives},
  volume={125},
  number={8},
  pages={087006},
  year={2017},
  doi = {10.1289/EHP1026}
}

@article{beginmodelling,
  title={Modelling seasonal mortality: An age--period--cohort approach},
  author={B{\'e}gin, Jean-Fran{\c{c}}ois and Boudreault, Mathieu and Landry, Thomas},
  journal={Insurance: Mathematics and Economics},
  pages={103162},
  year={2025},
  publisher={Elsevier}
}

@misc{min2025assessing,
  title={Assessing Climate-Driven Mortality Risk: A Stochastic Approach with Distributed Lag Non-Linear Models},
  author={Min, Jiacheng and Li, Han and Nagler, Thomas and Li, Shuanming},
  note={Preprint at \url{https://arxiv.org/pdf/2506.00561}},
  year={2025}
}

@article{stmfnote,
  title={The short-term mortality fluctuation data series, monitoring mortality shocks across time and space},
  author={Jdanov, Dmitri A. and Galarza, Ainhoa Alustiza and Shkolnikov, Vladimir M. and Jasilionis, Domantas and N{\'e}meth, L{\'a}szl{\'o} and Leon, David A. and Boe, Carl and Barbieri, Magali},
  journal={Scientific Data},
  volume={8},
  number={1},
  pages={235},
  year={2021},
  publisher={Nature Publishing Group UK London},
  doi = {10.1038/s41597-021-01019-1}
}

@article{li2022joint,
  title={Joint extremes in temperature and mortality: A bivariate {POT} approach},
  author={Li, Han and Tang, Qihe},
  journal={North American Actuarial Journal},
  volume={26},
  number={1},
  pages={43--63},
  year={2022},
  publisher={Taylor \& Francis},
  doi = {10.1080/10920277.2020.1823236}
}

@article{robben2025short,
  title={The short-term association between environmental variables and mortality: evidence from {E}urope},
  author={Robben, Jens and Antonio, Katrien and Kleinow, Torsten},
  journal={Journal of the Royal Statistical Society Series A: Statistics in Society},
  year={2025},
  publisher={Oxford University Press UK},
  doi={10.1093/jrsssa/qnaf052}
}

@article{jordan2019evaluating,
  title={Evaluating probabilistic forecasts with {scoringRules}},
  author={Jordan, Alexander and Kr{\"u}ger, Fabian and Lerch, Sebastian},
  journal={Journal of Statistical Software},
  volume={90},
  pages={1--37},
  year={2019}
}

@article{wood2016smoothing,
  title={Smoothing parameter and model selection for general smooth models},
  author={Wood, Simon N and Pya, Natalya and S{\"a}fken, Benjamin},
  journal={Journal of the American Statistical Association},
  volume={111},
  number={516},
  pages={1548--1563},
  year={2016},
  publisher={Taylor \& Francis},
  doi={10.1080/01621459.2016.1180986}
}

@article{gneiting2007strictly,
  title={Strictly proper scoring rules, prediction, and estimation},
  author={Gneiting, Tilmann and Raftery, Adrian E},
  journal={Journal of the American statistical Association},
  volume={102},
  number={477},
  pages={359--378},
  year={2007},
  publisher={Taylor \& Francis}
}

@misc{camsdata,
  author = {{Institut national de l'environnement industriel et des risques (Ineris)} and {Aarhus University} and {Norwegian Meteorological Institute (MET Norway)} and {J\"ulich Institut f\"ur Energie- und Klimaforschung (IEK)} and {Institute of Environmental Protection – National Research Institute (IEP-NRI)} and {Koninklijk Nederlands Meteorologisch Instituut (KNMI)} and {METEO FRANCE} and {Nederlandse Organisatie voor toegepast-natuurwetenschappelijk onderzoek (TNO)} and {Swedish Meteorological and Hydrological Institute (SMHI)} and {Finnish Meteorological Institute (FMI)} and {Italian National Agency for New Technologies} and {Energy, Sustainable Economic Development (ENEA)} and {Barcelona Supercomputing Center (BSC)}},
  title = {{CAMS} {E}uropean air quality forecasts, {ENSEMBLE} data. },
  publisher = {{C}opernicus {A}tmosphere {M}onitoring {S}ervice ({CAMS}) {A}tmosphere {D}ata {S}tore ({ADS})},
  howpublished= {\url{https://ads.atmosphere.copernicus.eu/cdsapp\#!/dataset/cams-europe-air-quality-reanalyses?tab=overview}},
  note = {Accessed 01-08-2023},
  year = {2022} 
}

@article{seklecka2017mortality,
  title={Mortality effects of temperature changes in the {U}nited {K}ingdom},
  author={Seklecka, Malgorzata and Pantelous, Athanasios A and O'Hare, Colin},
  journal={Journal of Forecasting},
  volume={36},
  number={7},
  pages={824--841},
  year={2017},
  publisher={Wiley Online Library},
  doi={10.1002/for.2473}
}

@book{pitacco2009modelling,
  title={Modelling longevity dynamics for pensions and annuity business},
  author={Pitacco, Ermanno},
  year={2009},
  publisher={Oxford University Press}
}

@book{wood2017generalized,
  title={Generalized additive models: An introduction with {R}},
  author={Wood, Simon N.},
  year={2017},
  publisher={CRC Press},
  doi = {10.1201/9781315370279}
}
\newpage

\begin{center}
    {\huge \bf Supplementary material: \\ “A penalized distributed lag non-linear Lee-Carter framework for regional weekly mortality forecasting”}
\end{center}

\appendix
\renewcommand{\thefigure}{\thesection.\arabic{figure}}
\renewcommand{\thetable}{\thesection.\arabic{table}}
\renewcommand{\theequation}{\thesection.\arabic{equation}}
\setcounter{figure}{0}
\setcounter{table}{0}
\setcounter{equation}{0}

\section{Sensitivity analysis of DLNM specifications} \label{app:sensitivitydlnm}
\added{We conduct a sensitivity analysis over alternative DLNM specifications to assess the robustness of our results to modeling choices. The final model specification used in the main paper is defined as follows.}

\added{For temperature, we employ a cross-basis function with a cubic B-spline for the exposure–response relationship (degree 3), with two internal knots placed at the 10th and 90th percentiles of the pooled temperature distribution across all regions and time periods. Boundary knots are defined using the global temperature range, with the lower boundary set at 5$^\circ$C below the observed minimum and the upper boundary at the observed maximum. We model the lag–response relationship using a natural cubic spline with an intercept, a maximum lag of 3 weeks, and internal knots at lags 0.5 and 1.5.}

\added{For influenza-like illness (ILI), the exposure–response relationship is specified as linear (first-degree polynomial), after scaling ILI incidence by its region-specific maximum to ensure comparability across regions. We again model the lag–response relationship using a natural spline with an intercept, a maximum lag of 7 weeks, and internal knots at lags 0.5 and 1.5.}

\added{To motivate these choices, we vary (i) the maximum lag in the temperature and ILI DLNMs, (ii) the functional form of the covariate–response relationships (including alternative spline bases and degrees), and (iii) the specification of the lag structure. Each alternative specification modifies one component at a time. We then refit our proposed weekly Lee-Carter model on each specification and compare the performance using BIC, which penalizes model complexity and favors more parsimonious specifications. The results are reported in Table~\ref{tabA:sens.dlnm}.}

\begin{table}[!ht]
\centering
\caption{\added{Sensitivity analysis of DLNM specifications We report the number of parameters, the log-likelihood value, the BIC value, and the increase in BIC when using alternative specifications.}}
\label{tabA:sens.dlnm}
\adjustbox{width = \textwidth}{
\begin{tabular}{p{6cm} p{7cm} p{1.8cm} p{2.2cm} p{2.2cm} p{2cm}}
\toprule
\textbf{Component} & \textbf{Specification} & \textbf{\# Par} & \textbf{LogLik} & \textbf{BIC} & $\boldsymbol{\Delta}$\textbf{BIC} \\
\midrule
\multicolumn{2}{l}{\textbf{Final model specification}} & 1\ 437 & -586\ 490 & 1\ 190\ 430 & $\bigtimes$\\
Max lag (temperature) & 3 weeks &  &  &  &  \\
Max lag (ILI) & 7 weeks &  &  &  &  \\
Temperature exposure--response & B-spline (cubic; knots at 10th, 90th perc.) &  &  &  &  \\
Temperature lag--response & Natural spline (knots at lags 0.5, 1.5) &   &  &  &  \\
ILI exposure--response & Linear (first-degree polynomial) &  &  &  &  \\
ILI lag--response & Natural spline (knots at lags 0.5, 1.5) &  &  &  &  \\
Boundary knots (temperature) & Global range (min $-5^\circ$C, max observed) &  &  &  &  \\
\midrule
\multicolumn{6}{l}{\textbf{Alternative specifications (one change at a time)}} \\
Max lag (temperature) & 4 weeks &  1\ 437 & -586\ 680 & 1\ 190\ 810 & 380 \\
                      & 5 weeks &  1\ 437 & -586\ 818 & 1\ 191\ 085 & 655\\
Max lag (ILI)         & 5 weeks &  1\ 437 & -586\ 503 & 1\ 190\ 457 & 26 \\
                      & 6 weeks &  1\ 437 & -586\ 497 & 1\ 190\ 444 & 14\\
                      & 8 weeks &  1\ 437 & -586\ 511 & 1\ 190\ 472 & 41\\
Temperature exposure--response & Natural spline (same knots) &  1\ 341 & -588\ 696 & 1\ 193\ 677 & 3\ 246 \\
                               & B-spline (quadratic) & 1\ 389 & -587\ 373 & 1\ 191\ 613 & 1\ 183 \\
                               & B-spline (quartic) & 1\ 485 & -586\ 444 & 1\ 190\ 921 & 491\\
                               & B-spline (knots at 10th, 50th, 90th perc.) & 1\ 485 & -586\ 495 & 1\ 191\ 022 & 591 \\ 
                               & B-spline (knots at 10th, 75th, 90th perc.) & 1\ 485 & -586\ 505 & 1\ 191\ 042 & 611\\
                               & P-spline (cubic, df = 5) & 1\ 437 & -588\ 563 & 1\ 194\ 575 & 4\ 145\\  
                               
ILI exposure--response & Quadratic (second degree polynomial) &  1\ 485 & -586\ 308 & 1\ 190\ 648 & 218  \\
                       & B-spline (linear, knots at 10th, 90th perc.) & 1\ 533 & -586\ 202 & 1\ 191\ 020 & 590 \\
                       & B-spline (quadratic, knots at 10th, 90th perc.) & 1\ 581 & -586\ 312 & 1\ 191\ 822 & 1\ 392\\
                       & B-spline (cubic, knots at 10th, 90th perc.) & 1\ 629 & -586\ 257 & 1\ 192\ 296 & 1\ 865\\
Lag--response (temperature, ILI) & B-spline &  1\ 581 & -586\ 481 & 1\ 192\ 160 & 1\ 729 \\
Lag-response (ILI) & Linear spline & 1\ 413 & -586\ 508 & 1\ 190\ 175 & -256\\
Boundary knots (temperature) & Global min  &  1\ 437 & -586\ 554 & 1\ 190\ 559 & 128\\
                             & Global min $-3^\circ$C &  1\ 437 & -586\ 519 & 1\ 190\ 488 & 57 \\
                             & Global min $-8^\circ$C &  1\ 437 & -586\ 527 & 1\ 190\ 504 & 74\\
                             & Global max $+3^\circ$C &  1\ 437 & -587\ 469 & 1\ 192\ 388 & 1\ 958\\
                             & Global max $+5^\circ$C & 1\ 437 & -588\ 362 & 1\ 194\ 174 & 3\ 743\\ \bottomrule
\end{tabular}}
\end{table}

\added{Overall, the results indicate that the selected specification achieves a good balance between fit and parsimony. All alternative DLNM specifications result in higher BIC values. In particular, extending the maximum lag for temperature does not improve model fit, suggesting that temperature effects are primarily short-term. Similarly, modifications to the spline specification for temperature and ILI lead to noticeable increases in BIC. Interestingly, setting the lower boundary knot of the temperature exposure–response curve slightly below the observed minimum improves model fit, indicating greater flexibility in capturing cold effects. This pattern is not observed for the upper boundary knot, where extending it beyond the observed maximum leads to a clear deterioration in performance.}

\section{Model comparison and interaction DLNM construction}
\label{app:modelcomparison}

\subsection{Construction of the interaction DLNM}
\label{app:interaction}

\added{Following the notation of Section~\ref{subsec:dlnm}, we construct a three-dimensional cross-basis function to capture non-linear, delayed, and interactive effects of temperature and influenza. First, each covariate is expanded using its respective basis functions. We then combine these expansions via a tensor product, and impose a distributed lag structure through a common lag basis applied to both covariates. This yields the following interaction DLNM:
\begin{align*}
g^{\text{int}}_r(\xi^{(1)}_{t,w,r}, \xi^{(2)}_{t,w,r})
&=
\sum_{\ell=0}^{L}
\sum_{j=1}^{v_{\xi_1}}
\sum_{m=1}^{v_{\xi_2}}
\sum_{k=1}^{v_{\ell}}
\eta^{\text{int}}_{jmk,r}
\cdot
b^{(1)}_j(\xi^{(1)}_{t,w-\ell,r})
\cdot
b^{(2)}_m(\xi^{(2)}_{t,w-\ell,r})
\cdot
c_k(\ell).
\end{align*}}

\added{As in Section~\ref{subsec:dlnm}, $b^{(1)}_j(\cdot)$ and $b^{(2)}_m(\cdot)$ denote the basis functions for the covariate dimensions, while $c_k(\ell)$ represents the basis functions for the common lag dimension. The index $\ell = 0,1,\dots,L$ denotes the lag structure, with $L$ the maximum lag considered. The parameters $\eta^{\text{int}}_{jmk,r}$ are region-specific coefficients associated with the interaction between the two covariates over the lag structure.}

\subsection{Performance comparison of alternative models}
\label{app:perftable}

\added{We consider five competing specifications of the model in Eq.~\eqref{eq:modelstructure}, which differ in their inclusion of distributed lag non-linear effects (DLNMs) for temperature and influenza, as well as their interaction. As in Section~\ref{sec:modelspecification}, we assume that the number of deaths $D_{x,t,w,r}$ follows a Negative Binomial (NB) distribution:
\begin{align*}
D_{x,t,w,r} \sim \text{NB}(E_{x,t,w,r} \cdot \mu_{x,t,w,r}, \phi_{x,r}),
\end{align*}
The corresponding specifications for the weekly force of mortality $\mu_{x,t,w,r}$ are:}
\begin{itemize}
    \item[\added{\textbf{(i)}}]\added{\textbf{Weekly Lee–Carter (baseline)}: Excludes all DLNM effects:
    \begin{align*}
        \log\mu_{x,t,w,r}^{(b)} = \alpha_{x,r} + \beta_x \kappa_{t,r} + \gamma_x \lambda_{w,r}.
    \end{align*}}

    \item[\added{\textbf{(ii)}}] \added{\textbf{Temperature DLNM}: Extends the baseline model by adding a distributed lag non-linear effect of temperature:
    \begin{align*}
        \log\mu_{x,t,w,r} = \log\mu_{x,t,w,r}^{(b)} + \delta_x f_r^{(1)}(\text{Tavg}_{t,w,r}).
    \end{align*}}

    \item[\added{\textbf{(iii)}}] \added{\textbf{Influenza DLNM}: Adds a distributed lag non-linear effect of influenza-like illnesses:
    \begin{align*}
        \log\mu_{x,t,w,r} = \log\mu_{x,t,w,r}^{(b)} + \epsilon_x f_r^{(2)}(\text{ILI}_{t,w,r}).
    \end{align*}}

    \item[\added{\textbf{(iv)}}] \added{\textbf{Temperature and Influenza DLNM}: Includes both main DLNM effects:
    \begin{align*}
        \log\mu_{x,t,w,r} = \log\mu_{x,t,w,r}^{(b)} + \delta_x f_r^{(1)}(\text{Tavg}_{t,w,r}) + \epsilon_x f_r^{(2)}(\text{ILI}_{t,w,r}).
    \end{align*}}

    \item[\added{\textbf{(v)}}] \added{\textbf{Full model with interaction DLNM}: Extends the previous specification by incorporating a two-dimensional cross-basis capturing temperature–influenza interaction effects (see Section~\ref{app:interaction} for the specification):
    \begin{align*}
        \log\mu_{x,t,w,r} = \log\mu_{x,t,w,r}^{(b)} + \delta_x f_r^{(1)}(\text{Tavg}_{t,w,r}) + \epsilon_x f_r^{(2)}(\text{ILI}_{t,w,r}) + \nu_x g_r(\text{Tavg}_{t,w,r}, \text{ILI}_{t,w,r}).
    \end{align*}}
\end{itemize}

\added{We estimate all models using an identical dispersion structure (\(\phi_{x,r} = \exp(\phi_x + \phi_r)\)) and subject to similar identifiability constraints as in Eq.~\eqref{eq:constraints}. To ensure comparability across specifications and reduce computational burden in this model comparison exercise, we fix the DLNM basis functions and do not include an additional penalisation term in the log-likelihood. We evaluate the model performance using the Bayesian Information Criterion (BIC). Table~\ref{tab:modelcomparison} summarises the results.}

\sisetup{detect-weight=true,detect-inline-weight=math}
\begin{table}[htbp]
\centering
\caption{\added{Comparison of model performance across five specifications. Lower BIC indicates better fit. The best model in each column is bolded.}}
\label{tab:modelcomparison}
\adjustbox{width = \textwidth}{
\begin{tabular}{l  S[table-format=7.3]  S[table-format=7.1]  S[table-format=7.1]  S[table-format=7.3] }
\toprule
\textbf{Model} & \textbf{\# Par} & \textbf{LogLik} & \textbf{BIC} & $\boldsymbol{\Delta}$\textbf{BIC} \\
\midrule
Weekly Lee-Carter (baseline)          & 1131 & -594513 & 1202760 & 0\\
+ Temperature DLNM                    & 1380 & -589082 & 1194922 & -7839\\
+ Influenza DLNM                      & 1188 & -592815 & 1200056 & -2704\\
+ Temperature and Influenza DLNM      & 1437 & -586490 & \bfseries 1190430 & \bfseries-12330\\
+ Temp, Infl and Interaction DLNM     & 1686 & \bfseries-586305 & 1193084 & -9677 \\
\bottomrule
\end{tabular}}
\end{table}

\added{The model including both temperature and influenza DLNMs, but excluding their interaction, achieves the lowest BIC value. This suggests that accounting for the separate effects of temperature and influenza improves model fit, while there is no sufficient evidence that their interaction provides additional explanatory power once model complexity is penalised via the BIC.}

\section{Calibration and evaluation of the spatial penalization framework} \label{app:smoothingpsi}
\begin{figure}[!ht]
    \centering
    \includegraphics[width=0.75\linewidth]{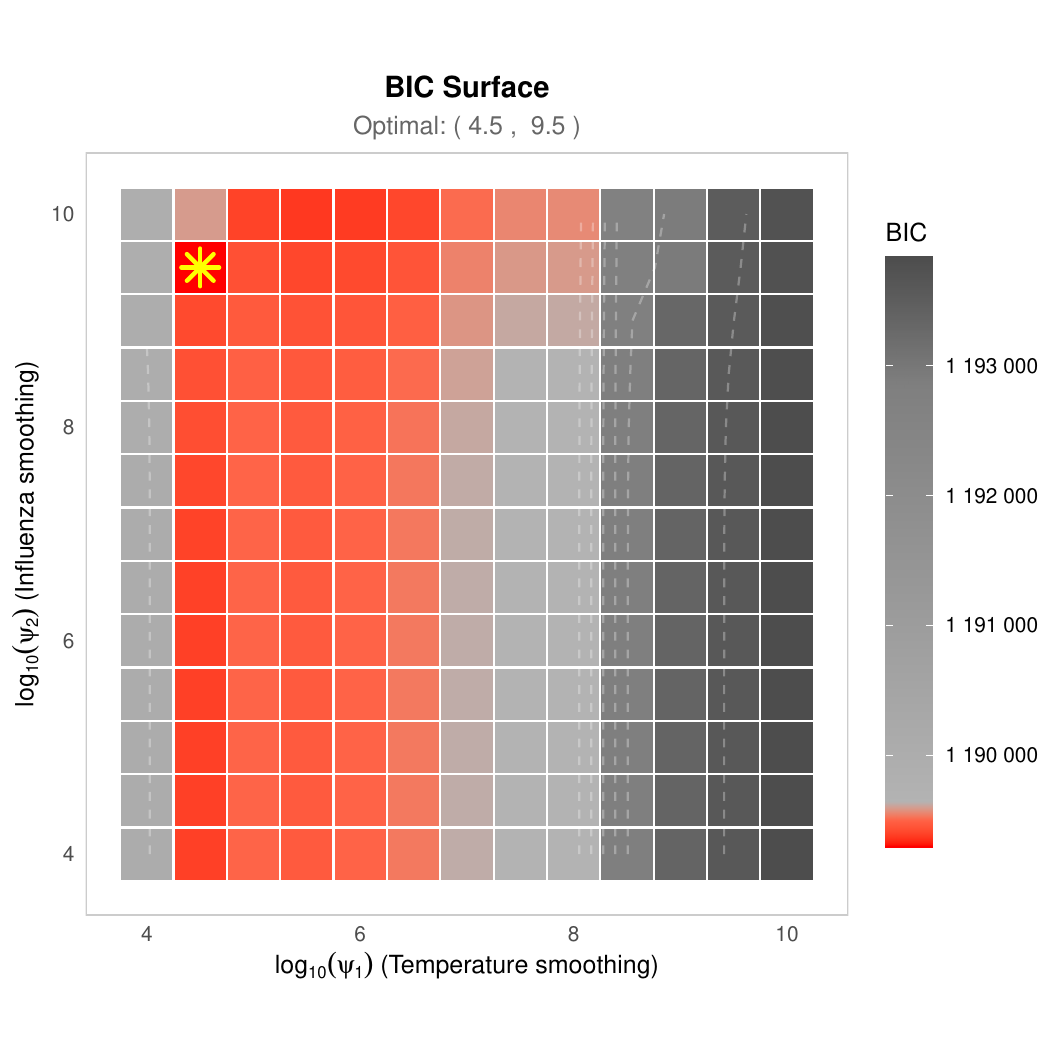}
    \caption{\added{BIC surface for smoothing parameter selection. The heatmap shows BIC values across a grid of $\log_{10}(\psi_1)$ and $\log_{10}(\psi_2)$ (ranging $10^4$ to $10^{10}$). Red indicates lower BIC (better fit), gray indicates higher BIC (poorer fit). The yellow star marks the optimal parameters $(\hat{\psi}_1, \hat{\psi}_2) = (10^{4.5}, 10^{9.5})$. \label{fig:bic_surface}}}
\end{figure}

\added{To assess the sensitivity of the model to the smoothing parameters $\psi_1$ and $\psi_2$ in Section~\ref{subsec:specdlnmcase}, we evaluate the BIC over the two-dimensional grid of candidate values $\psi_i \in \{10^4, 10^{4.5}, 10^5, ..., 10^{10}\}$, where $i = 1$ refers to the smoothing parameter for the temperature DLNM and $i = 2$ to the one of the influenza DLNM. Figure~\ref{fig:bic_surface} visualizes the resulting BIC surface over the $(\psi_1, \psi_2)$-grid. The surface shows a minimum at the values $\psi_1 = 10^{4.5}$ and $\psi_2 = 10^{9.5}$. This suggests that the data provide sufficient information to identify an appropriate level of smoothing for both DLNM components.}

\added{To further assess the role of the spatial penalization in Eq.~\eqref{eq:penloglik}, we estimate a restricted version of the model without penalization, i.e., $\psi_1 = \psi_2 = 0$, and compare it to the proposed specification with optimal smoothing parameters $\psi_1 = 10^{4.5}$ and $\psi_2 = 10^{9.5}$. The log-likelihood of the penalized model equals $-586\,517.8$, which is slightly lower than that of the unpenalized model ($-586\,490.4$). However, the effective number of degrees of freedom decreases substantially from 1437 in the unpenalized model to 1339 in the penalized model, corresponding to a reduction of nearly 100 degrees of freedom. Accounting for this difference in complexity, the BIC improves from $1\,190\,430$ in the unpenalized case to $1\,189\,295$ for the penalized specification, indicating a better trade-off between fit and parsimony when spatial penalization is included.}

\added{Importantly, the predictive performance remains largely unchanged. For both the penalized and unpenalized specifications, we recalibrate the model using data from 1990--2014 and generate out-of-sample forecasts for the period 2015--2019, following the same procedure as in Section~\ref{subsec:performance} under the observed covariate setting (Model 6). Comparing the resulting point forecasts in terms of RMSE and MAE yields very similar values across all regions (see Table~\ref{tab:rmse_mae_spatial}). This indicates that the spatial penalty primarily acts as a regularization mechanism that stabilizes the spatial structure and reduces model complexity, without affecting the model's predictive accuracy.}

\begin{table}[!ht]
\centering
\caption{\added{Predictive performance comparison between the unpenalized and penalized models across regions, based on RMSE and MAE. Lower values indicate better predictive accuracy. \label{tab:rmse_mae_spatial}}}
\adjustbox{width = \textwidth}{
\begin{tabular}{l@{\hspace{2em}}SS@{\hspace{2em}}SS}
\toprule
Region  & \text{RMSE: Unpenalized} & \text{RMSE: Penalized} & \text{MAE: Unpenalized} & \text{MAE: Penalized} \\
\midrule
Ile-de-France & 10.62 & 10.63 & 7.67 & 7.66 \\ 
  Centre - Val de Loire & 5.50 & 5.50 & 3.91 & 3.91 \\ 
  Bourgogne-Franche-Comté & 6.32 & 6.30 & 4.37 & 4.36 \\ 
  Normandie & 5.79 & 5.80 & 4.25 & 4.25 \\ 
  Hauts-de-France & 8.13 & 8.12 & 5.85 & 5.83 \\ 
  Grand Est & 8.52 & 8.51 & 5.91 & 5.90 \\ 
  Pays de la Loire & 6.49 & 6.49 & 4.62 & 4.62 \\ 
  Bretagne & 6.49 & 6.50 & 4.60 & 4.60 \\ 
  Nouvelle-Aquitaine & 9.53 & 9.53 & 6.66 & 6.66 \\ 
  Occitanie & 9.16 & 9.13 & 6.40 & 6.39 \\ 
  Auvergne-Rhône-Alpes & 11.17 & 11.25 & 7.31 & 7.35 \\ 
  Provence-Alpes-Côte d’Azur & 8.82 & 8.81 & 6.11 & 6.10 \\
\bottomrule
\end{tabular}}
\end{table}

\section{Parameter uncertainty via the observed Fisher information} \label{App:parameter.uncertainty.Fisher}
\added{As an alternative to the parametric bootstrap procedure described in Section~\ref{subsec:calibration}, we obtain approximate confidence intervals for the estimated parameter vector by using the inverse of the observed Fisher information matrix, following \cite{brouhns2002measuring}. The observed Fisher information is defined as the negative Hessian of the penalized log-likelihood evaluated at the parameter vector $\hat{\boldsymbol{\theta}}_{\hat{\psi}_1, \hat{\psi}_2}$:
\begin{equation} \label{eq:fisherinfo}
\widehat{\mathrm{Var}}(\hat{\boldsymbol{\theta}}_{\hat{\psi}_1, \hat{\psi}_2}) = \left[ -\boldsymbol{H}_{\boldsymbol{\theta}}\left( \hat{\boldsymbol{\theta}}_{\hat{\psi}_1, \hat{\psi}_2} \right) \right]^{-1},
\end{equation}
where $\boldsymbol{H}_{\boldsymbol{\theta}}(\cdot)$ denotes the Hessian matrix of the penalized log-likelihood in Eq.~\eqref{eq:penloglik} with respect to $\boldsymbol{\theta}$. We compute all second-order derivatives in an analytical way to ensure computational efficiency.}

\begin{figure}[!ht]
    \centering
    \includegraphics[width=\linewidth]{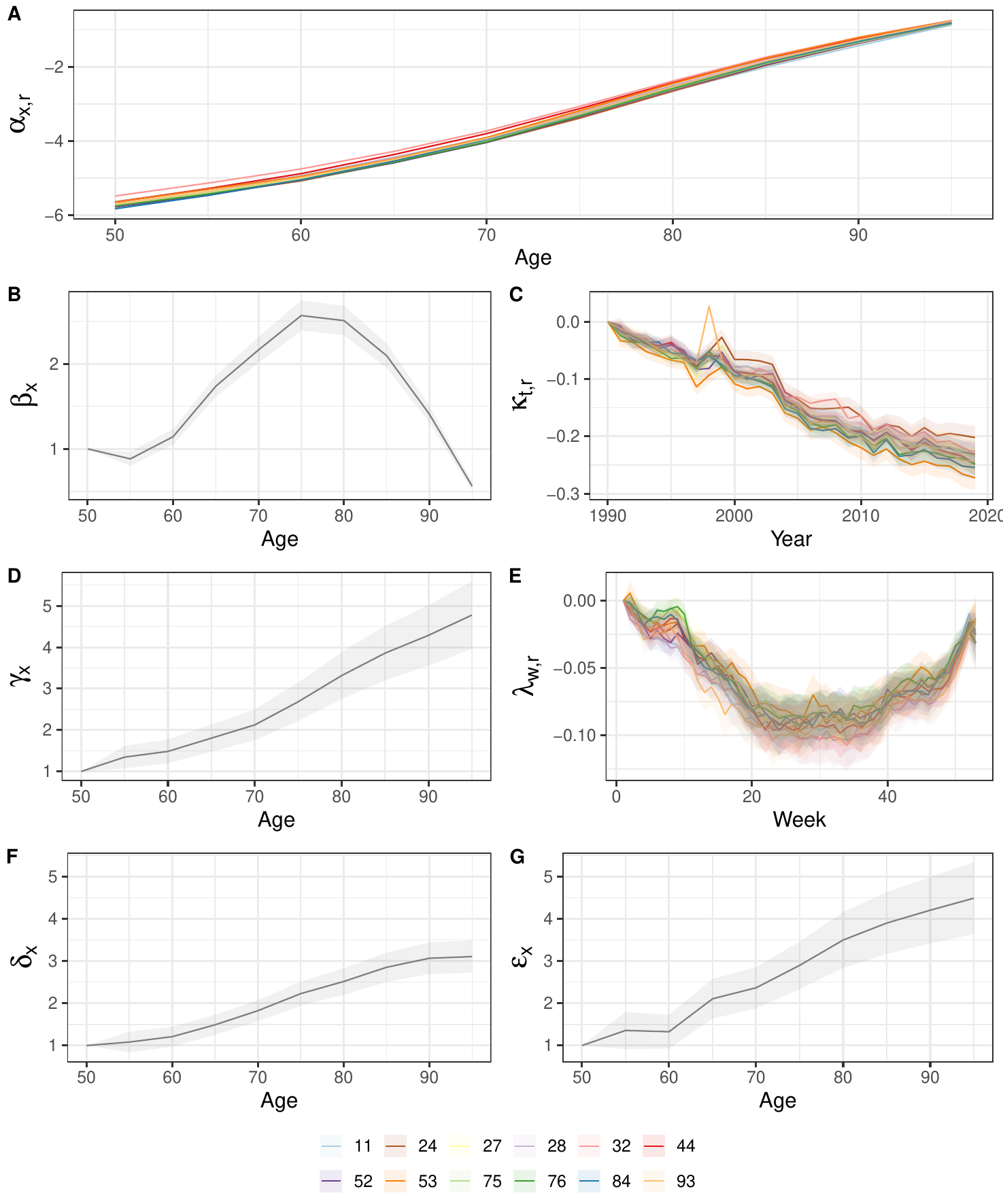}
    \caption{\added{Estimated parameters from the model: $\hat{\alpha}_{x,r}$ (A), $\hat{\beta}_x$ (B), $\hat{\kappa}_{t,r}$ (C), $\hat{\gamma}_x$ (D), $\hat{\lambda}_{w,r}$ (E), $\hat{\delta}_x$ (F), and $\hat{\epsilon}_x$ (G). Female data, French administrative regions, 5-year age groups $50-54$, $55-59$, ..., $95+$, and years $1990-2019$. Confidence intervals are based on the Hessian of the penalized negative binomial log-likelihood. The age-specific parameters, with the exception of $\hat{\alpha}_{x,r}$, do not vary by region $r$ and are shown in black.
    \label{figA:param1}}}
\end{figure}

\added{Figure~\ref{figA:param1} shows the resulting parameter estimates with confidence intervals derived from the inverse Hessian. Compared to the bootstrap-based confidence intervals reported in Figure~\ref{fig:param1}, the intervals for panels (F) and (G) are somewhat narrower. This is because the Hessian-based approach approximates uncertainty conditional on the stage-1 parameters, rather than propagating uncertainty through the full joint estimation procedure. As a result, it provides a computationally convenient but slightly less conservative approximation of parameter uncertainty. For this reason, we report the parametric bootstrap intervals in the main text and treat the Hessian-based results as a robustness check.}

\section{Comparison of (S)ARIMA and (S)ARIMAX specifications for influenza and the latent mortality index} \label{app:compareSARIMAX}

\added{Temperature is an exogenous driver that may influence the transmission and intensity of influenza activity, but not vice versa \citep{lowen2007influenza,lowen2014roles}. Both temperature and influenza activity can, in turn, affect the evolution of the latent mortality index $\kappa_{t,r}$ \citep{seklecka2017mortality}. This motivates a causal ordering among the three time-dependent processes and the use of Seasonal Autoregressive Integrated Moving Average models with eXogenous regressors (SARIMAX) to describe the stochastic components in our framework. A SARIMAX$(p,d,q)(P,D,Q)_s$ model for a time series $y_t$ is given by \citep{hyndman2008automatic}:
\begin{equation}\label{eqA:sarimax}
\begin{aligned} 
y_{t} &= \mathbf{x}_t^\top \boldsymbol{\beta} + u_t, \\
\Phi(B^s)\,\phi(B)\,(1-B^s)^D(1-B)^d \,u_t &= c + \Theta(B^s)\,\theta(B)\,\omega_t,
\end{aligned}
\end{equation}
where $\mathbf{x}_t$ denotes the vector of exogenous regressors and $\boldsymbol{\beta}$ the corresponding coefficients.}

\added{For influenza-like illness (ILI) and the latent mortality index $\kappa_{t,r}$, we compare specifications with and without external regressors. For ILI, we consider SARIMA$(1,0,1)(1,1,1)_{52}$ and SARIMAX$(1,0,1)(1,1,1)_{52}$ with temperature as an external regressor. For $\kappa_{t,r}$, we compare ARIMA$(0,1,1)$ with ARIMAX$(0,1,1)$ including annual averages of temperature and ILI as external regressors, in line with the annual time scale of $\kappa_{t,r}$. Temperature itself is modeled using SARIMA$(1,0,1)(1,1,1)_{52}$ without external regressors, as it is treated as exogenous. The choices for the non-seasonal, seasonal AR, differencing, and MA orders in the (S)ARIMA(X) processes are discussed in Section~\ref{subsec:performance}. Table~\ref{tabA:compSARIMAX} reports the BIC values for both comparisons per region.} 

\begin{table}[!ht]
\centering
\caption{\added{BIC values for ILI and $\kappa_{t,r}$: without vs. with external regressors. \label{tabA:compSARIMAX}}}
\adjustbox{width = \textwidth}{
\begin{tabular}{l@{\hspace{2em}}SS@{\hspace{2em}}SS}
\toprule
Region  & \text{ILI: SARIMA} & \text{ILI: SARIMAX} & \text{$\kappa_{t,r}$: ARIMA} & \text{$\kappa_{t,r}$: ARIMAX} \\
\midrule
Ile-de-France & \bfseries 1597.69 & 1604.77 & \bfseries -146.98 & -145.92 \\ 
Centre - Val de Loire & \bfseries 2482.96 & 2488.62 & \bfseries -132.76 & -131.10 \\ 
Bourgogne-Franche-Comté & \bfseries 2129.77 & 2136.43 & -132.02 & \bfseries -132.06 \\ 
Normandie & \bfseries 2138.77 & 2145.67 & \bfseries -149.22 & -144.05 \\ 
Hauts-de-France & \bfseries 2637.45 & 2644.33 & \bfseries -153.09 & -147.40 \\ 
Grand Est & \bfseries 1668.31 & 1675.44 & \bfseries -151.80 & -146.77 \\ 
Pays de la Loire & \bfseries 2620.25 & 2626.53 & \bfseries -138.13 & -137.22 \\ 
Bretagne & \bfseries 2171.63 & 2178.75 & \bfseries -144.67 & -139.72 \\ 
Nouvelle-Aquitaine & \bfseries 2012.10 & 2018.73 & \bfseries -145.88 & -141.88 \\ 
Occitanie & \bfseries 2020.19 & 2025.62 & \bfseries -146.21 & -140.26 \\ 
Auvergne-Rhône-Alpes & \bfseries 1835.47 & 1842.54 & \bfseries -150.21 & -143.92 \\ 
Provence-Alpes-Côte d’Azur & \bfseries 2181.45 & 2187.27 & \bfseries -109.93 & -104.17 \\ 
\bottomrule
\end{tabular}}
\end{table}

\added{For ILI, the SARIMA specification without temperature as an external regressor yields lower BIC values for all regions. This suggests that temperature does not improve the fit of influenza dynamics in our setting. For $\kappa_{t,r}$, the ARIMA specification without external regressors performs better for 11 out of 12 regions, with only Bourgogne-Franche-Comté showing a negligible improvement for ARIMAX. We therefore retain the SARIMA for ILI and the ARIMA for $\kappa_{t,r}$ without external regressors.}

\section{Goodness-of-fit under Poisson assumption} \label{app:gofPoisson}
A typical assumption in the actuarial mortality modeling literature is that the death counts follow a Poisson distribution. In the case of weekly deaths, this is often too restrictive because the Poisson distribution does not account for overdispersion. In this section, we demonstrate that, for our data and model, the Poisson specification is less adequate. We therefore assume instead that:
\begin{equation} \label{eqA:modeldist}
    D_{x,t,w,r} \sim \text{Poisson}\left( E_{x,t,w,r} \cdot \mu_{x,t,w,r} \right),
\end{equation}
with the same structure for the force of mortality as in Eq.~(3.2) of the main paper:
\begin{align} \label{eqA:modelstructure}
    \log \mu_{x,t,w,r} = \alpha_{x,r} + \beta_x \kappa_{t,r} + \gamma_x \lambda_{w,r} + \delta_x f_r^{(1)} \left( \text{Tavg}_{t,w,r}\right) + \epsilon_x f_r^{(2)} \left( \text{ILI}_{t,w,r}\right),
\end{align}
with $x \in \mathcal{X}$, $t\in \mathcal{T}$, $w \in \mathcal{W}$, and $r \in \mathcal{R}$.

Model calibration follows the same procedure as in Section 4 of the main paper: we maximize the log-likelihood using a Newton-type optimization method. The Poisson log-likelihood is given by
\begin{align*}
    l_{\text{poi}}(\boldsymbol{\theta}) = \displaystyle \sum_{x,t,w,r} \left[ d_{x,t,w,r} \, \log (\mu_{x,t,w,r} \cdot E_{x,t,w,r} ) - \mu_{x,t,w,r} \cdot E_{x,t,w,r} - \log \Gamma( d_{x,t,w,r} + 1)\right],
\end{align*}
and we apply the same spatial penalty structure as in the main paper. 

\begin{figure}[!ht]
    \centering
    \includegraphics[width=0.95\linewidth]{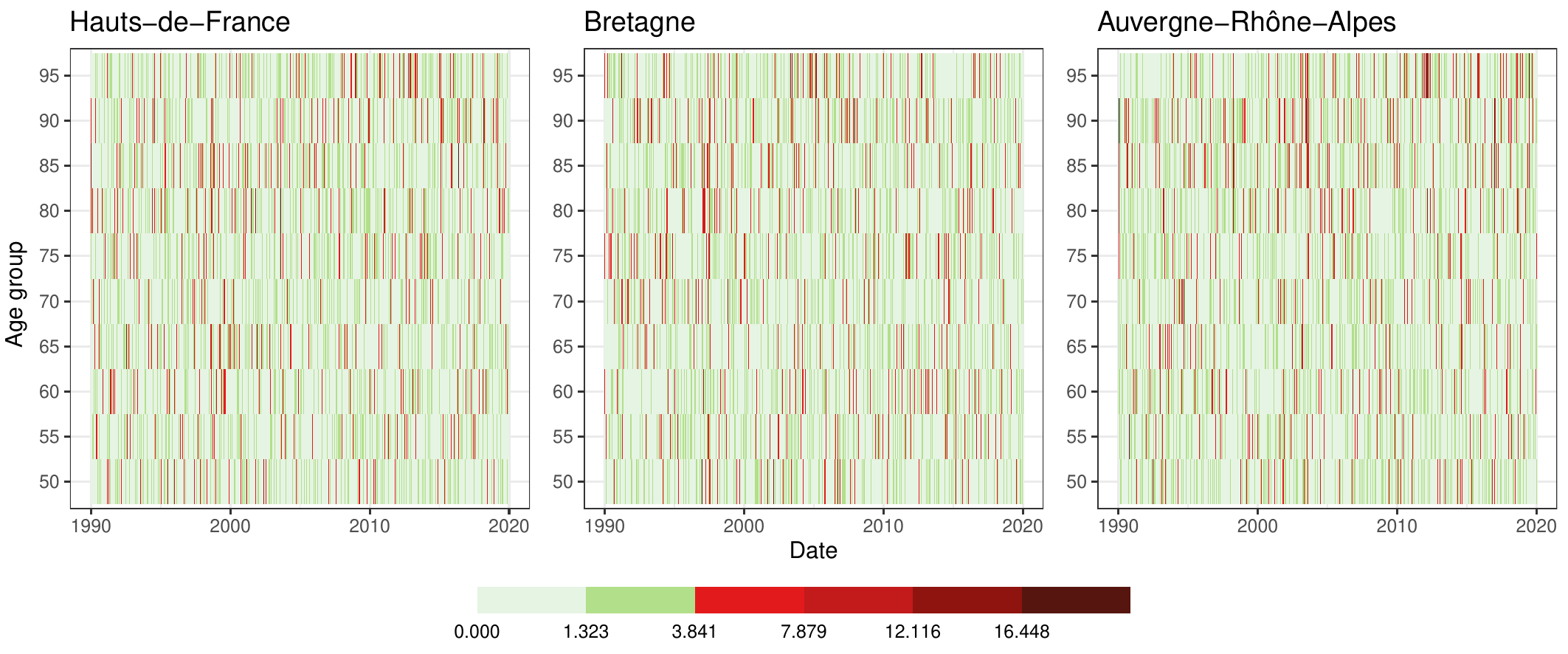}
    \caption{\added{Heat map of the squared Pearson residuals $\rho_{x,t,w,r}^2$ across age and time under the Poisson assumption. We show the results for Hauts-de-France (left), Bretagne (middle), and Auvergne-Rhône-Alpes (right). Green cells indicate areas with a good fit, while red and black cells correspond to areas with a poor fit ($\rho_{x,t,w,r}^2 > \chi_1^2$).}}
    \label{fig:pearsonPOI}
\end{figure}

Under the Poisson distributional assumption, Figure~\ref{fig:pearsonPOI} visualizes the squared Pearson residuals for Hauts-de-France, Bretagne, and Auvergne-Rhône-Alpes. Similarly to the main paper, the model specified in Eq.\eqref{eqA:modeldist} and~\eqref{eqA:modelstructure} is correctly specified if each $\rho_{x,t,w,r}^2$ is asymptotically distributed as a chi-square random variable with one degree of freedom ($\chi^2_1$). Consequently, for each region, around $5\%$ of the cells are expected to exceed the 95th percentile of the $\chi^2_1$-distribution, i.e., $\rho_{x,t,w,r}^2 > 3.841$. For the three regions shown, the observed proportions are \added{$6.09\%$, $6.17\%$, and $6.54\%$}, respectively. For the overall model’s goodness-of-fit, the total sum of the squared Pearson residuals equals \added{209\,990.4}, which is substantially larger than the corresponding 95th percentile of 187\,487.7. This suggests that, for our data and model, the negative binomial distribution, with overdispersion captured by $\phi_{x,r}$, provides a more appropriate fit.

\section{Robustness to temperature specification}
\label{app:temp_robustness}
\added{To evaluate the sensitivity of the results to the chosen temperature measure, we use four alternative weekly temperature measures, within the model specification of Eq.~\eqref{eq:modelstructure}, all constructed from daily mean temperatures: (i) the weekly mean (average of daily mean temperatures), (ii) the weekly maximum (maximum of daily mean temperatures), (iii) the weekly minimum (minimum of daily mean temperatures), and (iv) the maximum three-day moving average of daily mean temperatures within each week. For each specification, we estimate the proposed model over the same full sample period 1990–2019, following Eqs.~\eqref{eq:modeldist} and~\eqref{eq:modelstructure}, but replacing the temperature measure accordingly.}

\added{To ensure comparability across the four specifications, all other components of the model are kept unchanged. In particular, the DLNM structure, basis functions, lag lengths, and spatial penalization remain identical to the original specification using the weekly average temperature, as outlined in Section~\ref{subsec:specdlnmcase}. In this way, we make sure that any differences in model performance can be attributed solely to the choice of temperature variable.}

\added{Table~\ref{tabA:ll_temp_robustness} reports the resulting log-likelihood values. The specification based on weekly mean temperature yields the highest log-likelihood, and thus the best in-sample fit among the considered specifications, followed by the three-day maximum average, weekly minimum, and weekly maximum temperature. Overall, these results suggest that the use of weekly mean temperature appears the best choice to capture the overall temperature–mortality relationship in this setting.}

\begin{table}[!ht]
\centering
\caption{Log-likelihood values under alternative temperature specifications. \label{tabA:ll_temp_robustness}}
\adjustbox{width = 0.6\textwidth}{
\begin{tabular}{l@{\hspace{2em}}S}
\toprule
Temperature specification & {Log-likelihood} \\
\midrule
Mean temperature & \bfseries -586517.8 \\
Maximum temperature & -588303.8 \\
Minimum temperature & -588029.3 \\
3-day maximum average &  -587405.2\\
\bottomrule
\end{tabular}}
\end{table}

\added{While the above analysis based on log-likelihood provides a measure of overall in-sample fit over the full observation period, it primarily reflects average model performance and does not specifically focus on the model’s ability to capture mortality responses under extreme temperature conditions. Since the main interest of this robustness exercise is to assess whether alternative temperature definitions lead to different conclusions in the tails of the distribution, we also evaluate the predictive accuracy during extreme temperature periods.}

\added{To this end, we define four sets of extreme observations at the regional level based on empirical temperature percentiles over the period 1990--2019. Specifically, we consider (i) moderate heat events defined as weeks in which at least one of the four temperature measures exceeds its 95th percentile, (ii) severe heat events defined using the 97.5th percentile, (iii) moderate cold events defined using the 5th percentile, and (iv) severe cold events defined using the 2.5th percentile. All thresholds are computed separately for each region.}

\added{For each subset of extreme observations, we compute out-of-sample predictive performance for all four temperature specifications considered in the main analysis. In particular, for each region and age group, we evaluate the RMSE and MAE between observed deaths $d_{x,t,w,r}$ and fitted deaths $E_{x,t,w,r}\hat{\mu}{x,t,w,r}$, where $E{x,t,w,r}$ denotes the exposure and $\hat{\mu}_{x,t,w,r}$ the estimated force of mortality from the respective model. The resulting performance measures are then averaged across regions for each age group, as well as overall.}

\added{Tables~\ref{tab:extreme_temp_performance} report the results across all four extreme temperature regimes. Overall, and across most of the (older) age groups, the specification based on weekly mean temperature consistently leads to the lowest RMSE and MAE. The three alternative specifications based on maximum, minimum, and three-day maximum temperatures perform worse, with larger differences in the moderate and severe heat case, and relatively smaller differences in the moderate and severe cold case. Overall, using the weekly mean temperature seems to be the best option as it leads to the best overall in-sample fit and the most accurate predictive performance under extreme temperature conditions.}

\begin{table}[!ht]
\centering
\begin{adjustbox}{max width=0.475\textwidth}
\begin{tabular}{l *{4}{S} *{4}{S}}
\toprule 
&\multicolumn{8}{c}{\textbf{Moderate heat (95th percentile) }} \\
\toprule 
 & \multicolumn{4}{c}{\textbf{RMSE}} & \multicolumn{4}{c}{\textbf{MAE}} \\
\cmidrule(lr){2-5} \cmidrule(lr){6-9}
Age  
  & {Mean} & {Max} & {Min} & {3-day} 
  & {Mean} & {Max} & {Min} & {3-day} \\
\midrule
50--54 & 2.82 & 2.84 & \bfseries 2.81 & 2.83 & 2.26 & 2.28 & \bfseries 2.25 & 2.27 \\ 
55--59 & \bfseries 3.33 & 3.38 & 3.33 & 3.36 & 2.67 & 2.69 & \bfseries 2.67 & 2.68 \\ 
60--64 & \bfseries 3.72 & 3.78 & 3.73 & 3.76 & \bfseries 3.00 & 3.03 & 3.00 & 3.03 \\ 
65--69 & \bfseries 4.61 & 4.72 & 4.64 & 4.67 & \bfseries 3.62 & 3.66 & 3.64 & 3.64 \\ 
70--74 & \bfseries 5.61 & 5.93 & 5.64 & 5.84 & \bfseries 4.33 & 4.48 & 4.37 & 4.44 \\ 
75--79 & \bfseries 6.96 & 7.77 & 7.13 & 7.40 & \bfseries 5.30 & 5.57 & 5.36 & 5.46 \\ 
80--84 & \bfseries 9.54 & 11.19 & 9.80 & 10.62 & \bfseries 7.29 & 7.80 & 7.44 & 7.63 \\ 
85--89 & \bfseries 11.54 & 14.10 & 11.97 & 13.42 & \bfseries 8.80 & 9.71 & 8.95 & 9.52 \\ 
90--94 & \bfseries 11.30 & 14.99 & 11.65 & 13.99 & \bfseries 8.56 & 9.67 & 8.72 & 9.34 \\ 
95+ & 7.76 & 9.82 & \bfseries 7.68 & 9.48 & \bfseries 5.95 & 6.57 & 5.95 & 6.46 \\ 
\midrule
Overall & \bfseries 6.72 & 7.85 & 6.84 & 7.54 & \bfseries 5.18 & 5.55 & 5.23 & 5.45 \\ 
\bottomrule
\end{tabular}
\end{adjustbox} \hspace{0.3cm}
\begin{adjustbox}{max width=0.475\textwidth}
\begin{tabular}{l *{4}{S} *{4}{S}}
\toprule 
&\multicolumn{8}{c}{\textbf{Severe heat (97.5th percentile) }} \\
\toprule 
 & \multicolumn{4}{c}{\textbf{RMSE}} & \multicolumn{4}{c}{\textbf{MAE}} \\
\cmidrule(lr){2-5} \cmidrule(lr){6-9}
Age  
  & {Mean} & {Max} & {Min} & {3-day} 
  & {Mean} & {Max} & {Min} & {3-day} \\
\midrule
50--54 & 2.85 & 2.89 & \bfseries 2.84 & 2.88 & 2.28 & 2.31 & \bfseries 2.27 & 2.31 \\ 
55--59 & \bfseries 3.39 & 3.46 & 3.39 & 3.43 & 2.73 & 2.75 & \bfseries 2.72 & 2.74 \\ 
60--64 & \bfseries 3.73 & 3.83 & 3.76 & 3.79 & \bfseries 3.01 & 3.07 & 3.02 & 3.06 \\ 
65--69 & \bfseries 4.82 & 4.99 & 4.88 & 4.92 & \bfseries 3.79 & 3.86 & 3.83 & 3.84 \\ 
70--74 & \bfseries 5.76 & 6.28 & 5.83 & 6.12 & \bfseries 4.36 & 4.62 & 4.41 & 4.55 \\ 
75--79 & \bfseries 7.45 & 8.75 & 7.72 & 8.18 & \bfseries 5.52 & 6.02 & 5.61 & 5.83 \\ 
80--84 & \bfseries 10.32 & 12.82 & 10.65 & 12.01 & \bfseries 7.73 & 8.61 & 7.84 & 8.39 \\ 
85--89 & \bfseries 12.71 & 16.53 & 13.28 & 15.53 & \bfseries 9.62 & 11.25 & 9.80 & 10.88 \\ 
90--94 & \bfseries 12.51 & 18.00 & 13.06 & 16.50 & \bfseries 9.32 & 11.33 & 9.54 & 10.74 \\ 
95+ & 8.49 & 11.53 & \bfseries 8.25 &  11.08 & 6.43 & 7.48 & \bfseries 6.36 & 7.32 \\  
\midrule
Overall & \bfseries 7.20 & 8.91 & 7.37 & 8.45 & \bfseries 5.48 & 6.13 & 5.54 & 5.97 \\ 
\bottomrule
\end{tabular}
\end{adjustbox}

\vspace{0.5cm}

\begin{adjustbox}{max width=0.475\textwidth}
\begin{tabular}{l *{4}{S} *{4}{S}}
\toprule 
&\multicolumn{8}{c}{\textbf{Moderate cold (5th percentile) }} \\
\toprule 
 & \multicolumn{4}{c}{\textbf{RMSE}} & \multicolumn{4}{c}{\textbf{MAE}} \\
\cmidrule(lr){2-5} \cmidrule(lr){6-9}
Age  
  & {Mean} & {Max} & {Min} & {3-day} 
  & {Mean} & {Max} & {Min} & {3-day} \\
\midrule
50--54 & 2.98 & 2.98 & \bfseries 2.98 & 2.98 & 2.37 & 2.37 & \bfseries 2.37 & 2.37 \\ 
55--59 & 3.46 & 3.46 & \bfseries 3.46 & 3.46 & 2.75 & 2.74 & 2.75 & \bfseries 2.74 \\ 
60--64 & 4.00 & 4.00 & 4.00 & \bfseries 4.00 & 3.17 & \bfseries 3.16 & 3.17 & 3.16 \\ 
65--69 & \bfseries 4.85 & 4.85 & 4.85 & 4.85 & 3.86 & 3.87 & \bfseries 3.86 & 3.87 \\ 
70--74 & 5.70 & 5.73 & \bfseries 5.70 & 5.72 & 4.54 & 4.56 & \bfseries 4.54 & 4.56 \\ 
75--79 & \bfseries 7.22 & 7.23 & 7.29 & 7.22 & 5.67 & 5.67 & 5.73 & \bfseries 5.66 \\ 
80--84 & \bfseries 9.76 & 9.81 & 9.91 & 9.82 & \bfseries 7.72 & 7.74 & 7.82 & 7.76 \\ 
85--89 & \bfseries 11.94 & 12.13 & 12.11 & 12.05 & \bfseries 9.33 & 9.43 & 9.49 & 9.37 \\ 
90--94 & \bfseries 11.42 & 11.54 & 11.76 & 11.46 & \bfseries 8.82 & 8.90 & 9.05 & 8.87 \\ 
95+ & \bfseries 8.07 & 8.10 & 8.21 & 8.07 & \bfseries 6.20 & 6.24 & 6.28 & 6.21 \\
\midrule
Overall & \bfseries 6.94 & 6.98 & 7.03 & 6.96 & \bfseries 5.44 & 5.47 & 5.50 & 5.46 \\ 
\bottomrule
\end{tabular}
\end{adjustbox} \hspace{0.3cm}
\begin{adjustbox}{max width=0.475\textwidth}
\begin{tabular}{l *{4}{S} *{4}{S}}
\toprule 
&\multicolumn{8}{c}{\textbf{Severe cold (2.5th percentile) }} \\
\toprule 
 & \multicolumn{4}{c}{\textbf{RMSE}} & \multicolumn{4}{c}{\textbf{MAE}} \\
\cmidrule(lr){2-5} \cmidrule(lr){6-9}
Age  
  & {Mean} & {Max} & {Min} & {3-day} 
  & {Mean} & {Max} & {Min} & {3-day} \\
\midrule
50--54 & \bfseries2.91 & 2.91 & 2.91 & 2.91 & 2.32 & 2.33 & \bfseries 2.32 & 2.32 \\ 
55--59 & 3.39 & 3.39 & \bfseries 3.39 & 3.39 & 2.70 & 2.70 & \bfseries 2.70 & 2.70 \\ 
60--64 & \bfseries 4.08 & 4.09 & 4.09 & 4.08 & 3.24 & \bfseries 3.23 & 3.24 & 3.24 \\ 
65--69 & \bfseries 4.97 & 4.98 & 4.98 & 4.98 & \bfseries 3.92 & 3.93 & 3.93 & 3.93 \\ 
70--74 & 5.89 & 5.93 & \bfseries 5.86 & 5.91 & 4.66 & 4.69 & \bfseries 4.65 & 4.68 \\ 
75--79 & 7.13 & 7.14 & 7.23 & \bfseries 7.13 & 5.68 & 5.69 & 5.76 & \bfseries 5.67 \\ 
80--84 & \bfseries 9.88 & 10.01 & 10.06 & 10.00 & \bfseries 7.95 & 8.02 & 8.08 & 8.03 \\ 
85--89 & \bfseries 12.43 & 12.69 & 12.55 & 12.60 & \bfseries 9.72 & 9.85 & 9.90 & 9.80 \\ 
90--94 & \bfseries 11.88 & 12.07 & 12.29 & 11.99 & \bfseries 9.26 & 9.40 & 9.50 & 9.37 \\ 
95+ & \bfseries 7.96 & 7.99 & 8.11 & 7.97 & \bfseries 6.10 & 6.11 & 6.20 & 6.12 \\ 
\midrule
Overall & \bfseries 7.05 & 7.12 & 7.15 & 7.10 & \bfseries 5.56 & 5.59 & 5.63 & 5.59  \\ 
\bottomrule
\end{tabular}
\end{adjustbox}
\caption{\added{Predictive performance (RMSE and MAE) across age groups under alternative temperature specifications, evaluated during extreme temperature periods (moderate heat, severe heat, moderate cold, severe cold). The four columns correspond to models based on weekly mean temperature, weekly maximum temperature, weekly minimum temperature, and the three-day maximum average temperature, respectively. The best-performing specification within each age group is highlighted in bold. \label{tab:extreme_temp_performance}}}
\end{table}

\end{document}